\documentclass[aps,prd,notitlepage,amsmath,amssymb,preprintnumbers,nofootinbib,longbibliography,superscriptaddress,10pt]{revtex4-1}
\usepackage{graphicx} 
\usepackage[utf8]{inputenc}
\usepackage[rgb,table,xcdraw]{xcolor}
\graphicspath{{pictures/}}

\usepackage{multirow}
\usepackage{hyperref}

\usepackage{csquotes}

\newcommand{\SMDR}{{\tt SMDR}}
\newcommand{\Mathematica}{{\tt Mathematica}}
\newcommand{\RGBeta}{{\tt RGBeta}}
\def\MSbar{$\overline{\mathrm{MS}}$}
\def\onshell{\textit{on-shell}}
\newcommand{\GeV}{\ensuremath{\text{GeV}}}

\newcommand\pow[1]{\cdot10^{#1}}
\newcommand\mytextcircled[1]{\textcircled{\raisebox{-.9pt} #1}} 

\begin{document}

\title{On the renormalization-group analysis of the SM: \\ loops, uncertainties, and vacuum stability}
\author{A.V. {\sc Bednyakov}}\email{bednya@jinr.ru}
\affiliation{%
	Joint Institute for Nuclear Research, Joliot-Curie, 6, Dubna 141980, Russia
}
\author{A.S. {\sc Fedoruk}}\email{fedoruk.as@phystech.edu}
\affiliation{%
Moscow Institute of Physics and Technology (National Research University), Dolgoprudny, Moscow region, 141701 Russia
}
\author{D.I. {\sc Kazakov}}\email{kazakovd@theor.jinr.ru}
\affiliation{%
	Joint Institute for Nuclear Research, Joliot-Curie, 6, Dubna 141980, Russia
}

\begin{abstract}
	Renormalization-group equations (RGE) is one of the key tools in studying high-energy behavior of the Standard Model (SM).  
	We begin by reviewing one-loop RGE for the dimensionless couplings of the SM and proceed to the state-of-the-art results.
 	Our study focuses on the RGE solutions at different loop orders. We compare not only the standard (``diagonal'') loop counting 
	when one considers gauge, Yukawa, and scalar self-coupling beta functions at the same order 
	but also ``non-diagonal'' ones, inspired by the so-called Weyl consistency conditions.
	We discuss the initial conditions for RGE (``matching'') for different loop configurations and study 
	the uncertainties of running couplings both related to the limited precision of the experimental 
	input (``parametric'') and the missing high-order corrections (``theoretical''). 
	As an application of our analysis we also estimate the electroweak vacuum decay probability and study 
	how the uncertainties in the running parameters affect the latter. We argue that ``non-diagonal'' 
	beta functions, if coupled with a more consistent ``non-diagonal'' matching, lead to larger theoretical uncertainty 
	than ``diagonal'' ones.
\end{abstract}

\maketitle
\section{Introduction\label{sec:intro}}

The discovery of the Higgs boson at the Large Hadron Collider (LHC) \cite{ATLAS:2012yve,CMS:2012qbp} has revived interest in analysing of the behavior of the Standard Model coupling constants at high energies based on the renormalization group (RG) equations. In particular, special attention has been paid to the behavior of the Higgs boson self-coupling constant and the related problem of the stability of the electroweak vacuum. An important role in this analysis is played by the beta functions calculated in higher orders of perturbation theory (PT), which define the dependence of the running parameters on the characteristic energy scale.

The two-loop renormalization group (RG) equations\footnote{Here and below it is assumed that the dimensional regularization \cite{tHooft:1972tcz} and the \MSbar\ renormalization scheme \cite{Bardeen:1978yd} are used.} for the dimensionless coupling constants in an arbitrary renormalizable theory were first considered more than 40 years ago in Refs.~\cite{Machacek:1983tz,Machacek:1983fi,Machacek:1984zw}. Using the formalism of auxiliary ("dummy") fields \cite{Martin:1993zk}, in the early 2000s the corresponding equations were supplemented by beta functions for the dimensional parameters \cite{Luo:2002ti}.
Since then, the two-loop RG analysis of the SM and its extensions has become a de-facto standard: several publicly available computer programs \cite{Staub:2013tta,Sartore:2020gou,Litim:2020jvl,Deppisch:2020aoj,Thomsen:2021ncy} have appeared that allow one to automatically find RG equations in arbitrary renormalizable models in four space-time dimensions and solve them numerically.
It is worth noting that only 15 years later, inaccuracies in the two-loop formulas were discovered, related, firstly, to the possible mixing in the scalar sector \cite{Bednyakov:2018cmx,Schienbein:2018fsw}, and, secondly, to the need for a more careful approach to deriving RG for dimensional parameters \cite{Schienbein:2018fsw}.

The next important step in the history of renormalization group calculations in the SM occurred in early 2012,  coinciding with the discovery of the Higgs boson at the Large Hadron Collider: in 2012–2013, a complete set of three-loop renormalization group equations in the SM was obtained. The gauge couplings were considered in Refs.~\cite{Mihaila:2012fm,Mihaila:2012pz,Bednyakov:2012rb}. Three-loop beta functions for the Yukawa couplings in the SM were first obtained by the authors of Refs.~\cite{Chetyrkin:2012rz,Bednyakov:2012en}. It is worth noting that it is at this loop level that the Yukawa couplings receive a nontrivial contribution \cite{Chetyrkin:2012rz} from the diagrams containing the Dirac matrix $\gamma_5$ which is ill-defined in the dimensional regularization (see, for example, reviews \cite{Jegerlehner:2000dz}). The parameters of the Higgs potential were considered in Refs.~\cite{Chetyrkin:2013wya,Bednyakov:2013eba}.

The leading four-loop contributions to the RG equation for the strong coupling constant in the SM were obtained in Refs.~\cite{Bednyakov:2015ooa,Zoller:2015tha}. Unlike three loops, this order introduces for the first time the ambiguity of the na\"{i}ve interpretation of the matrix $\gamma_5$ (see, e.g., Ref.~\cite{Jegerlehner:2000dz}), which consists in simultaneously using its anticommutativity with other $\gamma_\mu$ and the cyclicity of traces involving it. This ambiguity was investigated in detail in Ref.~\cite{Bednyakov:2015ooa} and arguments were given in favor of the expression, which was subsequently and independently confirmed in Ref.~\cite{Poole:2019txl}. The reasoning behind the latter was based on the idea of a gradient renormalization-group flow and the use of the so-called \textit{Weyl consistency conditions} \cite{Osborn:1991gm}, which are a consequence of the Abelian nature of local scale transformations. The authors of Ref.~\cite{Poole:2019txl} related the ambiguities in the gauge sector arising at four loops to unambiguous three-loop contributions to the RG functions of the Yukawa couplings. Thanks to this, all beta functions of the SM gauge sector in the fourth order of PT \cite{Davies:2019onf} were found almost immediately. This approach turned out to be easily scalable to arbitrary renormalizable theories \cite{Bednyakov:2021qxa,Davies:2021mnc,Steudtner:2024teg}.
As for the four-loop beta functions for the Yukawa couplings and scalar self interactions in the SM, only partial results are known in the literature (see, e.g., Refs.~\cite{Martin:2015eia,Chetyrkin:2016ruf} and recent Ref.~\cite{Steudtner:2024teg}). It is also worth mentioning that RG computations in quantum chromodynamics (QCD) and theories without gauge interactions have advanced significantly further: six- \cite{Kompaniets:2017yct,Bednyakov:2021ojn} and even seven-loop \cite{Schnetz:2022nsc} results are available for the $\phi^4$ theory, the state-of-the-art five-loop QCD beta functions were obtained in Ref.~\cite{Baikov:2016tgj,Herzog:2017ohr,Luthe:2017ttc,Luthe:2017ttg}, while the four-loop RG functions \cite{Zerf:2017zqi} in the Gross-Neveu-Yukawa model were recently extended to five loops in Ref.~\cite{Gracey:2025aoj}.

This paper reviews current multi-loop RG calculations in the SM and largely follows the ideas presented in Ref.~\cite{Bednyakov:2015sca}. Our main goal is to analyze different orders in perturbation theory with special attentions to non-trivial loop ordering proposed in Ref.~\cite{Antipin:2013sga}.  
In Sec.~\ref{sec:SM_and_RGE} we introduce the SM Lagrangian and carry out a detailed study of one-loop RGEs aiming to illustrate possible RG flows in a (sub)space of the SM couplings.
To solve RGE for phenomenological relevant scenarios, one needs to specify initial conditions. The latter should be extracted from experiment by relating at a certain loop level the \MSbar\ parameters  at a fixed scale to a set of observables (``matching'' procedure\footnote{In the abuse of notation, we use ``matching'' to denote two related but different concepts: the SM parameters in the \MSbar\ scheme can be extracted from the measured experimental observables or from the known values of the \MSbar\ parameters of effective low-energy theory, in which heavy particles are decoupled.}). Two-loop RGE with one-loop matching were studies in the early 1990s \cite{Arason:1991ic}. Since then, many authors have contributed to the calculation of relevant experimental quantities at the two-loop level and beyond. There are several approaches on the market to carry out matching that we review in Section.~\ref{sec:running_par_ew}.   
We provide a detailed study how the initial conditions at the electroweak scale depend on the loop orders and analyze the corresponding uncertainties in Sec.~\ref{sec:initial_conditions}.
The vacuum stability issue is discussed in Sec.~\ref{sec:stability}, in which we analyze the running of the Higgs boson self-coupling and estimate the vacuum decay probability.
Our conclusions can be found in Sec.~\ref{sec:conclusions}. Some details regarding our numerical computations can be found in Appendix.~\ref{app:smdr}. For convenience, we also provide the expressions for the SM beta functions in Appendix.~\ref{app:bigbigformulas}.

\section{SM Lagrangian and ``running'' parameters\label{sec:SM_and_RGE}}
The SM Lagrangian can be schematically represented as follows:
\begin{align}
	\mathcal{L}_{\text{SM}} & = 
	\mathcal{L}_{\text{Gauge}}(g_s, g, g')  
	+ \mathcal{L}_{\text{Yukawa}}(y_u, y_d, y_l) 
	+ \mathcal{L}_{\text{Higgs}}(\lambda, m_\Phi^2) 
	,\label{eq:sm_lag_full}
\end{align}
The gauge $SU(3)_c\times SU(2)_W \times U(1)$ interactions of SM particles with vector bosons are given by $\mathcal{L}_{\text{Gauge}}$ and are characterized by three gauge constants $g_s$, $g$, and $g'$, respectively.
The interactions of SM fermions with the Higgs field are described by Yukawa matrices for up ($y_u$) and down ($y_d$) quarks, and for charged leptons\footnote{In general, one can add $y_\nu$ coupling to the right-handed neutrinos, but in this paper we neglect this quantity.} ($y_l$). The matrix nature of these parameters allows one to correctly describe the mixing and CP violation in the quark sector. However, here we restrict ourselves to diagonal Yukawa matrices and the case when only $y_f \equiv y_t, y_b$ for the top and bottom quarks are assumed to be nonzero. The Lagrangian $\mathcal{L}_{\text{Higgs}}$ describes the self-interactions of the Higgs doublet $\Phi$ and contains a single dimensional parameter $m_\Phi^2$:
\begin{align}
	- \mathcal{L}_{\text{Higgs}}(\lambda,m_\Phi^2) = V_{\text{Higgs}}(\lambda,m^2_\Phi) & = m^2_\Phi \Phi^\dagger \Phi + \lambda  (\Phi^\dagger \Phi)^2 
	\label{eq:Higgs_potential}\\
											    & \Rightarrow \frac{m^2_\Phi}{2} (v + h)^2 + \frac{\lambda}{4} (v+h)^4, \quad \Phi = (0, \phi/\sqrt2)^T_{\text{unitary gauge}}
	\label{eq:Higgs_potential_unitary}
\end{align}
where in eq.~\eqref{eq:Higgs_potential_unitary} the tree Higgs potential $V_{\text{Higgs}}$ is written in the unitary gauge in terms of the Higgs field $\phi = v + h$ with the vacuum expectation value $\langle \phi \rangle = v$. The latter occurs in the case $m^2_\Phi < 0$ when the ground state (the electroweak vacuum) breaks the original gauge symmetry to the electromagnetic subgroup $SU(2)_W \times U(1) \to U(1)_{\text{em}}$. Due to spontaneous symmetry breaking, elementary particles of the SM acquire masses (the Englert–Brout–Higgs–Guralnik–Hagen–Kibble mechanism \cite{Englert:1964et,Higgs:1964pj,Guralnik:1964eu}).

 When calculating quantum corrections, ultraviolet (UV) divergences arise, which are eliminated by means of local counterterms. In the framework of renormalizable quantum field theories, the counterterms follow the form of the original Lagrangian and allow one to determine the ``bare'' fields and coupling constants of the model. The latter depend on the parameters of the utilized regularization and diverge when the latter is removed. The choice of counterterms is ambiguous and determines the relationship between bare and renormalizable quantities (the renormalization scheme).

 The renormalized SM parameters in the \MSbar\ scheme, which we will henceforth denote collectively as\footnote{The $5/3$ coefficient corresponds to the $SU(5)$-normalization of the weak hypercharge.} $(16 \pi^2) a = \{5/3 \cdot g'^2,g^2,g_s^2, y^2_f, \lambda\}$ and $m_\Phi^2$, 
 implicitly depend on an arbitrary renormalization scale $\mu$ and this dependence is described by the renormalization group  equations of the form 
\begin{align}
	\dot{a} \equiv \frac{d a}{d t}  & =  \beta_a, \qquad a = \{ a^i \}, \qquad t = \ln \mu^2/\mu_0^2,
	\label{eq:RG_gen}
\end{align}
where the beta functions $\beta_a$ can be calculated using perturbation theory and define a vector field in the space of coupling constants $\{ a^i\}$. An important feature of calculating $\beta_a$ in the \MSbar\ scheme is the ability to neglect effects associated with spontaneous symmetry breaking, i.e., to use massless fields corresponding to $v=0$.
Here and in what follows we will discuss the dimensionless couplings $a$ that are used to construct the standard perturbation theory. It is assumed that the corresponding RG equation can also be written for the dimensional parameter $m_\Phi^2$. In the \MSbar\ scheme, $\beta_a$ do not depend on $m_\Phi^2$ and are functions of $a$ only.

The observables (cross-sections, decay probabilities, etc.) should not depend on the choice of $\mu$, and two sets of parameters $a(\mu_1)$ and $a(\mu_2)$ are considered equivalent (specifying the same physical situation) if solving \eqref{eq:RG_gen} with the initial condition $a(\mu_1)$ we obtain $a(\mu_2)$, and vice versa. Thus, the observables are assumed to depend on the integral curve of \eqref{eq:RG_gen}. Issues concerning the initial conditions for \eqref{eq:RG_gen} are discussed in Section~\ref{sec:running_par_ew}.

\subsection{One-loop RG equations}

For the sake of illustration\footnote{And inspired by the studies of Refs.~\cite{Kazakov:1975mc,Kazakov:1982xt}}, we discuss the properties of the renormalization group flow given by the leading one-loop contributions to $\beta_a$. We first consider the approximation in which only $x\equiv a_3 = g_s^2/(16\pi^2)$, $y\equiv a_t = y_t^2/(16 \pi^2)$, and $z \equiv a_\lambda = \lambda/(16 \pi^2)$ are nonzero. The corresponding RG equations are:
\begin{equation}\label{eq:dot00111}
\left\{
	\begin{array}{lcl}
	\dot{x} &= &-7x^2, \\[6pt]
	\dot{y} &= &y\left(\dfrac92 y-8x\right), \\[12pt]
	\dot{z} &= &6z\left(2z+y\right)-3y^2.  
\end{array}
\right.
\end{equation}

It is easy to see that the self-coupling $z$ does not contribute to the one-loop\footnote{As will be shown below, the Yukawa couplings can first appear in the gauge beta function starting from two loops. The contributions to the beta function of the Yukawa and gauge constants proportional to the scalar self-coupling first appear in two and three loops, respectively.} beta functions for the gauge ($x$) and Yukawa ($y$) couplings. At the same time, the Yukawa (and electroweak gauge) interactions of the Higgs fields lead to contributions to $\dot z$ that are not proportional to $z$.

The first equation describes the behavior of the strong coupling in QCD with six quark flavors \cite{Gross:1973id,Politzer:1973fx}. Its solution with the initial condition $x_0 = a_3(\mu_0)$ defined at $\mu=\mu_0$ is well-known
\begin{align}
    x (\mu) = \frac{x_0}{1 + 7 x_0 \ln \frac{\mu^2}{\mu^2_0}}
    \label{eq:as_sol_1L}
\end{align}
and corresponds to the asymptotic freedom ($a_3(\mu) \to 0$ when $\mu \to \infty$). In this case, $x = 0$ is a fixed point of the equation corresponding to free quarks and gluons.

\begin{figure}[h]
  \centering
  \includegraphics{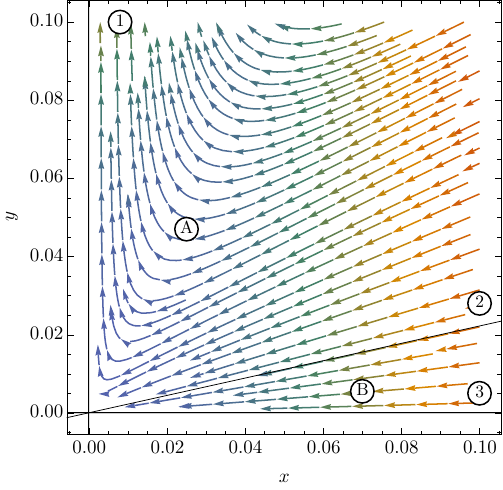}
\caption{Renormalization group flow in variables $x$ (gauge) and $y$ (yukawa) defined by the first two equations \eqref{eq:dot00111}. The flow is directed towards the UV region. Linear phase trajectories \mytextcircled{1} $x=0$, \mytextcircled{2} $y = 2/9 x$, and \mytextcircled{3} $y=0$ are marked dividing the phase plane into two regions \mytextcircled{A} and \mytextcircled{B}.}
\label{fig:00111x3y}
\end{figure}

The solution of the second equation with the boundary condition $y_0 \equiv a_t(\mu_0)$ for $x_0 \neq 0$ can be represented as
\begin{align}
y(\mu) = \frac{y_0 \left(x/x_0\right)^{\tfrac{8}{7}}  }{1 - \frac{9}{2}(y_0/x_0) \big( 1 - \left(x/x_0\right)^{\tfrac{1}{7}}\big)},
    \label{eq:at_sol_1L}
\end{align}
where $x=x(\mu)$ is given by expression \eqref{eq:as_sol_1L}. Taking the limit $x_0 \to 0$ in \eqref{eq:at_sol_1L}, we get the solution
\begin{align}
	y(\mu) = \frac{y_0}{1 - \frac{9}{2} y_0 \ln \frac{\mu^2}{\mu^2_0}}, \qquad \text{for} \quad x = x_0 = 0,
	\label{eq:at_zero_a3}
\end{align}
leading to a ``Landau pole'' in the ultraviolet region. Note also that when choosing $y_0 = 2/9 \cdot x_0$, the solution \eqref{eq:at_sol_1L} satisfies the relation
\begin{align}
	y = \frac{2}{9} x_0 \cdot \left(\frac{x}{x_0}\right)^{\tfrac{8}{7}-\tfrac{1}{7}} = \frac{2}{9} x.
	\label{eq:sep_y29x}
\end{align}
Figure~\ref{fig:00111x3y} shows the RG flow in the $(x,y)$ plane defined by the first two equations \eqref{eq:dot00111}. Separatrices are indicated and represented by straight lines $x=0$, $y=0$ and $y=2/9\cdot x$ passing through the origin (the Gaussian fixed point). The separatrices divide the phase plane into two regions. As can be seen from the figure, if we choose the initial point in the region \textcircled {\raisebox{-0.9pt} {A}}, in the ultraviolet asymptotics it will ``be attracted'' to the curve $x=0$ corresponding to the solution \eqref{eq:at_zero_a3} for which $y$ increases and eventually reaches the Landau pole. If the initial point is located in the region \textcircled {\raisebox{-0.9pt} {B}}, then the renormalization-group flow attracts it to the separatrix $y=0$ corresponding to the solution \eqref{eq:as_sol_1L}. The separatrix $y=2/9 \cdot x$ is unstable, since any small deviations of the initial point from it will eventually lead the phase curve to the attraction region either of $x=0$ or $y=0$.

\begin{figure}[h]
  \centering
	  \includegraphics[width = 0.9\textwidth]{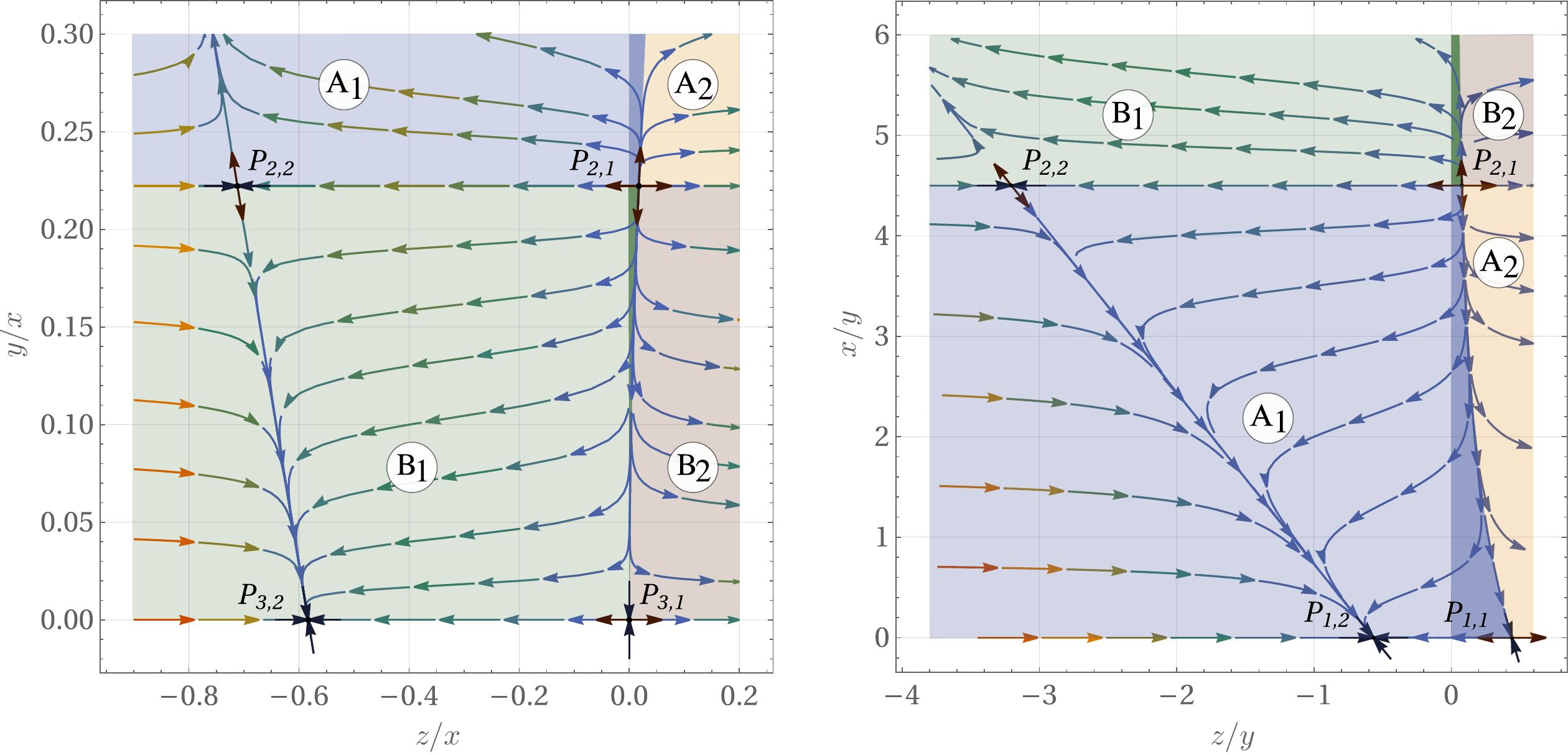}  
\caption{The direction field for equations \eqref{eq:RGE_YZ} for $x\neq 0$ (left) and \eqref{eq:RGE_XZ} for $y\neq0$ (right).
The singular points $P_{i,j}$ corresponding to the rays passing through the origin in the phase space $(x,y,z)$ of the system \eqref{eq:dot00111} are indicated. The projection of the rays $P_{1,i}$ onto the plane $(x,y)$ is the separatrix \mytextcircled{1} in Fig.~\ref{fig:00111x3y}, while $P_{2,i}$ give rise to the separatrix \mytextcircled{2}, and $P_{3,i}$ to  the separatrix \mytextcircled{3} . Regions $\text{A}_{1,2}$ correspond to the condition $9 y>2 x$ \mytextcircled{A}, and $\text{B}_{1,2}$ to the condition $9 y< 2 x$ \textcircled {\raisebox{-0.9pt} {B}}, (see Fig.~\ref{fig:00111x3y}). \label{fig:1L_XZ_YZ}}
\end{figure}

\begin{figure}[t]
  \centering
  \begin{tabular}{ccc}
	  \includegraphics{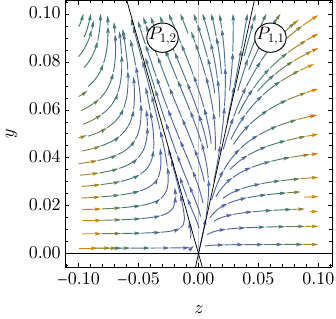} &
	  \includegraphics{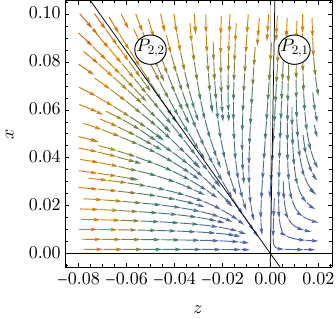} &
	  \includegraphics{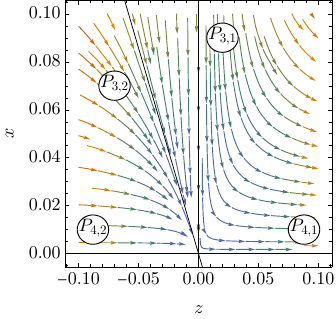}
  \end{tabular}
 
  \caption{Phase curves in the plane $x=0$ (left), $y=2/9 x$ (center), and $y=0$ (right) (see the corresponding separatrices \mytextcircled{1}, \mytextcircled{2}, and \mytextcircled{3} in Fig.~\ref{fig:00111x3y}) for the one-loop equations \eqref{eq:dot00111}. The rays $P_{i,j}$ corresponding to the singular points in Fig.~\ref{fig:1L_XZ_YZ} and defined by equations \eqref{eq:XZ_P_1_1} - \eqref{eq:YZ_P_3_2} are indicated. \label{fig:sep_combo}}
\end{figure}

Let us now consider all three equations. Since the direction field (in the one-loop approximation) \eqref{eq:dot00111} is homogeneous, to study the neighborhood of a Gaussian fixed point $x=y=z=0$ it is convenient to use the coordinates $(x,Y,Z)$ where $Y\equiv y/x$ and $Z\equiv z/x$ ($x \neq 0)$ or $(X,y,\tilde Z)$, where $X\equiv x/y$ and $\tilde Z \equiv z/y$ ($y\neq 0)$.
Using \eqref{eq:dot00111}, one can obtain RG equations for $Y,Z$
\begin{align}
	\dot Y & = \frac{Y}{2} \left(9 Y - 2\right) \cdot x,
	\nonumber\\
	\dot Z & = \left(12 Z^2  + Z (6 Y + 7) - 3 Y^2\right) \cdot x, 
	\label{eq:RGE_YZ}
\end{align}
and $X,\tilde Z$:
\begin{align}
	\dot X & =  X( 2 X - 9) \cdot \frac{y}{2} ,
	\nonumber\\
	\dot {\tilde Z} & = (24 \tilde Z^2 + \tilde Z (3 + 16 X) - 6 )\cdot \frac{y}{2} . 
	\label{eq:RGE_XZ}
\end{align}
The RHSs of \eqref{eq:RGE_YZ} and \eqref{eq:RGE_XZ} define the direction fields (see Fig.~\ref{fig:1L_XZ_YZ}). The singular points of these fields correspond to the linear phase curves \eqref{eq:dot00111} in the original space $(x,y,z)$ passing through the origin (Fig.~\ref{fig:sep_combo})
\begin{align}
	P_{1,1}: && X^* = 0 ,&&& \tilde Z^* = -\frac{1}{16}\left(1 - \sqrt{65}\right) && \rightarrow && x = 0 , && z = -\frac{y}{16} \left(1 - \sqrt{65}\right), \label{eq:XZ_P_1_1} \\
	P_{1,2}: && X^* = 0 ,&&& \tilde Z^* = -\frac{1}{16}\left(1 + \sqrt{65}\right) && \rightarrow && x = 0 , && z = -\frac{y}{16} \left(1 + \sqrt{65}\right), \label{eq:XZ_P_1_2} \\
	P_{2,1}: && Y^* = \frac{2}{9},&&& Z^* = -\frac{1}{72}\left(25 - \sqrt{689}\right) && \rightarrow && y = \frac{2}{9} x, && z = -\frac{x}{72} \left(25 - \sqrt{689}\right), \label{eq:YZ_P_2_1} \\
	P_{2,2}: && Y^* = \frac{2}{9},&&& Z^* = -\frac{1}{72}\left(25 + \sqrt{689}\right) && \rightarrow && y = \frac{2}{9} x, && z = -\frac{x}{72} \left(25 + \sqrt{689}\right), \label{eq:YZ_P_2_2} \\
	P_{3,1}: && Y^* = 0, &&& Z^* = 0 && \rightarrow && y = 0, && z=0, \label{eq:YZ_P_3_1}\\
	P_{3,2}: && Y^* = 0,&&& Z^* = -\frac{7}{12} && \rightarrow && y = 0, && z = -\frac{7x }{12}. \label{eq:YZ_P_3_2}  
\end{align}
Fixed points $P_{1,j}$ correspond to the separatrix \mytextcircled{1} ($x=0$), $P_{2,j}$ to the separatrix \mytextcircled{2} ($y=2/9 x$), and $P_{3,j}$ -- to the separatrix \mytextcircled{3} ($y=0$) in Fig.~\ref{fig:00111x3y}.
In addition to the lines indicated, there is a solution \eqref{eq:dot00111} for which $x=y=0$ and $z\neq0$:
\begin{align}
    z(\mu) = \frac{z_0}{1 - 12 z_0 \ln \frac{\mu^2}{\mu^2_0} }. 
    \label{eq:alam_at_zero_a3_ay}
\end{align}
We will denote the corresponding phase curves as $P_{4,1}$ ($z>0$ and increases indefinitely in the UV) and $P_{4,2}$ ($z<0$ and approaches zero from below in the UV). Note that the trajectory $x=z=0$ lying entirely on the $y$ axis does not exist, since on this axis $\dot z = -3 y^2 \neq 0$.

The ray $P_{1,1}$ ($P_{1,2}$) lies in the $(y,z)$ plane and corresponds to the synchronous growth of $y$ and $z$ $(-z)$ in the UV region. On the rays $P_{2,i}$ and $P_{3,i}$ all coupling constants tend to zero in the UV region along the corresponding directions, and along $P_{2,1}$ and $P_{3,1}$ the self-interaction constant of the Higgs boson $z$ approaches zero from above, and along $P_{2,2}$ and $P_{3,2}$ -- from below. Thus, along all $P_{i,2}$ the values of $z$ are negative and, consequently, lead to a potential \eqref{eq:Higgs_potential} that is unbounded from below, i.e., to possible instability of the vacuum.

The lines described by stable nodes $P_{1,2}$ and $P_{3,2}$ attract trajectories starting in the regions\footnote{Note that both $\text{A}_1$ and $\text{B}_1$ have subregions corresponding to $z>0$ (marked darker in Fig.~\ref{fig:1L_XZ_YZ}). However, in both cases the RG flow ``leads'' $z$ to the negative region.} $\text{A}_1$ ($y>2/9 x$) and $\text{B}_1$ ($y<2/9 x$), respectively (see Fig.~\ref{fig:1L_XZ_YZ}).  
Along the ray $P_{1,2}$ we have $x=0$ and $y$ increases without bound, while $z$ is negative and decreases without bound proportionally to $y$ \eqref{eq:XZ_P_1_2}. 
Along the ray $P_{3,2}$ the Yukawa constant $y=0$, while $z$ tends to zero from below according to \eqref{eq:YZ_P_3_2}. 
If the initial points are in the regions $\text{A}_2$ ($y$ increases in the UV) and $\text{B}_2$ ($y$ tends to zero in the UV), then $z$ is positive and increases without bound.

The points $P_{1,1}$, $P_{2,2}$ and $P_{3,1}$ are saddle points for the direction fields \eqref{eq:RGE_YZ} and \eqref{eq:RGE_XZ}. Therefore, in the space $(x,y,z)$ there exist two-dimensional surfaces bounded by the ray $P_{2,1}$ and the ray of the corresponding saddle. These surfaces are defined by the lines in Fig~\ref{fig:1L_XZ_YZ} connecting the unstable node $P_{2,1}$ with one of the saddle points\footnote{For example, for the line connecting $P_{2,1}$ and $P_{2,2}$, in the $(x,y,z)$ space we have a surface defined parametrically as $x = t, y=2/9 t, z = Z t$, where $t>0$, and $-(\sqrt{689}+25) < 72 Z < (\sqrt{689}-25)$.}. Despite the fact that the RG flow does not take the point beyond these surfaces, the latter are unstable and small deviations of the initial value lead to the phase trajectory being attracted to either $P_{1,2}$ or to $P_{3,2}$.

The analysis performed above demonstrates the general properties of the one-loop renormalization group flow in the three-dimensional $(x,y,z) \equiv (a_3, a_t,a_\lambda)$ subspace of all SM parameters. Obviously, the above picture is modified when taking into account higher orders of perturbation theory in the expressions for $\beta_a$, as well as the (mutual) impact of the remaining coupling constants.

\begin{figure}[t]
  \centering
  \begin{tabular}{ccc}
	  \includegraphics{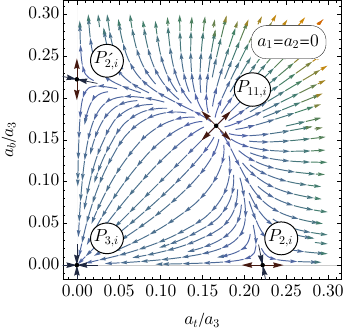} &
	  \includegraphics{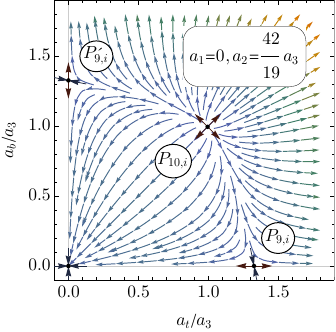} &
	  \includegraphics{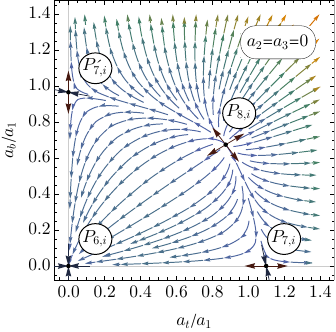}
  \end{tabular}
      \caption{Direction fields for the cases $a_1=a_2=0$ (left), $a_1=0$, $a_2=42/19 a_3$ (middle) and $a_2=a_3=0$ (right), demonstrating the stability of some rays \eqref{eq:rays_1L_SM} under varying the initial conditions for the Yukawa coupling constants of the $t$- and $b$-quarks. The pictures are qualitatively similar: $a_t=a_b=0$ is a UV stable node and $a_t=a_b\neq0$ is unstable. The rays on which one of the Yukawa constants vanishes turn out to be saddles. Note the absence of a linear trajectory passing through $a_\lambda = 0$ for the case $a_1=0$, $a_2=42/19 a_3$. This means that only when $a_1 = a_2 = a_3 = 0$ does the negative initial $a_\lambda$ tend from below to zero in the UV. Otherwise, $a_\lambda \to + \infty$.} \label{fig:sep_ab_at}
\end{figure}

For example, taking into account the renormalization group equations for the electroweak gauge couplings $a_1, a_2$ and the Yukawa coupling $a_b$ of the bottom quark
\begin{equation}\label{eq:dot111111}
  \left\{
  \begin{array}{l}
    \dot{a}_1 = \dfrac{41 a_1^2}{10}, \\[6pt]
    \dot{a}_2 = -\dfrac{19}{6} a_2^2,\\[6pt]
    \dot{a}_3 = -7 a_3^2,\\[6pt]
    \dot{a}_t = a_t \left(\dfrac{9 a_t}{2}-8 a_3-\dfrac{17 a_1}{20}-\dfrac{9 a_2}{4}+\dfrac{3 a_b}{2}\right),\\[6pt]
    \dot{a}_b = a_b \left(\dfrac{9 a_b}{2}-8 a_3-\dfrac{a_1}{4}-\dfrac{9 a_2}{4}+\dfrac{3 a_t}{2}\right),\\[6pt]
    \dot{a}_\lambda = 12 a_\lambda^2 +6 (a_t + a_b) a_\lambda 
    -\dfrac{9 a_\lambda}{10} (a_1 + 5 a_2)
    -3 (a_t^2 + a_b^2) + \dfrac{27 a_1^2}{400}+\dfrac{9 a_1 a_2}{40} +\dfrac{9 a_2^2}{16},
  \end{array}
  \right.
\end{equation}
we obtain additional linear trajectories passing through the Gaussian fixed point. For convenience, we list here all the rays that arise:
\begin{align}
	P^{(')}_{1,i}:&& &a_1=a_2=a_3=0, &&a_b=0, && a_{\lambda }= -\frac{1}{16} \left(1\mp\sqrt{65}\right) a_t && \dot a_t = \frac{9}{2} a_t^2 , \\
	P^{(')}_{2,i}:&& &a_1=a_2=0 && a_b=0, a_t= \frac{2 a_3}{9}, && a_{\lambda }= -\frac{1}{72} \left(25\mp\sqrt{689}\right) a_3, && \dot a_3 = -7 a_3^2 , \\
	P_{3,1}:&& &a_1=a_2=0 && a_t=a_b=0, && a_{\lambda }=0, && \dot a_3 = - 7 a_3^2  , \\
	P_{3,2}:&&&a_1=a_2=0&& a_t=a_b=0, && a_{\lambda }= -\frac{7 a_3}{12} && \dot a_3 = - 7 a_3^2  , \\
	P_{4,i}:&&&a_1=a_2=a_3=0,  && a_t=a_b=0, && && \dot a_\lambda = 12 a_\lambda^2, \\
        P_{5,1}:&&		&a_1=a_2=a_3=0, && a_b= a_t, &&  a_{\lambda }= \frac{a_t}{2}, && \dot a_t = 6 a_t^2 , \\
	P_{5,2}:&&	&a_1=a_2=a_3=0, && a_b= a_t, && a_{\lambda }= -a_t,  && \dot a_t = 6 a_t^2, \\
	P_{6,i}:&&&a_2=a_3=0, && a_t=a_b=0, && a_{\lambda }= \frac{1}{120} \left(25\pm4 \sqrt{34}\right) a_1, && \dot a_1 = \frac{41}{10} a_1^2, \\
	P_{7,i}:&&&a_2=a_3=0 && a_b=0,  a_t= \frac{11 a_1}{10}, &&  a_{\lambda }= -\frac{1}{120} \left(8\mp\sqrt{4339}\right) a_1, && \dot a_1 = \frac{41}{10} a_1^2 , \\
	P'_{7,i}:&&&a_2=a_3=0&& a_t=0,  a_b= \frac{29 a_1}{30}, &&  a_{\lambda }= -\frac{1}{120} \left(4\mp\sqrt{3299}\right) a_1, && \dot a_1 = \frac{41}{10} a_1^2 
		, \\
	P_{8,i}:&&&a_2=a_3=0, && a_b= \frac{27 a_1}{40}, a_t = \frac{7}{8} a_1 && a_{\lambda }= -\frac{1}{240} \left(43\mp\sqrt{19111}\right) a_1, && \dot a_1 = \frac{41}{10} a_1^2, \\
	P^{(')}_{9,i}:&& &a_1=0,a_2= \frac{42 a_3}{19}, && a_b = 0, a_t= \frac{227 a_3}{171}, && a_{\lambda }= -\frac{1}{684} \left(143\mp\sqrt{119402}\right) a_3, && \dot a_3 = - 7 a_3^2 , \\
	P_{10,i}:&&&a_1=0, a_2= \frac{42}{19}a_3, && a_b= a_t = \frac{227}{228} a_3, && a_{\lambda }= -\frac{1}{456} \left(171\mp\sqrt{84671}\right) a_3, && \dot a_3 = - 7 a_3^2,  \\
	P_{11,i}:&&&a_1=a_2=0, &&  a_b= a_t = \frac{a_3}{6}, && a_{\lambda }= -\frac{1}{24} \left(9\mp\sqrt{89}\right) a_3, && \dot a_3 = - 7 a_3^2. 
	\label{eq:rays_1L_SM}
\end{align}
Note that since $a_2$ and $a_3$ appear in the one-loop equations \eqref{eq:dot111111} for $\dot{a}_t$ with the same coefficients as for $\dot{a}_b$, and $a_t$ and $a_b$ appear in $\dot{a}_\lambda$ also with the same coefficients, then the rays $P'_{i,1-2}$ ($i=1,2,9$) lying in the plane $a_1=0$ (the only gauge interaction distinguishing between up and down quarks is ``switched off''), for which $a_t = 0$ and $a_b \neq 0$, can be obtained from the corresponding $P_{i,1-2}$ by simple substitution $a_t \leftrightarrow a_b$. It is also easy to see that if the Abelian gauge coupling on the ray is nonzero $a_1 \neq 0$, then the non-Abelian ones are necessarily zero $a_2 = a_3 = 0$ ($P^{(')}_{6-8,i}$) and, conversely, if $a_3\neq0$ on the curve, then $a_1 = 0$ (rays $P^{(')}_{2-3,i}$, $P^{(')}_{9-11,i}$). This is obviously related to different UV asymptotics of the Abelian and non-Abelian gauge couplings in the SM. In Fig.~\ref{fig:sep_ab_at}, as an illustration of the stability of linear trajectories with nonzero values of the Yukawa constants, the corresponding direction fields are presented for the cases when $a_1=a_2=0$ (left) and $a_2=a_3 = 0$ (right).

As in the three-dimensional case, we can consider the system of equations for the gauge and Yukawa coupling constants separately. Rays $P_{i,j}$ with the same first index have the same projection onto the corresponding subspace defined by $a_\lambda = 0$. Considering the renormalization group flow in the plane formed by this projection and the $a_\lambda$ axis, the phase trajectories for the cases $P_{5-8,i}$ will qualitatively coincide with those shown in Fig.~\ref{fig:sep_combo} (left), and for cases $P_{9-11,i}$ --- with those shown in Fig.~\ref{fig:sep_combo} (middle). In this case, the phase curves on these surfaces are ``repelled'' from $P_{i,1}$ and ``attracted'' to $P_{i,2}$. On the rays $P_{i,1}$ the self-interaction coupling is positive, and on the rays $P_{i,2}$ (except $P_{6,2}$) it is negative. Along $P_{5-8,i}$ the nonzero coupling constants increase indefinitely (in absolute value) in the UV asymptotics (Landau pole), and along $P_{9-11,i}$ they tend to zero (asymptotic freedom).

As a demonstration of the analysis of the behavior of curves using separatrices, let us consider, for example, a realistic point \cite{Bednyakov:2015sca} 
\begin{equation}
\begin{array}{l}
    a_1^e=1.35555\pow{-3},\, a_2^e=2.66006\pow{-3},\, a_3^e=8.60063\pow{-3}, \\
    a_t^e=5.53811\pow{-3},\, a_b^e=1.84306\pow{-6},\, a_\lambda^e=8.05123\pow{-4},
\end{array}
  \label{eq:fixed_initial_condition}
\end{equation}
corresponding to the running parameters at the benchmark scale $\mu_0 = 173.22$ \GeV.
\begin{figure}[t]
  \centering
  \includegraphics[width = 15cm]{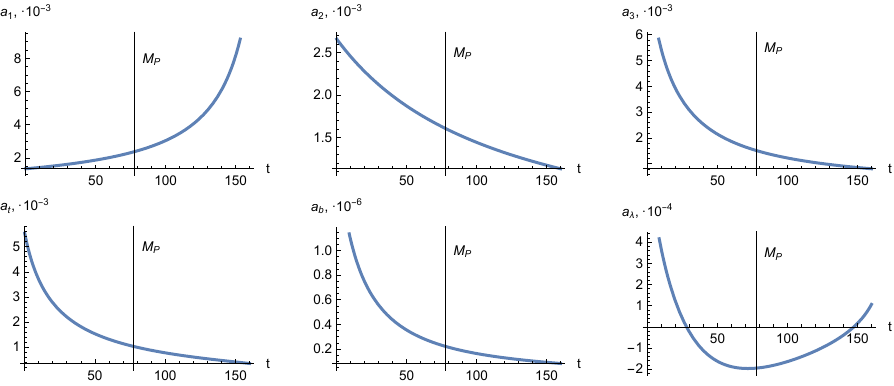}
  \caption{Solution of the one-loop RGE \eqref{eq:dot111111} through \eqref{eq:fixed_initial_condition}. The line $M_P$ denotes $t$, corresponding to the Planck scale ($1.2\cdot10^{19}~\GeV$).}\label{fig:11111example}
\end{figure}
The solution of the differential equations passing through the point \eqref{eq:fixed_initial_condition}
has the form given in Fig.~\ref{fig:11111example}, in which we also consider values of $t$ far beyond the Planck scale $M_P = 1.2 \cdot 10^{19}$ \GeV. 

Let us try to explain this behavior. Initially $a_1$ is small, and $a_t$ is significantly nonzero, so we can expect the behavior of the solution to follow the trajectories of the curves near $P_{2,i}$ or $P_{9,i}$ and far from $P_{3,i}$ or $P_{4,i}$ or $P_{6,i}$ (where $a_t\equiv0$).  However, the value 
$$
\dfrac{a_2^e}{a_3^e} \approx 0.3093
$$
signalizes that we shouldn't analyze the initial behavior of the curve in terms of $P_{9,i}$ ($a_2/a_3 = 42/19 \approx2$) but rather in terms of $P_{2,i}$ ($a_2 \equiv 0$). 
When projected to $(a_t, a_3)$ and $(a_\lambda,a_3)$, the initial condition
is ``above'' $P_{2,i}$ and ``between'' $P_{2,1}$ and $P_{2,2}$, respectively, where we define 
$$
\begin{array}{c}
     \text{``above'' } P_{2,i} = \left( \dfrac{a_t}{a_3} > \dfrac29  \right) \\
     \text{``between'' } P_{2,1} \text{ and } P_{2,2} = \left( -\dfrac{1}{72}\left(25 + \sqrt{689}\right)  < \dfrac{a_\lambda}{a_3} < -\dfrac{1}{72}\left(25 - \sqrt{689}\right)  \right),
\end{array}$$
since
$$
\begin{array}{l}
  \dfrac{a_t^e}{a_3^e}-\dfrac29\approx 0.425 > 0 \\[8pt]
  \dfrac{a_\lambda^e}{a_3^e}+\dfrac{1}{72}\left(25 - \sqrt{689}\right)\approx-0.076 < 0. 
\end{array}
$$

\begin{figure}
    \centering
    \includegraphics{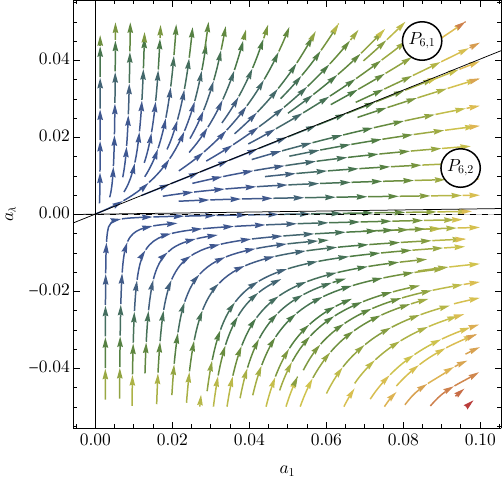}
    \caption{Phase space at $a_2=a_3=a_t=a_b=0$ for $(a_1,a_\lambda)$. Solid lines represent separatrices, dashed - axes (namely abscissa $a_\lambda=0$).}
    \label{fig:P6}
\end{figure}

As can be seen from Figs.~\ref{fig:00111x3y}~and~\ref{fig:sep_combo} (center), this region resembles $\text{A}_1$ from Fig.~\ref{fig:1L_XZ_YZ} and is characterized by $a_3$ tending to zero in the UV with an initial decrease of $a_t$ and an initial decrease of \emph{positive} $a_\lambda$ until the latter eventually becomes negative attracted by $P_{1,2}$.
The problem comes with the later synchronous growth of $a_t$ and $|a_\lambda|$ ($P_{1,2}$). But after a certain point when $a_1$ increases due to the positive one-loop beta function
($\dot a_1 \propto a_1^{\,2}$) reaching $a_1 > a_2,a_3,a_t$, the trajectory approaches $P_{6,2}$ (a separatrix, where only $a_1$ and $a_\lambda$ are nonzero) with positive $a_\lambda$ in the deep UV  from ``below'' in the $(a_1,a_\lambda)$ projection (see Fig.~\ref{fig:P6}): 
$$
\text{``under'' } P_{6,2} = \left( \dfrac{a_\lambda}{a_1} < \dfrac{25-4 \sqrt{34}}{120} \approx0.014  \right).
$$
This can be explained by the fact that growing $a_1$ gives a larger negative contribution to the one-loop $a_t$ beta function, and a (leading) positive contribution to $\beta_\lambda$. As a consequence, we see the absence of a later growth of $a_t$ (curve stays near $P_{6,2}$ as it is the separatrix of attraction, on which $a_t\equiv0$). Indeed, at $\mu = M_P$ we have
\begin{align*}
	a_1(M_P) & =2.87\pow{-3},\, &a_2(M_P)& =1.48\pow{-3},\, &a_3(M_P) & =\phantom{-}1.28\pow{-3}, \\
	a_t(M_P) & =0.82\pow{-3},\, &a_b(M_P)& =1.74\pow{-7},\, &a_\lambda(M_P)& =-1.73\pow{-4}
\end{align*}
with $a_1$ being the largest coupling and $a_\lambda/a_1 < 0$. As the curve approaches $P_{6,2}$, it results in the asymptotics $a_\lambda(\mu) \approx 0.014\: a_1(\mu)$ (see Fig.\ref{fig:P6}), which gives us a rapid growth of positive $a_\lambda$ (given the rapid growth of $a_1$) after $M_P$.

Such behavior of integral curves is, in general, typical for curves from a large neighborhood of \eqref{eq:fixed_initial_condition}. What really changes is the minimum value of $a_\lambda(\mu)$, which can be negative, as for our choice \eqref{eq:fixed_initial_condition}, resulting in the instability of the vacuum (see section \ref{sec:stability}). It is important to note, however, that some effects, including the unlimited increase of $a_1$ and $a_\lambda$ (Landau pole), arise already at scales larger than the Planck scale (to the right of the line $M_P$ in Fig. \ref{fig:11111example}), and can not be trusted without accounting for quantum-gravity effects.

\subsection{Higher orders and constraints on beta functions}

To characterize the PT orders used in solving \eqref{eq:RG_gen}, it is convenient to specify a triple of numbers $L_g,L_y,L_\lambda$ (or $L_g L_y L_\lambda$) corresponding to the $L_g$-loop beta function for the gauge constants $g,g',g_s$, the $L_y$-loop beta function for the Yukawa constants $y_t, y_b$, and the $L_\lambda$-loop beta function for the Higgs field self-interaction constant $\lambda$. 

An independent loop counting for gauge, Yukawa and Higgs coupling constants is of interest in connection with questions concerning the general properties of the renormalization group flow in four-dimensional theories (see, for example, Refs.~\cite{Osborn:1991gm,Jack:1990eb,Jack:2013sha}). In particular, it is assumed that there exists a so-called $\tilde A$-function that depends on the dimensionless running parameters $a^i$ and varies monotonically along the RG flow ($a$-theorem \cite{Cardy:1988cwa}). It can be shown that this function satisfies the Osborn equation \cite{Osborn:1991gm}
\begin{align}
	\frac{\partial \tilde A}{\partial a^i} \equiv \beta_{i} = \chi_{ij} \beta^{i} \Rightarrow \frac{d \tilde A}{d t} = \beta^i \chi_{ij} \beta^j, \qquad \beta^i = \chi^{ij} \frac{\partial \tilde A}{\partial a^j},  
	\label{eq:osborn_eq}
\end{align}
where $\chi_{ij} = \chi_{ij}(a)$ is a ``metric''\footnote{Positive definiteness of $\chi_{ij}$ is equivalent to the monotonicity condition of $A$ along the RG flow ($a$-theorem).} in the space of coupling constants $a^i$, $\chi^{ij}$ is the corresponding inverse matrix, and $\beta^i$ is the beta function of $a^i$, which enters into \eqref{eq:RG_gen}. The expression for $\chi_{ij}$ can be found in perturbation theory \cite{Osborn:1991gm}. The diagonal contributions to $\chi_{ij}$ arise starting from one, two, and three loops for the case of gauge, Yukawa, and Higgs couplings, respectively. For the SM, one can write 
\begin{align}
	\chi_{ij} & = \text{diag}\left( \frac{1}{a_1^2}, \frac{3}{a^2_2}, \frac{8}{a^2_3}, \frac{2}{a_{t}}, \frac{2}{a_{b}}, 4\right) + \ldots,
		  & \chi^{ij} & = \text{diag}\left( a_1^2, \frac{a^2_2}{3}, \frac{a^2_3}{8}, \frac{a_t}{2}, \frac{a_b}{2}, \frac{1}{4}\right) + \ldots.
	\label{eq:metric_leading}
\end{align}
Since the mixed derivatives $\partial_i \partial_j \tilde A = \partial_j \partial_i \tilde A$ are equal to each other (the integrability condition), equations \eqref{eq:osborn_eq} impose restrictions \cite{Poole:2019kcm} on the possible form of beta functions:
\begin{align}
   \frac{\partial \beta_i}{\partial a^j} =  \frac{\partial \beta_j}{\partial a^i} 
   \Rightarrow \frac{\partial }{\partial a^j} (\chi_{ik} \beta^k) =  \frac{\partial }{\partial a^i} (\chi_{jk} \beta^k). 
   \label{eq:wcc}
\end{align}
As can be easily seen from the explicit form of \eqref{eq:metric_leading}, $\chi_{ij}$ ``mixes'' the PT orders when going from $\beta^i$ to $\beta_i$ and allows us to link ``cross contributions'' to the beta functions of different (types of) coupling constants in different loops \cite{Antipin:2013sga,Poole:2019kcm}. For example, taking \eqref{eq:metric_leading} into account allows us to relate one-loop contributions of the form $a_\lambda a_{1}$ to the beta function of the Higgs coupling $\beta_\lambda$ with a three-loop contribution of the form $a_{1}^3 a_\lambda$ to the beta function of the gauge constant $a_{1}$. Similarly, the one-loop contribution $a_t a_\lambda$ to $\beta_\lambda$ uniquely determines the two-loop contribution of the form $a_t^2 a_\lambda$ to the beta function of the Yukawa constant of the top quark $\beta_t$.

If the $\tilde A$-function is known up to $\mathcal{O}(a^L)$ terms, then to satisfy the constraints \eqref{eq:wcc} it is necessary to know the beta functions of the gauge, Yukawa, and Higgs constants up to $L,(L-1),(L-2)$ loops, respectively. For example, the expression \cite{Freedman:1998rd,Jack:2014pua} 
\begin{align}
	\tilde A &  = \tilde A_{0} + \left[ \frac{41}{10} a_1 - \frac{19}{2} a_2 - 56 a_3 \right]  + \ldots
	\label{eq:A_2l}
\end{align}
together with \eqref{eq:metric_leading} completely determines the one-loop contributions to the SM gauge beta functions. It was this observation that formed the basis for the argument that a consistent analysis of the SM vacuum stability should be carried out in the (``non-diagonal'') $321$ loop configuration \cite{Antipin:2013sga} rather than in the ``diagonal'' $333$ one known at that time.
In Sec.~\ref{sec:stability} we discuss the vacuum stability issue and compare the results obtained for different loop configurations. In particular, we extend the study of \cite{Antipin:2013sga} to the 432 case.

Before proceeding to this analysis, let us discuss the issue related to the choice of the integral curve for the RG equations \eqref{eq:RG_gen} that corresponds to the physical situation observed in the experiment. Namely, let us briefly describe the problem of finding, within the SM framework, the boundary conditions $a(\mu_0)$ at some scale $\mu_0$. Note that the approach consistent with the formal PT assumes that the initial conditions for the RG equations \eqref{eq:RG_gen} also depend on the order in which $\beta_a$ are calculated.

\section{``Running'' parameters of the SM at the electorweak scale\label{sec:running_par_ew}} 

To determine the values of the parameters entering into the Lagrangian \eqref{eq:sm_lag_full}, it is necessary to consider a set of observable (in experiments) quantities, the minimum number of which is equal to the number of model parameters, and find (``extract'') the latter from a comparison (``matching'') of theoretical predictions with experimental data.

The following quantities are usually used as such a set of (pseudo-)observables (in the RG study of the SM): the fine structure constant $\alpha$, the strong coupling constant in quantum chromodynamics\footnote{Defined in the \MSbar\ scheme within the effective field theory (EFT) with five quark flavors.} $\alpha_s^{(5)}$, the $Z$-boson mass $M_Z$, the Higgs boson mass $M_h$, the SM fermion masses (quarks and leptons) $M_f$, and the Fermi constant $G_F$. These quantities are related to the SM parameters via the formulas of the form
\begin{align}
	(4 \pi) \alpha & = \frac{g^2 g'^2}{g^2 + g'^2}[1 + \bar \delta_\alpha], & (4\pi) \alpha^{(5)}_s & = g_s^2 [1 + \bar\delta_s] &  G_F & = \frac{1}{\sqrt{2} v^2} [1 + \bar \delta r], \nonumber\\
	M_Z^2 & = \frac{(g^2 + g'^2) v^2}{4}[1 + \bar \delta_Z], & M_h^2 & = 2 \lambda v^2 [1 + \bar \delta_h] ,  & M_f & = \frac{y_f v}{\sqrt 2} [1 + \bar\delta_f].
	\label{eq:SM_rel_pars_obs}
\end{align}
Here $\bar\delta_i$ correspond to the corrections calculated in perturbation theory. In relations \eqref{eq:SM_rel_pars_obs} the vacuum expectation value of the Higgs field in the leading order can be defined\footnote{See the discussion below} in terms of the Lagrangian parameters as $v = \sqrt{|m_\Phi^2|/\lambda}$. Neglecting perturbative corrections ($\bar\delta_i=0$), it is easy to solve equations \eqref{eq:SM_rel_pars_obs} in the tree-level approximation:
\begin{align}
			       a_1 & = \frac{5}{3}\frac{M_Z^2 G_F}{4\sqrt{2}\pi^2} \left[ 1 -  \sqrt{1 - \frac{2 \sqrt{2} \pi \alpha}{G_F M_Z^2}}\,\right],
			      & a_2& = \frac{M_Z^2 G_F}{4\sqrt{2}\pi^2} \left[ 1 + \sqrt{1 - \frac{2 \sqrt{2} \pi \alpha }{G_F M_Z^2}}\,\right],
			      & a_3 & = \frac{\alpha_s}{4\pi}, 
\\
	a_t& 
	= \frac{M^2_t G_F}{4 \sqrt{2} \pi^2},
	    & a_\lambda& 
	    =  \frac{M_h^2 G_F}{16\pi^2\sqrt 2}, 
	    & v^2 & =  \frac{1}{\sqrt{2} G_F},
	    \label{eq:matching_tree}
\end{align}
and thereby express all the predictions of the SM written as functions of the SM Lagrangian parameters in terms of a specified set of measured quantities (PDG2024) \cite{ParticleDataGroup:2024cfk}:
\begin{align}\label{eq:pdg2024_1}
  \alpha &= 7.2973525693(11) \pow{-3},
	 & \alpha_S^{(5)}(M_Z) &= 0.1180(9)
	 & G_{F} &= 1.1663788(6) \pow{-5}\, \GeV^{-2}, 
\nonumber\\
      M_Z &= 91.1880(20)\, \GeV, 
	  &M_h &= 125.20(11)\, \GeV, 
	&M_t &= 172.57(29)\, \GeV. 
\end{align}
In the leading order we obtain:
\begin{align}
	a_1 = 1.2 \cdot 10^{-3},
	&&
	a_2 = 2.7 \cdot 10^{-3},
	&&
	a_3 = 9.4 \cdot 10^{-3},
	&&
	a_t = 6.2 \cdot 10^{-3},
	&&
	a_\lambda = 8.1 \cdot 10^{-4},
	&& 
	v = 246~\GeV.
	\label{eq:tree_level_matching}
\end{align}

Note that the choice of pseudo-observables is ambiguous and the specific set \eqref{eq:pdg2024_1} is largely determined by the precision of the corresponding measurements. For example, as an alternative to the electromagnetic constant $\alpha$ (or, for example, $G_F$), the mass of the $W$-boson can be used:
\begin{align}
	M_W^2 = \frac{g^2 v^2}{4} [ 1 + \bar \delta_W],
	\qquad 
      M_W = 80.3692(133)\, \GeV, 
	\label{eq:MW_as_exp_par}
\end{align}
which, however, is experimentally obtained with worse precision than $M_Z$.

Taking into account quantum corrections $\bar \delta_i$ can modify and significantly complicate expressions \eqref{eq:matching_tree}. This raises the questions in what scheme, at what scale, and with what accuracy the numerical values of the parameters \eqref{eq:tree_level_matching} are found. In particular, it is possible to \emph{define} the scheme so that all (or some) $\bar\delta_i = 0$ (so-called \onshell\ renormalization and its analogues). However, this is not always convenient when studying the model at significantly different energy scales.

As already mentioned above, we assume that the calculations are carried out with dimensional regularization and the \MSbar\ scheme, in which $\bar \delta_i = \bar \delta_i (a(\mu), \mu)$ are functions of the running coupling constants $a(\mu)$ and the renormalization scale $\mu$. Thus, formulas \eqref{eq:SM_rel_pars_obs} can be interpreted as \emph{conversion formulas} relating the \MSbar\ parameters (right-hand sides) to the parameters defined in the \onshell\ scheme (left-hand sides). In all relations \eqref{eq:SM_rel_pars_obs}, except for the case of a running strong coupling constant $\alpha_s^{(5)} \equiv \alpha_s^{(5)}(\mu)$, the left-hand sides (physical masses, fine structure constant, and Fermi constant) are assumed to be independent of the arbitrary renormalization scale $\mu$.

In the case of strong coupling, the ``measured quantity'' is the value of $\alpha_s^{(5)}(M_Z)$ defined in the $SU(3)_c \times U(1)_{em}$ gauge theory (QCD $\times$ QED) with five quark flavors (see for more details the review~\cite{ParticleDataGroup:2024cfk} and references therein). In this effective field theory, the relatively heavy top quark, Higgs boson, and electroweak gauge $W$- and $Z$-bosons are formally absent (``decoupled''), and the EFT is therefore considered as a low-energy approximation of the SM. When studying processes in which the characteristic scales are much smaller than the masses of the indicated particles, the latter cannot be produced as physical states. The leading quantum corrections from the corresponding virtual particles\footnote{Calculated within the SM in the \MSbar-like schemes} can be ``absorbed'' into the renormalization of the running coupling and are taken into account by going from $\alpha_s(\mu) \equiv g^2_s(\mu)/(4\pi)$ defined in the SM to $\alpha_s^{(5)}(\mu)$ defined in five-flavor QCD $\times$ QED. In this case, $\bar \delta_s$ in \eqref{eq:SM_rel_pars_obs} are determined by ``matching'' the full SM and five-flavor effective theory (see Refs.~\cite{Chetyrkin:2005ia,Schroder:2005hy} for four-loop relations in pure QCD, and \cite{Bednyakov:2014fua,Martin:2018yow} for leading electroweak effects). This boils down to choosing such running parameters of the latter that the EFT predictions reproduce the SM predictions at low energies, when corrections  suppressed by powers of heavy masses can be neglected. This approach is justified by the fact that due to the confinement phenomenon, the strong constant $\alpha_s$ cannot be extracted by considering, for example, a nonrelativistic process at small momentum transfer involving physical quarks. Therefore, it is convenient to use the running coupling $\alpha_s^{(5)}(\mu = Q)$ as a QCD parameter in processes involving strong interactions and characterized by the scale $M_b \lesssim Q\lesssim M_Z$.
Obviously, the electromagnetic coupling $\alpha^{(5)}$ can be defined in the same way.

A similar situation is observed in the case of quark masses (see also the discussion in Ref.~\cite{Huang:2020hdv}). Although perturbation theory can relate the ``physical'' (pole) quark masses to the real part of the pole in the propagator of the corresponding field, this quantity is poorly defined outside PT and cannot be extracted from experiments with an accuracy better than $\mathcal{O}(\Lambda_{\text{QCD}})$. Therefore, running quark masses in the \MSbar\ scheme are often used as QCD parameters. For example, instead of the pole mass of the $b$-quark $M_b = 4.78(6)\, \GeV$, it is customary to specify the quantity $\mu_b \equiv m_b(m_b)$ that is a solution to the implicit equation \cite{ParticleDataGroup:2024cfk}
 \begin{align}
     \mu_b = m_b(\mu_b), \qquad \mu_b = 4.183(7)\, \GeV
     \label{eq:mub}
 \end{align}
 with $m_b(\mu)$ being the running $b$-quark mass in five-flavor QCD $\times$ QED (see, e.g., Ref.~\cite{Herren:2017osy} and references therein for the QCD part, and Ref.~\cite{Bednyakov:2016onn,Martin:2018yow} for the electroweak part). In the case of the $t$-quark, there is also a problem of self-consistent definition of the corresponding mass. However, it is believed that since the top quark decays faster than it can form a bound state, its pole mass can be treated as a physically measurable quantity\footnote{There are also alternative definitions of quark masses (see, for example, the discussion in Ref.~\cite{ParticleDataGroup:2024cfk}).}.

As mentioned earlier, when considering the theoretical predictions for observables in terms of the running model parameters $a(\mu)$, the implicit dependence of the latter on $\mu$ compensates for the explicit \emph{logarithmic} dependence of the corrections $\bar\delta_i = \bar \delta_i(\mu)$ on this quantity. As for the ``decoupling'' relations between the \MSbar\ parameters in the effective (QCD) and more fundamental (SM) theories (such as $\alpha_s^{(5)}(\mu)$ and $g^2_s(\mu)/(4\pi)$, respectively), there remains on the right-hand side of \eqref{eq:SM_rel_pars_obs} a residual dependence on $\mu$, which is exactly given by the RG equations for the left-hand sides.

Thus, the knowledge of the left-hand sides of \eqref{eq:SM_rel_pars_obs} allows one to find \emph{numerical values} of the running parameters in the \MSbar\ scheme at an arbitrary scale without solving the RG equations. However, in practice, the quantum corrections $\bar\delta_i$ and the beta functions $\beta_a$ are calculated in perturbation theory and usually only a finite number of terms of the corresponding series is known. Therefore, the choice of $\mu$ turns out to be important, since it (at least partially) allows one to avoid the occurrence in the approximate expression for quantum corrections of potentially large powers of logarithms $\ln(M^2/\mu^2)$, where $M^2$ depends on the specific observable and can correspond to the mass of some particle or the square of some momentum. Using the RG equations for the running parameters of the SM allows one to rearrange the PT series by summing such logarithms to all orders. A typical situation is when the knowledge of the $L$-th order for $\beta_a$ and the $(L-1)$-loop expression for some observable allows one, by going from the scale $\mu = \mu_0$ to the scale $\mu=M$ using \eqref{eq:RG_gen}, to sum up to all orders both the leading $a(\mu_0)^n \ln^n(M^2/\mu_0^2)$ and the subleading logarithmic (N${}^{L-1}$LL) contributions of the form $a(\mu_0)^{L+n-1} \ln^n (M^2/\mu_0^2)$.

It is expected that the more terms of the PT series for $\bar\delta_i$ and $\beta_a$ are known, the less sensitive the matching procedure is to the choice of the scale $\mu$. This fact provides us with the standard method for determining the theoretical uncertainty\footnote{Estimating theoretical uncertainties is a non-trivial task, especially in the electroweak sector (see, e.g., Refs.~\cite{Freitas:2016sty,Freitas:2019bre}). Here we follow a simple criterion based on the variation of the renormalization scale.} when extracting running parameters $a(\mu_0)$ at some fixed scale $\mu_0$: we estimate the difference between the values of $a(\mu_0)$ found by inverting formulas \eqref{eq:SM_rel_pars_obs} directly at $\mu=\mu_0$ and the values of $a(\mu_0)$ obtained by solving the RG equations \eqref{eq:RG_gen} with the boundary conditions found from \eqref{eq:SM_rel_pars_obs} at $\acute\mu=2\mu_0$ and $\acute\mu=\mu_0/2$.

Note that when taking into account the radiative corrections in \eqref{eq:SM_rel_pars_obs}, which relate the fine structure constant $\alpha$ to the running SM parameters at the scale $\mu_0\geq M_Z$, it is necessary to compute the contribution of hadrons, which is a non-trivial task (see, e.g., the review in Ref.~\cite{ParticleDataGroup:2024cfk}). The results of calculations by different groups are usually represented as an average value \cite{ParticleDataGroup:2024cfk}
\begin{align}
	 \Delta\alpha_{had}^{(5)}(M_Z) &= 0.02783(6), 
	 \label{eq:alpha_had}
\end{align}
where the correction is assumed to be added to the running fine structure constant $\alpha^{(5)}(\mu)$ considered together with the strong coupling constant as parameters of the effective QED $\times$ QCD theory with five quark flavors at the scale $\mu = M_Z$. This value is used together with \eqref{eq:pdg2024_1} as input data for the RG analysis in the SM.

Finally, it should be mentioned a peculiarity  related to the different treatment of the vacuum expectation value of the Higgs field. Even in the \MSbar\ scheme, the particle masses cannot be (completely) neglected, and the ground state, in the vicinity of which one uses perturbation theory, is determined by the minimum of the \emph{effective}, not tree, potential $V_{\text{eff}}(\phi)$
\begin{align}
   V_{\text{eff}}(\phi) = V_{\text{Higgs}}(\phi) + \Delta V (\phi),
   \label{eq:eff_pot}
\end{align}  
with $\Delta V(\phi)$ corresponding to quantum corrections. Minimization of $V_{\text{eff}}(\phi)$ \eqref{eq:eff_pot}
\begin{align}
	\langle h\rangle = \left.\frac{d V_{\text{eff}}}{d \phi}\right|_{\phi=v} = m^2_\Phi v + \lambda v^3 + \left.\frac{ d \Delta V(\phi)}{d \phi} \right|_{\phi = v} = 0
			\label{eq:zero_tad}
\end{align}
in the leading order results in the expression $v_{\text{tree}} = \sqrt{|m_\Phi^2|/\lambda}$, already mentioned above. For the consistent application of PT, it is necessary to ensure that the vacuum expectation value of the quantum field is zero $\langle h \rangle=0$ in each order under consideration. There are different ways to implement the condition \eqref{eq:zero_tad} in PT \cite{Actis:2006ra}, 
which affects the interpretation of $v$ in the formulas \eqref{eq:SM_rel_pars_obs}. For example, one can assume that $v=v_{\text{tree}} + \Delta v$ and threat the latter as perturbation, i.e., include it as contributions to $\bar \delta_i$. 
In this case, $v_{\text{tree}}$ enters into the definition of the tree-level masses of quantum fields, and  
$\Delta v$ is found implicitly as a function of $v_{\text{tree}}$ in each order of PT from the condition of cancellation of ``tadpole'' diagrams appearing due to loop contributions, \eqref{eq:zero_tad} (a scheme of Fleischer and Jegerlehner or FJ-scheme \cite{Fleischer:1980ub}):
\begin{align}
	2 \lambda v_\text{tree}^2 \Delta v \left[ 1 + \frac{3}{2} \frac{\Delta v}{v_\text{tree}} + \frac{1}{2} \frac{\Delta v^2}{v_\text{tree}^2}\right] + \left.\frac{d \Delta V}{ d \phi} \right|_{\phi = v_{\text{{tree}} + \Delta v}} = 0
\end{align}
The advantage here is the explicit gauge independence of the observables expressed in terms of the Lagrangian parameters \eqref{eq:sm_lag_full}. The disadvantage is that the expressions for $\Delta v$ may numerically contain significant contributions proportional to powers of $M_t^4/(M_W^2 M_h^2) \sim 9$, where $M_t$, $M_W$, and $M_h$ are the masses of the top quark, $W$ boson, and Higgs boson, respectively. The approach is used by many authors (see, e.g., Refs.~\cite{Fleischer:1980ub,Jegerlehner:2001fb,Jegerlehner:2002em,Kniehl:2014yia,Kniehl:2015nwa}) and is implemented in the \texttt{mr} program \cite{Kniehl:2016enc}.

An alternative is the ``tadpole-free'' scheme \cite{Bohm:1986rj,Denner:1991kt} widely used in his calculations by S. Martin~\cite{Martin:2014cxa,Martin:2015rea,Martin:2015lxa,Martin:2016xsp,Martin:2018yow}. The perturbation theory in this approach is constructed under the assumption that $v$ satisfying \eqref{eq:zero_tad} enters into the quadratic SM Lagrangian and the leading terms $(\bar \delta_i = 0)$ in \eqref{eq:SM_rel_pars_obs} involve $v$ instead of $v_\text{tree}$. 
Given the condition \eqref{eq:zero_tad}, at each order of PT one can express $v^2_{\text{tree}}$ (or, equivalently, $m^2_{\Phi}$) as a function of $v$ \cite{Martin:2019lqd}
\begin{align}
	\frac{m_\Phi^2}{\lambda} \equiv v^2_{\text{tree}}  = v^2 - \frac{1}{\lambda v } \left.\frac{d \Delta V}{d \phi}\right|_{\phi = v}.
		\label{eq:Martin_scheme}
\end{align}
It is well known that, unlike the value of the potential $V_{\text{eff}}(\phi = v)$ at extrema, the value of $V_{\text{eff}}(\phi)$ for an arbitrary $\phi$, as well as the vacuum expectation value $v$ itself, are gauge-dependent quantities \cite{Jackiw:1974cv}. Therefore, proving the gauge independence of observables expressed in the ``tadpole-free'' scheme via $v$ is a non-trivial task and is based on the Nielsen identities \cite{Nielsen:1975fs,Nielsen:2014spa}. Nevertheless, the advantage here is that the large contributions $\Delta v$ arising in the FJ scheme at high PT orders are effectively taken into account by switching from $v_{\text{tree}}$ to $v$ in the Lagrangian. The most common gauge used in calculations within the ``tadpole-free'' scheme is the Landau one. In this paper, we routinely use the \SMDR\ program \cite{Martin:2019lqd} , which implements this approach. It is worth noting that a hybrid approach, which uses the advantages of both treatments, is also discussed in the literature (see, for example, Ref.~\cite{Dittmaier:2022maf}).

In the next section we consider the RG equations \eqref{eq:RG_gen} for the most important SM parameters and investigate the corresponding theoretical and experimental uncertainties in determining the running SM parameters in different orders of PT at the electroweak scale. We take the latter to be $\mu_0 = 173.22 $ GeV. We analyze not only the state-of-the-art expressions for the RG equations and relations of the form \eqref{eq:SM_rel_pars_obs} but also the cases when we neglect some loop contributions.

\section{Initial conditions for RG equations and different PT orders\label{sec:initial_conditions}}

Let us now try to investigate the dependence of the initial conditions for the renormalization group equations (i.e. the values of $a_1, a_2, a_3, a_t, a_b, a_\lambda$ at the chosen fixed scale $\mu_0 = 173.22~\GeV$) on the loop configuration. To carry out matching of the \onshell\ parameters to the \MSbar\ variables, the modified version of the \SMDR\ library was used. As the initial set of pseudo-observables, following the choice in \SMDR, we will use the PDG2024 \cite{ParticleDataGroup:2024cfk} data given in eq.~\eqref{eq:pdg2024_1},\eqref{eq:mub} and \eqref{eq:alpha_had}.

To denote perturbation theory orders, we also used the same notation $L_g L_y L_\lambda$, assuming that the initial values of the gauge, Yukawa, and Higgs constants are in the approximation of one loop smaller ($L-1$). As already mentioned, this approach allows one to solve the renormalization group equations in the $L_g L_y L_\lambda$ order self-consistently.

By default, the \SMDR\ code uses the most accurate known expressions for relations of the form \eqref{eq:SM_rel_pars_obs} and the RG equations \cite{Martin:2019lqd}. Unlike the author \cite{Martin:2019lqd} of \SMDR, who also studied the scale dependence of the running SM parameters in different PT orders, we focus on various combinations of $L_gL_yL_\lambda$ and try to consistently and quantitatively estimate the theoretical uncertainties for each of the considered cases. The modified version of \SMDR\  allows one to iteratively solve equations of the form \eqref{eq:SM_rel_pars_obs} in different PT orders\footnote{See Appendix~\ref{app:smdr} for details and subtleties} and find boundary conditions for the corresponding RG equations. The latter were obtained independently by means of the \RGBeta\ package \cite{Thomsen:2021ncy} (see Appendix~\ref{app:bigbigformulas} for explicit expressions) and solved using \Mathematica.

\subsection{Theoretical uncertainties of \MSbar\ parameters\label{sec:th_unc_couplings}}

\begin{figure}[!ht]
\centering
\begin{tabular}{cc}
	\includegraphics[width = 0.475\textwidth]{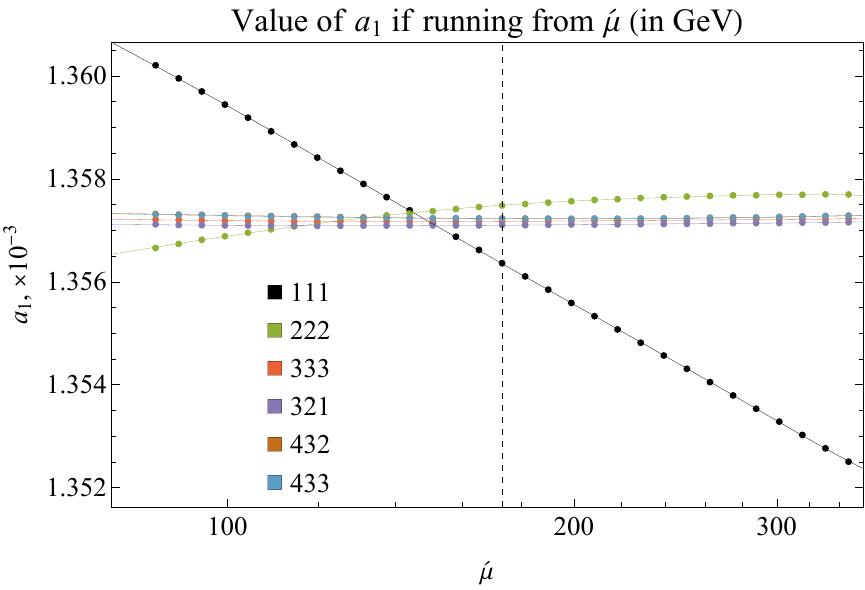} &
	\includegraphics[width = 0.485\textwidth]{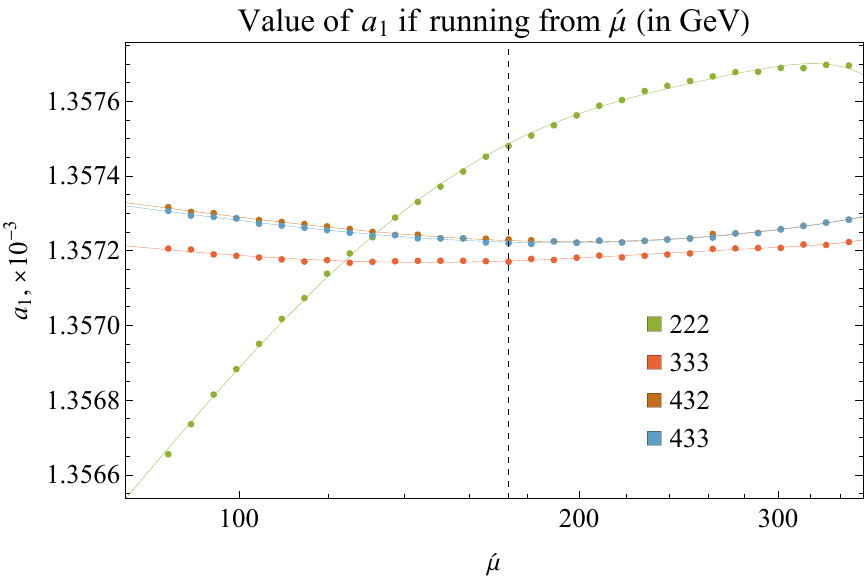} \\
	\includegraphics[width = 0.475\textwidth]{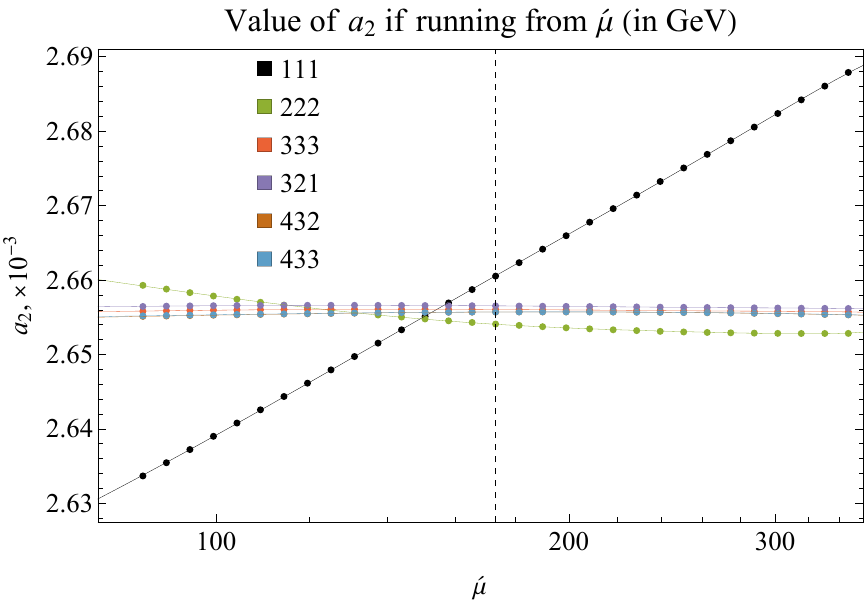} &
	\includegraphics[width = 0.485\textwidth]{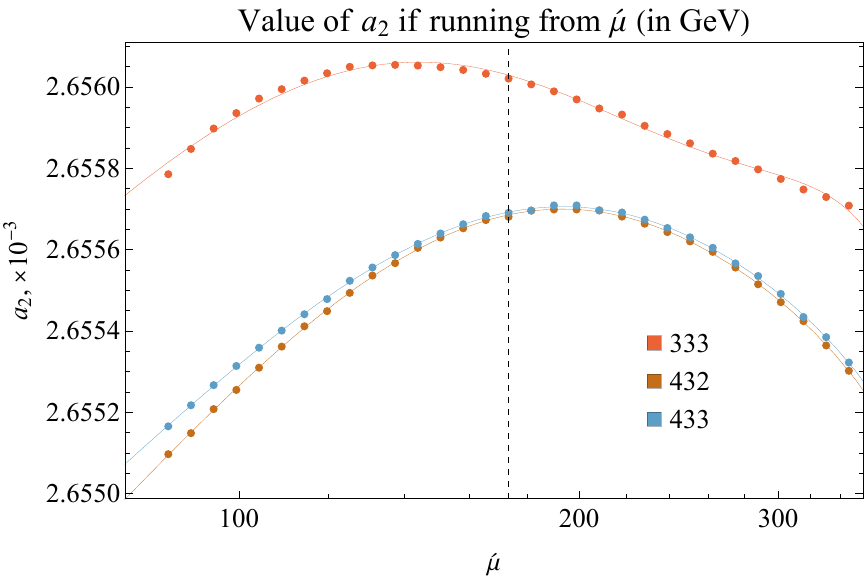} 
\end{tabular}
\caption{Running electroweak gauge couplings $a_{1-2}^{\;\to\mu_0}$ at $\mu_0=173.22\,\GeV$ as functions of the matching scale $\acute\mu$, at which the \MSbar\ parameters are extracted from the \onshell\ ones. The dependence on $\acute\mu \in (\mu_0/2,\mu_0\cdot2)$ gives an estimate of the theoretical uncertainty of $a_{1-2}(\mu_0)$. 
\label{fig:gauge_weak_th_unc}}
\end{figure}

\begin{figure}[!ht]
\centering
\begin{tabular}{cc}
	\includegraphics[width = 0.475\textwidth]{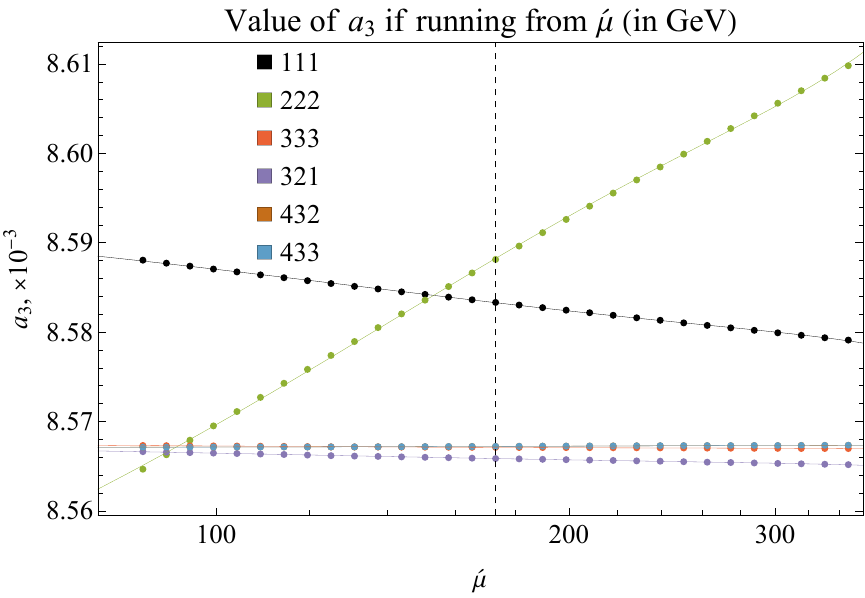} &
	\includegraphics[width = 0.485\textwidth]{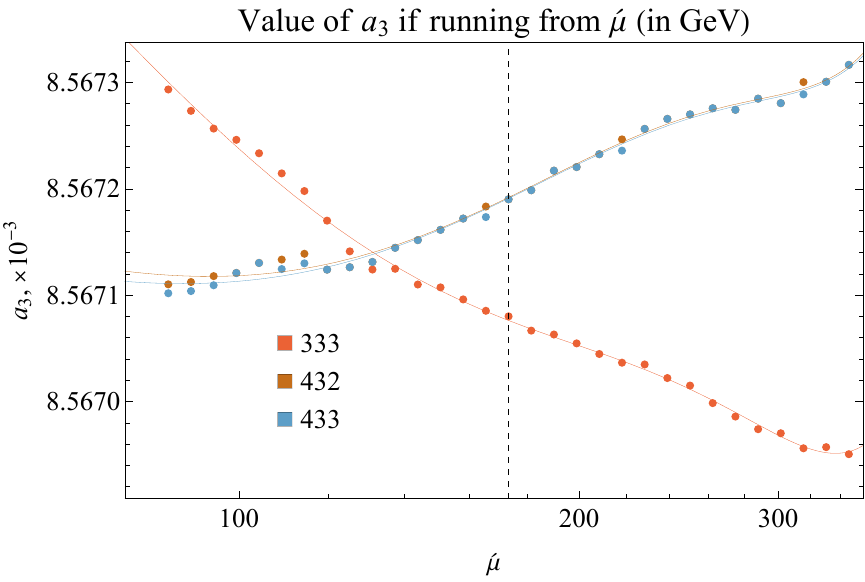} \\
\end{tabular}
\caption{Running strong coupling $a_{3}^{\;\to\mu_0}$ at $\mu_0=173.22\,\GeV$ as functions of the matching scale $\acute\mu$, at which the \onshell\ parameters are recalculated in the \MSbar\ scheme. The dependence on $\acute\mu \in (\mu_0/2,\mu_0\cdot2)$ gives an estimate of the theoretical uncertainty of $a_{3}(\mu_0)$. 
\label{fig:gauge_strong_th_unc}}
\end{figure}

\begin{figure}[t]
\centering
\begin{tabular}{cc}
	\includegraphics[width = 0.475\textwidth]{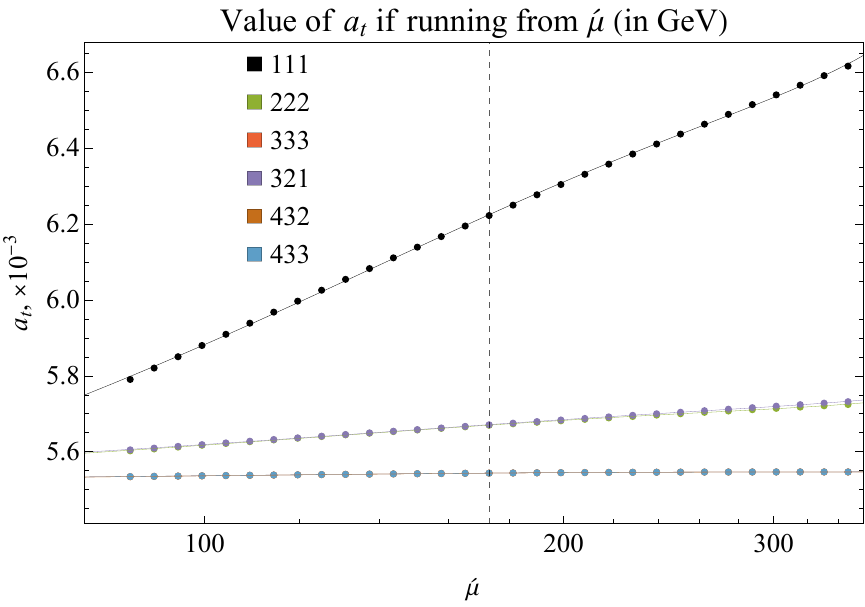} &
	\includegraphics[width = 0.485\textwidth]{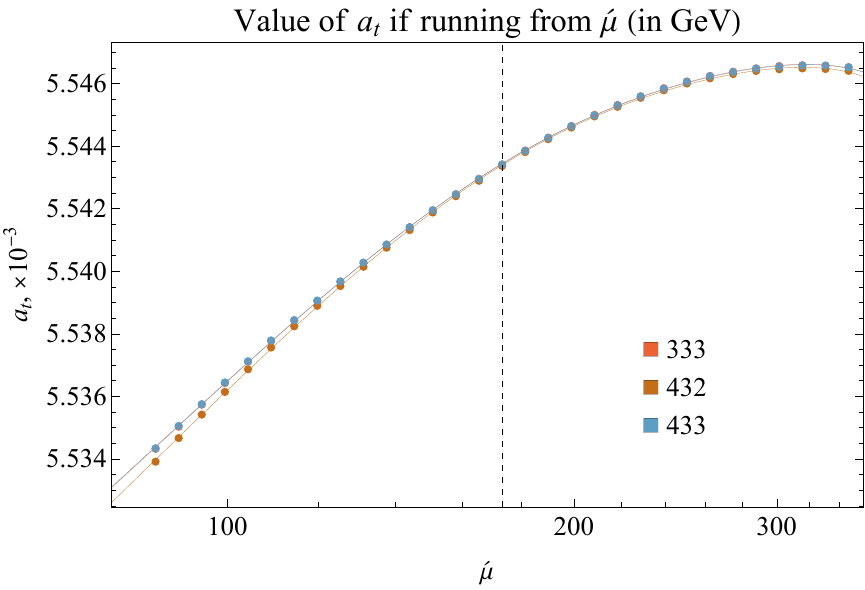} \\
	\includegraphics[width = 0.475\textwidth]{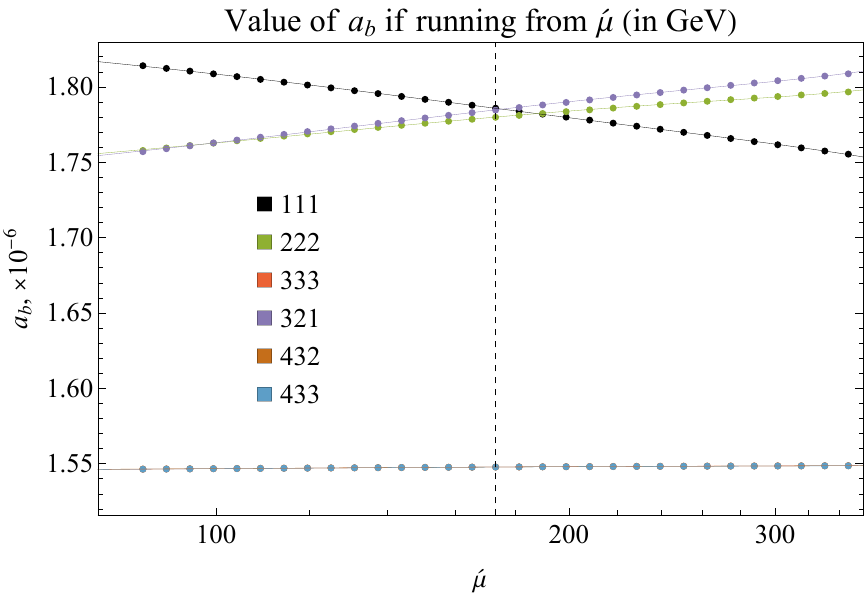} &
	\includegraphics[width = 0.485\textwidth]{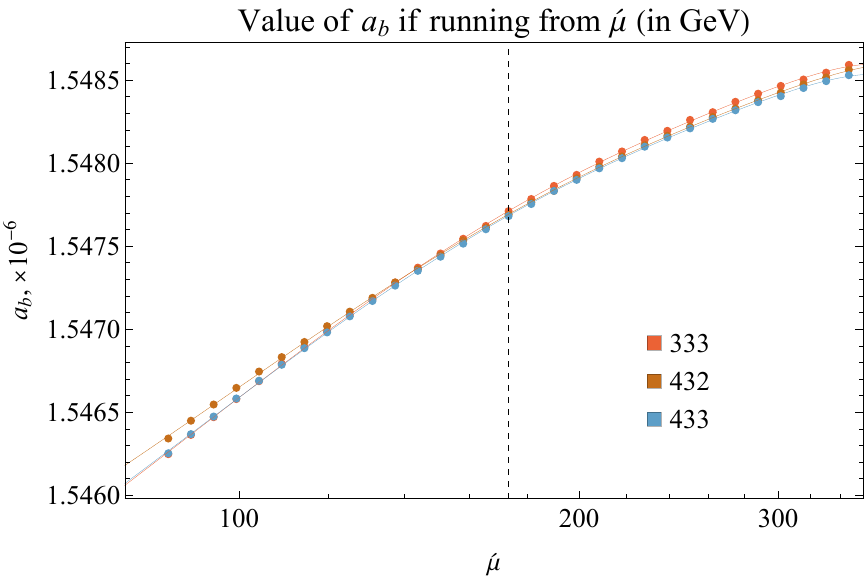} 
\end{tabular}
\caption{Running Yukawa coupling constants $a_{t,b}^{\:\to\mu_0}$ at $\mu_0=173.22\,\GeV$ as functions of the matching $\acute\mu$ scale, at which the \onshell\ parameters are recalculated in the \MSbar\ scheme. The dependence on $\acute\mu \in (\mu_0/2,\mu_0\cdot2)$ gives an estimate of the theoretical uncertainty of the values of $a_{t,b}(\mu_0)$. \label{fig:yukawa_th_unc}}
\end{figure}

\begin{figure}[t]
\centering
  \includegraphics[width = 0.475\textwidth]{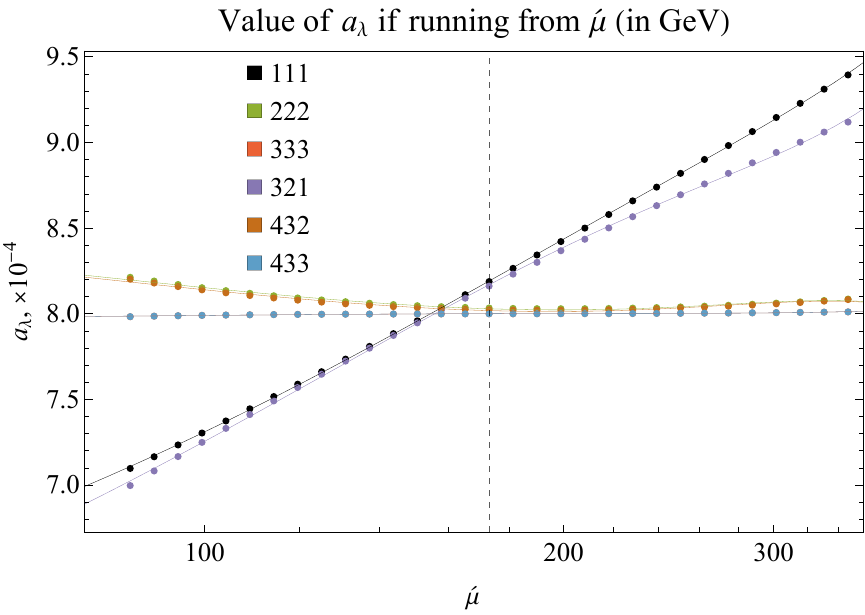}
  \includegraphics[width = 0.485\textwidth]{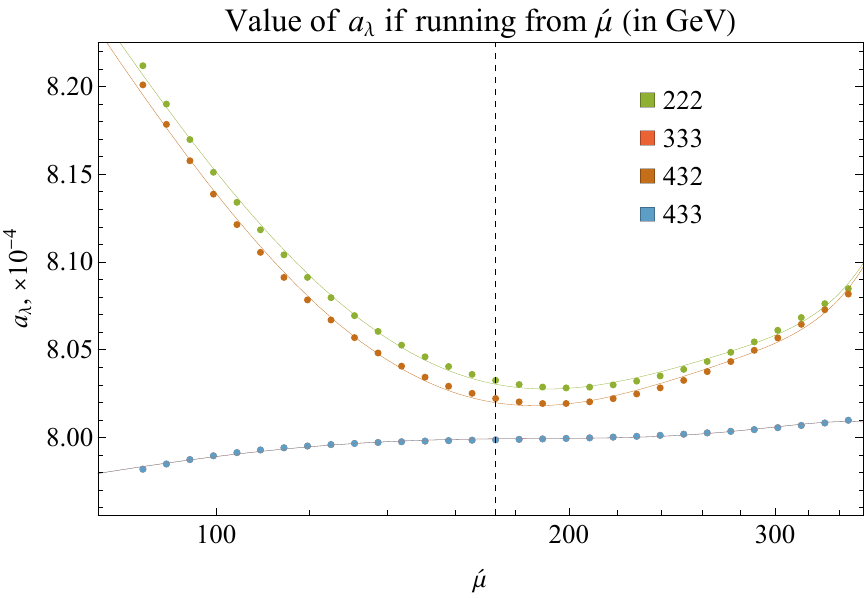}
  \caption{Running self-coupling constant $a_{\lambda}^{\,\to\mu_0}$ at $\mu_0=173.22\,\GeV$ as a function of the scale $\acute\mu$, on which the \onshell\ parameters are recalculated in the \MSbar\ scheme. The dependence on $\acute\mu \in (\mu_0/2,\mu_0\cdot2)$ gives an estimate of the theoretical uncertainty of the value $a_\lambda(\mu_0)$. \label{fig:lambda_th_unc}}
\end{figure}

To estimate theoretical uncertainty, we select several values of $\acute\mu$ near the fixed electroweak scale\footnote{Specifically, we consider the set $\{\mu_0\cdot2^{-1},\,\mu_0\cdot2^{-14/15},\, ... \mu_0,\, ... \mu_0\cdot2^{14/15},\,\mu_0\cdot2\}$ (``logarithmic grid'').} $\mu_0=173.22\,\GeV$. In the given order $L_g L_y L_\lambda$, we extract the \MSbar\ parameters from pseudoobservables using a modified version of the original \SMDR\ function {\tt Fit\_Inputs()} at each value of $\acute\mu$. 
Then we solve the corresponding RGE in the SM on the interval $(\acute\mu,\mu_0)$ obtaining the \MSbar\ parameters at $\mu_0$, which we denote as $a^{\to\mu_0}(\acute \mu)$.  
Figures~\ref{fig:gauge_weak_th_unc}-\ref{fig:lambda_th_unc} show the dependence of $a^{\to\mu_0}(\acute\mu)$ treated as functions of $\acute \mu$ in the various PT orders considered.
The difference between $a^{\to\mu_0}(\acute\mu)$ obtained for different $\acute\mu$ can serve as an estimate of theoretical uncertainty in the given order. 
In addition, the figures show how the ``central value'' $a^{\to\mu_0}(\mu_0)$ obtained at $\acute\mu=\mu_0$ shifts with increasing number of loops.

Obviously, $a^{\to\mu_0}(\mu_0)$ is identical to a direct calculation of {\tt Fit\_Inputs()} at the scale $\mu_0$ (without RG running).  As we see, the dependence $a^{\to\mu_0}(\acute\mu)$ is not necessarily monotonic, although it is quite smooth. In an idealized situation when all PT contributions are taken into account, $a^{\to\mu_0}(\acute\mu)$ should not depend on $\acute\mu$, at which the extraction of running \MSbar\ parameters from \onshell\ quantities occurs. Such a tendency is indeed observed in the plots: with an increase in the number of loops, the dependence on $\acute\mu$ weakens.


In particular, we can see that the theoretical uncertainties of some parameter exhibit significant dependence only on the accounted loops for \emph{that same} parameter and weakly depend on the loops of the others. For example, in Fig.~\ref{fig:yukawa_th_unc} different loop configurations are grouped by the number of loops for the Yukawa couplings taken into account: $111$ stays aside from others being the only considered configuration with one loop RGE for $a_t$ and $a_b$, then there is almost identical behavior of $222$ and $321$ with two loops for the Yukawa couplings, and the last group consisting of $333$, $432$, and $433$ with three-loop Yukawa RGEs. This same logic is applicable to all other \MSbar\ couplings. One can also see the similar grouping of the theoretical uncertainty bands in Fig.~\ref{fig:bc_uncertanties} below.

Now, to quantify the theoretical uncertainty $a(\mu_0)$, we take the maximum and minimum values, ($\text{max}_\text{th}$ and $\text{min}_\text{th}$), of the considered parameters $a^{\to\mu_0}$ (see Table.~\ref{tab:theorerr}).
As expected, the theoretical uncertainty turned out to be smaller the more terms of the PT series are taken into account.

\begin{table}[h]
\centering
\begin{tabular}{|c|c|c|c|c|c|c|c|}
  \hline \rowcolor[HTML]{EFEFEF}
  \multicolumn{2}{|c|}{} & $111$ & $222$ & $333$ & $321$ & $432$ & $433$ \\[6pt]
  \hline
   & $\text{max}_\text{th}$ & $1.3602$ & $1.3577$ & $1.3572$ & $1.3572$ & $1.3573$ & $1.3573$ \\
 \multirow{-2}{*}{$a_1, \,\pow{-3}$} & $\text{min}_\text{th}$ & $1.3525$ & $1.3567$ & $1.3572$ & $1.3571$ & $1.3572$ & $1.3572$ \\ \hline
  & $\text{max}_\text{th}$ & $2.6879$ & $2.6593$ & $2.6561$ & $2.6566$ & $2.6557$ & $2.6557$ \\
 \multirow{-2}{*}{$a_2, \,\pow{-3}$} & $\text{min}_\text{th}$ & $2.6337$ & $2.6528$ & $2.6557$ & $2.6562$ & $2.6551$ & $2.6552$ \\ \hline
  & $\text{max}_\text{th}$ & $8.5880$ & $8.6098$ & $8.5673$ & $8.5666$ & $8.5673$ & $8.5673$ \\
 \multirow{-2}{*}{$a_3, \,\pow{-3}$} & $\text{min}_\text{th}$ & $8.5791$ & $8.5646$ & $8.567$ & $8.5651$ & $8.5671$ & $8.5671$ \\ \hline
  & $\text{max}_\text{th}$ & $6.6149$ & $5.7240$ & $5.5466$ & $5.7316$ & $5.5465$ & $5.5466$ \\
 \multirow{-2}{*}{$a_t, \,\pow{-3}$} & $\text{min}_\text{th}$ & $5.7901$ & $5.6021$ & $5.5343$ & $5.6048$ & $5.5339$ & $5.5343$ \\ \hline
  & $\text{max}_\text{th}$ & $1.8140$ & $1.7965$ & $1.5486$ & $1.8086$ & $1.5486$ & $1.5485$ \\
 \multirow{-2}{*}{$a_b, \,\pow{-6}$} & $\text{min}_\text{th}$ & $1.7553$ & $1.7577$ & $1.5462$ & $1.7569$ & $1.5463$ & $1.5463$ \\ \hline
  & $\text{max}_\text{th}$ & $9.3931$ & $8.2117$ & $8.0097$ & $9.1179$ & $8.2008$ & $8.0097$ \\
 \multirow{-2}{*}{$a_\lambda, \,\pow{-4}$} & $\text{min}_\text{th}$ & $7.0959$ & $8.0282$ & $7.9818$ & $6.9961$ & $8.0193$ & $7.9819$\\[6pt]
 \hline
    \end{tabular}
    \caption{Theoretical uncertainties of the initial conditions for the renormalization group equations (values at the electroweak scale $\mu_0 = 173.22~\GeV$) depending on the loop configurations}
    \label{tab:theorerr}
\end{table}

\subsection{Parametric uncertainties\label{sec:parametric_uncertainties}}
In this subsection we discuss the parametric uncertainties of the initial values of the \MSbar\ parameters at fixed $\mu_0=173.22 \, \GeV$ originating from the experimental errors in \onshell\ quantities given in Eqs.~\eqref{eq:pdg2024_1}, \eqref{eq:mub}, and \eqref{eq:alpha_had}.

\begin{figure}[t]
\centering
  \includegraphics[width = 0.45\textwidth]{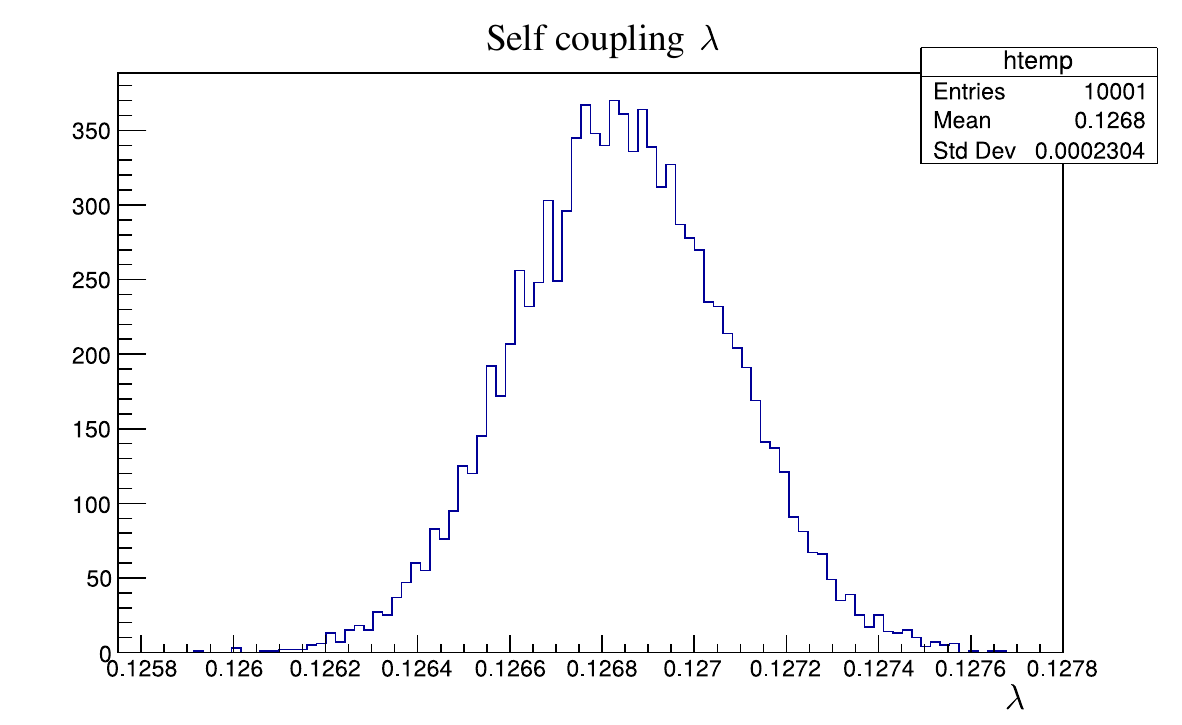}
  \includegraphics[width = 0.45\textwidth]{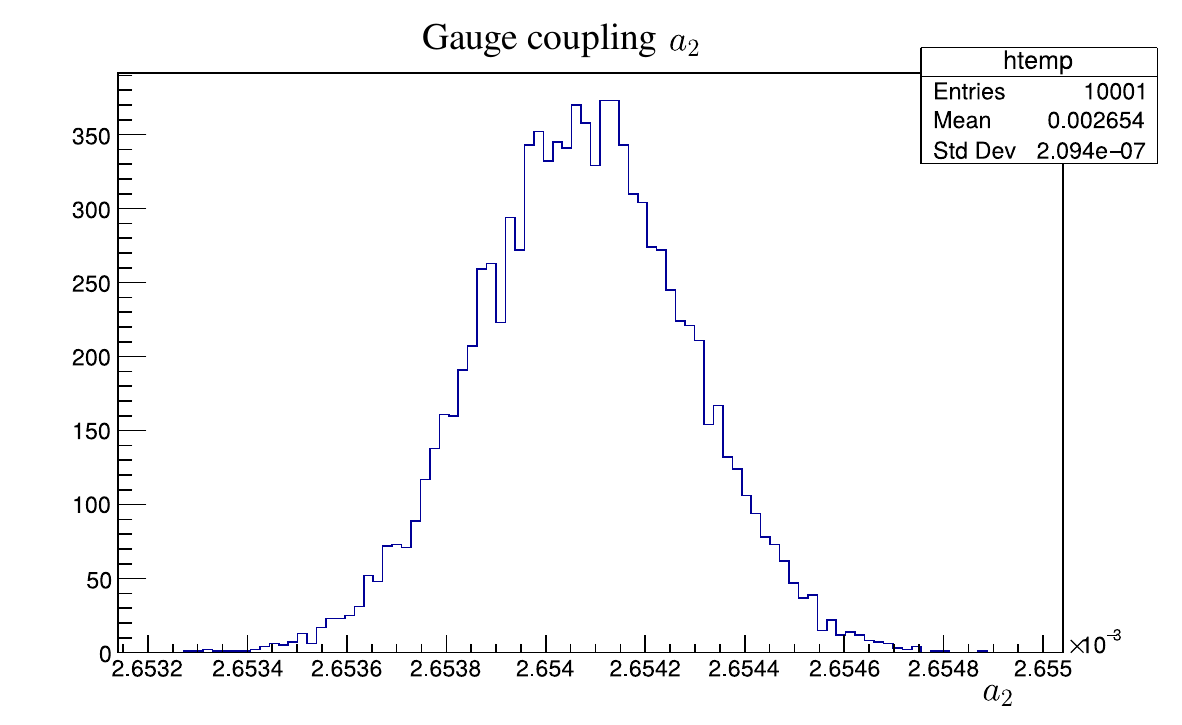}
  \caption{Examples of random generation: $10$k random values of $\lambda$ and $a_2$ in $222$ \label{fig:random_example}}
\end{figure}

These uncertainties can be estimated by generating a random sample of pseudoobservables according to the normal distributions defined by central values and experimental errors of the \onshell\ quantities. By computing the \MSbar\ parameters for each member of the sample by means of \SMDR\footnote{At certain loop level}, we can obtain distributions for the former, which turns out be very close to the normal ones (see, e.g., Fig.~\ref{fig:random_example}). The standard deviation of the obtained \MSbar\ sample gives us the size of parameteric uncertainties, while the means should correspond to the central values.   

This ``statistical'' approach turns out to be very time-consuming and we switch to another one based on approximate \emph{polynomial} relations between \onshell\ and \MSbar\ parameters inspired by  Ref.~\cite{Alam:2022cdv} and obtained by a fitting procedure. 
\begin{table}[ht]
\centering
\begin{tabular}{|l| c| c| c| c| c| }
\hline \rowcolor[HTML]{EFEFEF}
 & {\scriptsize$\hat\delta_h \equiv \dfrac{M_h}{M_{h(0)}} - 1$} & {\scriptsize$\hat\delta_t \equiv \dfrac{M_t}{M_{t(0)}} - 1$} & {\scriptsize$\hat\delta_S \equiv \dfrac{\alpha_S^{(5)}(M_Z)}{\alpha_S^{(5)}(M_Z)_{(0)}} - 1$} & {\scriptsize$\hat\delta_a \equiv \dfrac{\Delta\alpha_{had}^{(5)}(M_Z)}{\Delta\alpha_{had}^{(5)}(M_Z)_{(0)}} - 1$} & {\scriptsize$\hat\delta_b \equiv \dfrac{m_b(m_b)}{m_b(m_b)_{(0)}} - 1$} \\
\hline
	2014 & $\approx 0.3\%$ & $\approx 0.3\%$ & $\approx 0.5\%$ & $\approx 0.4\%$  & $\approx 0.71\%$ \\
2024 & $\approx 0.1\%$ & $\approx 0.2\%$  & $\approx 0.76\%$ & $\approx 0.2\%$ & $\approx 0.16\%$ \\
\hline
\end{tabular}
\caption{Dimensionless deviations and their typical values from PDG2014~\cite{ParticleDataGroup:2014cgo} and PDG2024~\cite{ParticleDataGroup:2024cfk}}
\label{tab:PDG_deviations}
\end{table}

We first introduce, for convenience, seven dimensionless deviations of the \onshell\ parameters from their average values:
five of them are given in Table.~\ref{tab:PDG_deviations} and have typical values\footnote{Corresponding to experimental relative uncertanties} of the order of a few per mil, while the other two\footnote{In our fits we do not consider the dependence on $\alpha$ due to the too small error of the latter and small dependence on it, so as not to complicate the fit.}
\begin{align*}
					    \hat\delta_Z = & \quad\dfrac{M_Z}{M_{Z(0)}} - 1 \approx 0.002 \%,
							   &&&
  \hat\delta_{GF} = &\;\dfrac{G_{F}}{G_{F(0)}} - 1 \approx 5\pow{-5}~\% 
\end{align*}
are assumed to be much smaller \cite{ParticleDataGroup:2024cfk}.

The fitting itself is performed on a grid of points, $5$ points on each axis corresponding to the above-mentioned deviations, i.e. $5^7 = 78125$ points, uniformly distributed on the intervals $(-5\sigma_i,5\sigma_i)$ for $\sigma_i$ being the uncertainty of an \onshell\ quantity $\mathcal{O}_i$, for which {\tt Fit\_Inputs()} is used. Further fitting is carryied out by the minimizer {\tt Minuit2} \cite{Hatlo:2005cj} of the {\tt ROOT} package \cite{rene_brun_2019_3895860} while also cross-validating fit results with {\tt Wolfram Mathematica}.

These \emph{fit} results will be routinely used later when converting \onshell\ into \MSbar\ in calculating the boundary values for RGEs. Thus, having initially done some work, we will obtain a simple, easily calculable (polynomial) approximations for the dependences of the \MSbar\ parameters on the \onshell\ quantities, which will further allow us to estimate the parametric uncertainties 
of the initial conditions (the values at $\mu=\mu_0=173.22\,\GeV$) for RG equations using standard formulas for estimating the uncertainty of a quantity analytically dependent on a quantity with uncertainty. 
\subsubsection{Approximate relations between \onshell\ and \MSbar}

Let us now move directly to choosing functions to fit. We first take $\lambda(\mu_0)$ and look for an approximate solution of the form:
\begin{align}    
	\lambda & = \lambda_0 \left( 1 + c^h_{\lambda} \hat\delta_h + c^t_{\lambda} \hat\delta_t + c^Z_{\lambda} \hat\delta_Z + c^S_{\lambda} \hat\delta_S + c^a_{\lambda} \hat\delta_a + c^{GF}_{\lambda} \hat\delta_{GF} + c^b_{\lambda} \hat\delta_b + c^{tt}_\lambda \hat\delta_t \hat\delta_t + c^{tS}_\lambda \hat\delta_t \hat\delta_S + c^{hh}_\lambda \hat\delta_h \hat\delta_h + c^{ht}_\lambda \hat\delta_h \hat\delta_t \right. \nonumber\\
		& \hspace{1.48cm}\left.
    + c^{SS}_\lambda \hat\delta_S \hat\delta_S + c^{hS}_\lambda \hat\delta_h \hat\delta_S + c^{ttt}_\lambda \hat\delta_t \hat\delta_t \hat\delta_t  + c^{ttS}_\lambda \hat\delta_t \hat\delta_t \hat\delta_S \right).
\label{eq:ansatzlambda}
\end{align}
This expression provides the polynomial dependence\footnote{Without the constant term, so that $\lambda_0$ corresponds to the ``central'' value of $\lambda(\mu_0)$, which we get for the central values of the \onshell\ parameters.} of the relative parametric error of $\lambda$ on the relative errors of the \onshell\ parameters. Also, the numerical values of the coefficients will allow us to determine the most and least significant contributions to each variable when shifting along the corresponding \onshell\ quantity.
The choice of specific terms for each fit is determined by the empirical selection of parameters, which, when added to the fit, cause a noticeable improvement in the accuracy of the latter. 
Fit results for $\lambda(\mu_0)$ ended up in Table.~\ref{tab:lambda_fit}.

\begin{table}[ht]\label{fitlambda}
\begin{center}
\begin{tabular}{|c|c|c|c|c|c|c|}
  \hline \rowcolor[HTML]{EFEFEF}
   & $11\mathbf{1}$ & $22\mathbf{2}$ & $33\mathbf{3}$ & $32\mathbf{1}$ & $43\mathbf{2}$ & $43\mathbf{3}$ \\ \hline
  $\lambda_0$ & $0.1293$ & $0.1268$ & $0.1263$ & $0.1289$ & $0.1267$ & $0.1263 $ \\[5pt]
 $c^h_\lambda$ & $2$ & $2.0273$ & $2.0345$ & $2.0044$ & $2.0318$ & $2.0345 $ \\[5pt]
 $c^t_\lambda$ & $-$ & $0.1364$ & $0.1$ & $-0.0167$ & $0.119$ & $0.1001 $ \\[5pt]
 $c^Z_\lambda$ & $-$ & $-0.0313$ & $-0.0325$ & $-0.0216$ & $-0.0309$ & $-0.0325 $ \\[5pt]
 $c^S_\lambda$ & $-$ & $-0.0058$ & $-0.0066$ & $7.321\pow{-4}$ & $-0.0073$ & $-0.0066 $ \\[5pt]
 $c^{tt}_\lambda$ & $-$ & $0.7871$ & $0.6565$ & $-0.0491$ & $0.7478$ & $0.6565 $ \\[5pt]
 $c^{tS}_\lambda$ & $-$ & $-0.0632$ & $-0.0799$ & $0.0028$ & $-0.0903$ & $-0.0799 $ \\[5pt]
 $c^{hh}_\lambda$ & $1$ & $1.0345$ & $1.0462$ & $1.0089$ & $1.0367$ & $1.0462 $ \\[5pt]
 $c^{ht}_\lambda$ & $-$ & $-0.1615$ & $-0.177$ & $-0.032$ & $-0.1576$ & $-0.1769 $ \\[5pt]
 $c^{SS}_\lambda$ & $-$ & $0.0018$ & $0.0022$ & $-1.024\pow{-4}$ & $0.0024$ & $0.0022 $ \\[5pt]
 $c^{hS}_\lambda$ & $-$ & $0.0068$ & $0.0098$ & $0.0014$ & $0.0098$ & $0.0098 $ \\[5pt]
 $c^{ttt}_\lambda$ & $-$ & $1.2429$ & $1.0136$ & $-0.0156$ & $1.2256$ & $1.0137$ \\[5pt]
 $c^{ttS}_\lambda$ & $-$ & $-0.1112$ & $-0.1842$ & $-$ & $-0.1841$ & $-0.1842$ \\[5pt]
 $c^a_\lambda$ & $-$ & $8.85\pow{-5}$ & $9.087\pow{-5}$ & $6.426\pow{-5}$ & $9.057\pow{-5}$ & $9.088\pow{-5} $ \\[5pt]
 $c^b_\lambda$ & $-$ & $3.975\pow{-5}$ & $3.265\pow{-5}$ & $7.405\pow{-6}$ & $3.157\pow{-5}$ & $3.265\pow{-5} $ \\[5pt]
 $c^{GF}_\lambda$ & $1$ & $0.9784$ & $0.9714$ & $0.9946$ & $0.9769$ & $0.9714 $ \\[5pt]
  \hline
\end{tabular}
\end{center}
\caption{Fitted coefficients for $\lambda$ introduced in Eq.~\eqref{eq:ansatzlambda}. The relative fit error \eqref{eq:fit_error} is $\delta_{err} \lesssim 10^{-6}$. The symbol ``$-$'' signifies that no sensitivity to such a term was found.\label{tab:lambda_fit}}
\end{table}

The accuracy of the produced fit was estimated on the same grid of points. Here $\delta_{err}$ is the average fit error on this grid, namely, the sum of absolute values of the difference between the value obtained by time-consuming evaluation of {\tt Fit\_Inputs()} and by the application of \eqref{eq:ansatzlambda} for each point of the grid divided by a number of points in the grid ($N=5^7$):
\begin{equation}
\delta_{err} = \dfrac1{\lambda_0} \cdot \dfrac1N \sum \left|\lambda^\text{(Fit\_Inputs)} - \lambda^\text{(fit)} \right|
\label{eq:fit_error}
\end{equation}

As will be seen later, the fit error is much smaller than the uncertainty $\lambda$ on this grid, which allows us to say that this fit is quite successful. It can be noted that with an increase in loops for $\lambda$, the fit error increases, which indicates a greater distortion from simple polynomial approximation. At the same time, the fitting process also spends more steps during minimization at higher loops. Also, those errors do not excess the precision that is provided by {\tt Fit\_Inputs()} (see the comments in Appendix \ref{app:smdr}).

Now we continue with the gauge couplings $g, g^\prime, g_s$. The data on which we perform the fitting is left unchanged. For $g(\mu_0)$ the fit was chosen of the form
\begin{equation}\label{eq:ansatzg}
    g = g_0 \left(1 + c^h_g \hat\delta_h + c^t_g \hat\delta_t + c^Z_g \hat\delta_Z + c^S_g \hat\delta_S + c^a_g \hat\delta_a + c^{GF}_g \hat\delta_{GF} + c^{tt}_g \hat\delta_t \hat\delta_t + c^{tS}_g \hat\delta_t \hat\delta_S\right)
\end{equation}
with the fit results given in Table~\ref{tab:fitg}.
\begin{table}[ht]
\begin{center}
\begin{tabular}{|c|c|c|c|c|c|c|}
  \hline \rowcolor[HTML]{EFEFEF}
   & $\mathbf{1}11$ & $\mathbf{2}22$ & $\mathbf{3}33$ & $\mathbf{3}21$ & $\mathbf{4}32$ & $\mathbf{4}33$ \\ \hline
 $g_0$ & $0.6482$ & $0.6474$ & $0.6476$ & $0.6477$ & $0.6476$ & $0.6476 $ \\[5pt]
 $c^{h}_g$ & $-$ & $-1.163\pow{-3}$ & $-1.067\pow{-3}$ & $-1.058\pow{-3}$ & $-1.065\pow{-3}$ & $-1.066\pow{-3} $ \\[5pt]
 $c^{t}_g$ & $5.13\pow{-4}$ & $0.0119$ & $0.0101$ & $0.0103$ & $0.01$ & $0.01 $ \\[5pt]
 $c^{Z}_g$ & $1.4379$ & $1.4204$ & $1.4237$ & $1.4235$ & $1.4236$ & $1.4236 $ \\[5pt]
 $c^{S}_g$ & $-4.99\pow{-5}$ & $-5.329\pow{-4}$ & $-5.309\pow{-4}$ & $-3.595\pow{-4}$ & $-6.355\pow{-4}$ & $-6.359\pow{-4} $ \\[5pt]
 $c^{tt}_g$ & $-2.503\pow{-4}$ & $0.0079$ & $0.0058$ & $0.0061$ & $0.0056$ & $0.0056 $ \\[5pt]
 $c^{tS}_g$ & $-$ & $-3.582\pow{-4}$ & $-1.224\pow{-3}$ & $-1.019\pow{-3}$ & $-1.412\pow{-3}$ & $-1.418\pow{-3} $ \\[5pt]
 $c^{a}_g$ & $-0.0066$ & $-0.0064$ & $-0.0064$ & $-0.0064$ & $-0.0064$ & $-0.0064 $ \\[5pt]
 $c^{GF}_g$ & $0.7199$ & $0.7123$ & $0.7135$ & $0.7134$ & $0.7134$ & $0.7134 $ \\[5pt]
  \hline
\end{tabular}
\end{center}
\caption{Fitted coefficients for $g$ defined in eq.~\eqref{eq:ansatzg}. The relative fit error $\delta_{err} \lesssim 5\pow{-7}$. The symbol ``$-$'' signifies that no sensitivity to such term was found.
\label{tab:fitg}}
\end{table}
For the $g^\prime(\mu_0)$ fit we use the form
\begin{equation}\label{eq:ansatzgp}
    g^\prime = g^\prime_0 \left(1 + c^h_{g^\prime} \hat\delta_h + c^t_{g^\prime} \hat\delta_t + c^Z_{g^\prime} \hat\delta_Z + c^S_{g^\prime} \hat\delta_S + c^a_{g^\prime} \hat\delta_a + c^{GF}_{g^\prime} \hat\delta_{GF} + c^{tt}_{g^\prime} \hat\delta_t \hat\delta_t \right)
\end{equation}
with the results presented in Table.~\ref{tab:fitgp}.
\begin{table}[ht]
\begin{center}
\begin{tabular}{|c|c|c|c|c|c|c|}
  \hline \rowcolor[HTML]{EFEFEF}
   & $\mathbf{1}11$ & $\mathbf{2}22$ & $\mathbf{3}33$ & $\mathbf{3}21$ & $\mathbf{4}32$ & $\mathbf{4}33$ \\ \hline
 $g^\prime_0$ & $0.3585$ & $0.3586$ & $0.3586$ & $0.3586$ & $0.3586$ & $0.3586 $ \\[5pt]
 $c^{h}_{g^\prime}$ & $-$ & $3.521\pow{-4}$ & $3.228\pow{-4}$ & $3.200\pow{-4}$ & $3.222\pow{-4}$ & $3.225\pow{-4} $ \\[5pt]
 $c^{t}_{g^\prime}$ & $-1.677\pow{-3}$ & $-0.0051$ & $-0.0045$ & $-0.0046$ & $-0.0045$ & $-0.0045 $ \\[5pt]
 $c^{Z}_{g^\prime}$ & $-0.4315$ & $-0.4276$ & $-0.4282$ & $-0.428$ & $-0.4283$ & $-0.4283 $ \\[5pt]
 $c^{S}_{g^\prime}$ & $1.631\pow{-4}$ & $3.095\pow{-4}$ & $3.072\pow{-4}$ & $2.554\pow{-4}$ & $3.389\pow{-4}$ & $3.391\pow{-4} $ \\[5pt]
 $c^{tt}_{g^\prime}$ & $8.164\pow{-4}$ & $-1.546\pow{-3}$ & $-9.489\pow{-4}$ & $-1.033\pow{-3}$ & $-8.68\pow{-4}$ & $-8.834\pow{-4} $ \\[5pt]
 $c^{a}_{g^\prime}$ & $0.0215$ & $0.0214$ & $0.0214$ & $0.0214$ & $0.0214$ & $0.0214 $ \\[5pt]
 $c^{GF}_{g^\prime}$ & $-0.2189$ & $-0.2173$ & $-0.2174$ & $-0.2174$ & $-0.2174$ & $-0.2174 $ \\[5pt]
  \hline
\end{tabular}
\end{center}
\caption{Fitted coefficients for $g^\prime$ defined in eq.~\eqref{eq:ansatzgp}. The relative fit error is $\delta_{err} \lesssim 5\pow{-7}$. The symbol ``$-$'' signifies that no sensitivity to such term was found.\label{tab:fitgp}}
\end{table}
We fit $g_s(\mu_0)$ by a polynomial 
\begin{equation}\label{eq:ansatzg3}
    g_s = g_{s0} \left(1 + c^h_{g_s} \hat\delta_h + c^t_{g_s} \hat\delta_t + c^Z_{g_s} \hat\delta_Z + c^S_{g_s} \hat\delta_S + c^{tt}_{g_s} \hat\delta_t \hat\delta_t + c^{tS}_{g_s} \hat\delta_t \hat\delta_S + c^{SS}_{g_s} \hat\delta_S \hat\delta_S + c^{SSS}_{g_s} \hat\delta_S \hat\delta_S \hat\delta_S\right)
\end{equation}
with the coefficients given in Table.~\ref{tab:fitg3}
\begin{table}[ht]
\begin{center}
\begin{tabular}{|c|c|c|c|c|c|c|}
  \hline \rowcolor[HTML]{EFEFEF}
   & $\mathbf{1}11$ & $\mathbf{2}22$ & $\mathbf{3}33$ & $\mathbf{3}21$ & $\mathbf{4}32$ & $\mathbf{4}33$ \\ \hline
 $g_{s0}$ & $1.1642$ & $1.1646$ & $1.1631$ & $1.163$ & $1.1631$ & $1.1631 $ \\[5pt]
 $c^{h}_{g_s}$ & $-$ & $3.44\pow{-5}$ & $3.172\pow{-5}$ & $3.375\pow{-5}$ & $3.176\pow{-5}$ & $3.179\pow{-5} $ \\[5pt]
 $c^{t}_{g_s}$ & $-0.0062$ & $-0.0068$ & $-0.0069$ & $-0.0069$ & $-0.0069$ & $-0.0069 $ \\[5pt]
 $c^{Z}_{g_s}$ & $-$ & $-$ & $1.971\pow{-4}$ & $1.967\pow{-4}$ & $2.055\pow{-4}$ & $2.054\pow{-4} $ \\[5pt]
 $c^{S}_{g_s}$ & $0.4563$ & $0.4568$ & $0.4545$ & $0.4543$ & $0.4546$ & $0.4545 $ \\[5pt]
 $c^{tt}_{g_s}$ & $0.0032$ & $0.0036$ & $0.0036$ & $0.0035$ & $0.0036$ & $0.0036 $ \\[5pt]
 $c^{tS}_{g_s}$ & $-0.0091$ & $-0.0104$ & $-0.0107$ & $-0.0104$ & $-0.0107$ & $-0.0107 $ \\[5pt]
 $c^{SS}_{g_s}$ & $-0.145$ & $-0.1444$ & $-0.1463$ & $-0.1466$ & $-0.1462$ & $-0.1462 $ \\[5pt]
 $c^{SSS}_{g_s}$ & $0.0689$ & $0.0689$ & $0.0691$ & $0.0689$ & $0.0690$ & $0.0691 $ \\[5pt]
  \hline
\end{tabular}
\end{center}
\caption{Fitted coefficients for $g_s$ defined in eq.~\eqref{eq:ansatzg3}. The relative fit error is $\delta_{err} \lesssim 10^{-7}$. The symbol ``$-$'' signifies that no sensitivity to such term was found.
\label{tab:fitg3}
}
\end{table}

Finally, let us repeat the same procedure for the Yukawa couplings $y_t(\mu_0), y_b(\mu_0)$. To perform fitting, we use the same data from the \SMDR\ output together with the following expressions:
\begin{align}
	y_t & = y_{t0} \left(1 + c^h_{y_t} \hat\delta_h + c^t_{y_t} \hat\delta_t + c^Z_{y_t} \hat\delta_Z + c^S_{y_t} \hat\delta_S + c^{GF}_{yt} \hat\delta_{GF} + c^{tt}_{y_t} \hat\delta_t \hat\delta_t + c^{ht}_{y_t} \hat\delta_h \hat\delta_t + c^{tS}_{y_t} \hat\delta_t \hat\delta_S + c^{SS}_{y_t} \hat\delta_S \hat\delta_S\right),\label{eq:ansatzyt} \\
	y_b & = y_{b_0} \left( 1 + c^b_{y_b} \hat\delta_b  + c^h_{y_b} \hat\delta_h  + c^t_{y_b} \hat\delta_t + c^Z_{y_t} \hat\delta_Z + c^S_{y_t} \hat\delta_S + c^{GF}_{yb} \hat\delta_{GF}  + c^{bb}_{y_b} \hat\delta_b \hat\delta_b + c^{bS}_{y_b} \hat\delta_b \hat\delta_S \right. \nonumber \\
	    & \hspace{1.55cm} \left. + c^{SS}_{y_b} \hat\delta_S \hat\delta_S + c^{tS}_{y_b} \hat\delta_t \hat\delta_S + c^{SSS}_{y_b} \hat\delta_S \hat\delta_S \hat\delta_S \right).
\label{eq:ansatzyb}
\end{align}
The numerical values of the coefficients can be found in Tables~\ref{tab:fityt} and \ref{tab:fityb}.
\begin{table}[ht]
\begin{center}
\begin{tabular}{|c|c|c|c|c|c|c|}
  \hline \rowcolor[HTML]{EFEFEF}
   & $1\mathbf{1}1$ & $2\mathbf{2}2$ & $3\mathbf{3}3$ & $3\mathbf{2}1$ & $4\mathbf{3}2$ & $4\mathbf{3}3$ \\ \hline
 $y_{t0}$ & $0.9912$ & $0.9462$ & $0.9356$ & $0.9463$ & $0.9356$ & $0.9356 $ \\[5pt]
 $c^{h}_{y_t}$ & $-$ & $-0.0033$ & $-0.0029$ & $-0.0032$ & $-0.0029$ & $-0.0029 $ \\[5pt]
 $c^{t}_{y_t}$ & $1$ & $1.0748$ & $1.0849$ & $1.0739$ & $1.0849$ & $1.0849 $ \\[5pt]
 $c^{Z}_{y_t}$ & $-$ & $-0.0123$ & $-0.0122$ & $-0.0121$ & $-0.0122$ & $-0.0122 $ \\[5pt]
 $c^{S}_{y_t}$ & $-$ & $-0.0458$ & $-0.0671$ & $-0.0454$ & $-0.0671$ & $-0.0671 $ \\[5pt]
 $c^{tt}_{y_t}$ & $-$ & $8.559\pow{-4}$ & $0.0118$ & $0.0016$ & $0.0115$ & $0.0126 $ \\[5pt]
 $c^{tS}_{y_t}$ & $-$ & $0.0304$ & $0.0268$ & $0.0297$ & $0.0268$ & $0.0268 $ \\[5pt]
 $c^{SS}_{y_t}$ & $-$ & $0.0029$ & $-0.0049$ & $0.0032$ & $-0.005$ & $-0.005 $ \\[5pt]
 $c^{ht}_{y_t}$ & $-$ & $-0.011$ & $-0.0118$ & $-0.0112$ & $-0.0118$ & $-0.0118$ \\[5pt]
 $c^{GF}_{y_t}$ & $0.5$ & $0.5018$ & $0.5016$ & $0.5017$ & $0.5016$ & $0.5016 $ \\[5pt]
  \hline
\end{tabular}
\end{center}
\caption{Fitted coefficients for $y_t(\mu_0)$ defined in eq.~\eqref{eq:ansatzyt}. The relative fit error is $\delta_{err} \lesssim 10^{-6}$. The symbol ``$-$'' signifies that no sensitivity to such term was found.
\label{tab:fityt}
}
\end{table}

\begin{table}[ht]
\begin{center}
\begin{tabular}{|c|c|c|c|c|c|c|}
  \hline \rowcolor[HTML]{EFEFEF}
   & $1\mathbf{1}1$ & $2\mathbf{2}2$ & $3\mathbf{3}3$ & $3\mathbf{2}1$ & $4\mathbf{3}2$ & $4\mathbf{3}3$ \\ \hline
 $y_{b0}$ & $0.0168$ & $0.0168$ & $0.0156$ & $0.0168$ & $0.0156$ & $0.0156 $ \\[5pt]
 $c^{b}_{y_b}$ & $1.1355$ & $1.1355$ & $1.1783$ & $1.1348$ & $1.1783$ & $1.1783 $ \\[5pt]
 $c^{h}_{y_b}$ & $-$ & $0.0021$ & $0.0021$ & $0.0021$ & $0.0021$ & $0.0021 $ \\[5pt]
 $c^{t}_{y_b}$ & $0.0020$ & $-0.0061$ & $-0.0068$ & $-0.0069$ & $-0.0068$ & $-0.0068 $ \\[5pt]
 $c^{Z}_{y_b}$ & $-0.0043$ & $-0.0152$ & $-0.0145$ & $-0.0152$ & $-0.017$ & $-0.0146 $ \\[5pt]
 $c^{S}_{y_b}$ & $-0.4614$ & $-0.4612$ & $-0.671$ & $-0.4563$ & $-0.6713$ & $-0.6711 $ \\[5pt]
 $c^{bb}_{y_b}$ & $0.0594$ & $0.0594$ & $0.0733$ & $0.0592$ & $0.0733$ & $0.0733 $ \\[5pt]
 $c^{bS}_{y_b}$ & $-0.2811$ & $-0.2809$ & $-0.3699$ & $-0.2781$ & $-0.37$ & $-0.37 $ \\[5pt]
 $c^{SS}_{y_b}$ & $-0.0588$ & $-0.059$ & $-0.2288$ & $-0.0538$ & $-0.2289$ & $-0.2289 $ \\[5pt]
 $c^{tS}_{y_b}$ & $-$ & $0.0043$ & $0.0058$ & $0.0046$ & $0.0058$ & $0.0059 $ \\[5pt]
 $c^{SSS}_{y_b}$ & $-0.0287$ & $-0.0287$ & $-0.1484$ & $-0.0256$ & $-0.0126$ & $-0.1485 $ \\[5pt]
 $c^{GF}_{y_b}$ & $0.4947$ & $0.492$ & $0.492$ & $0.492$ & $0.492$ & $0.492 $ \\[5pt]
  \hline
\end{tabular}
\end{center}
\caption{Fitted coefficients for $y_b(\mu_0)$ defined in eq.~\eqref{eq:ansatzyb}. The relative fit error is $\delta_{err} \lesssim 10^{-5}$. The symbol ``$-$'' signifies that no sensitivity to such term was found.
\label{tab:fityb}
}
\end{table}

The obtained polynomial expressions provide an efficient way to calculate the \MSbar\ variables at $\mu_0 = 173.22 \,\GeV$ much faster than the more rigorous \SMDR\ tool. At the same time, all the fits give a fit error $\delta_\text{err}$ significantly smaller than the relative parametric error $\delta_\text{par}$ (the calculation of which is described in the next subsection).

\subsubsection{Using fits for initial conditions}

Given simple expressions for converting \onshell\ to \MSbar, we can calculate the initial values of the \MSbar\ variables at the electroweak scale together with the parametric uncertainty originating from the limited precision of the experimental input. As average values, we use the corresponding parameters from the fits with the subscript $0$:
\begin{align*}
	a_1(\mu_0) & = \dfrac{5g_0^{\prime\;2}}{48\pi^2},
	   & a_2(\mu_0)  &= \dfrac{g^{\:2}_0}{16\pi^2} ,
	   & a_3(\mu_0)  &= \dfrac{g_{s0}^{\;\:2}}{16\pi^2}, \\[6pt] 
	      a_t(\mu_0) &= \dfrac{y_{t0}^{\;\,2}}{16\pi^2},
	   &  a_b(\mu_0) &= \dfrac{y_{b0}^{\;\,2}}{16\pi^2},
	    & a_\lambda(\mu_0) &= \dfrac{\lambda_0}{16\pi^2}.
\end{align*}
To estimate parametric uncertainty of these quantities, we use the standard ``propagation of error'' formalism applied to the non-linear functions of the parameters $f(\mathcal O_i)$, i.e., we linearise  in \onshell\ quantities $\mathcal O_i$ around the corresponding central values $\mathcal O_i^{(0)}$, and for uncorrelated $\mathcal{O}_i$ we get the absolute error of $f$ as 
\begin{align}
	\Delta f(\mathcal O_i) = \sqrt{\sum\limits_{i} \left(\left. \frac{\partial f}{\partial \mathcal O_i}\right|_{\mathcal O_i = \mathcal O_i^{(0)}} \cdot \Delta \mathcal O_i \right)^2} = \sqrt{\sum\limits_{i} \left(\left. \mathcal O_i^{(0)} \cdot \frac{\partial f}{\partial \mathcal O_i}\right|_{\mathcal O_i = \mathcal O_i^{(0)}} \cdot \delta_i \right)^2},
\end{align}
where $\Delta \mathcal O_i = \sigma_i$ here is the standard deviation of the corresponding measured \onshell\ quantity, and $\delta_i = \Delta \mathcal O_i / \mathcal O_i^{(0)}$ is the  relative error. 
Due to the definitions of the variables $\hat\delta_i$ that we have chosen, we have their average values zero, and their absolute errors take the form\footnote{Important: an \emph{absolute} error of $\hat\delta_i$ is a \emph{relative} error of $\mathcal O_i$, which is $\delta_i$}:
$$
\begin{array}{c}
     \Delta \hat\delta_i = \delta_i\\
     \Delta(\hat\delta_i\hat\delta_j) = \Delta(\hat\delta_i\hat\delta_j\hat\delta_k) = 0.
\end{array}
$$

This allows us 
to calculate the absolute parametric errors of the \MSbar\ parameters  using the found coefficients of the fits:
\begin{align*}
    \Delta g &= g_0 \sqrt{ \left(c^h_g \delta_h \right)^2 + \left(c^t_g \delta_t \right)^2 + \left( c^Z_g \delta_Z \right)^2 + \left(c^S_g \delta_S \right)^2 + \left( c^a_g \delta_a \right)^2 + \left( c^{GF}_g \delta_{GF} \right)^2 }\\
    \Delta g^\prime &= g^\prime_0 \sqrt{\left( c^h_{g^\prime} \delta_h \right)^2 + \left( c^t_{g^\prime} \delta_t \right)^2 + \left( c^Z_{g^\prime} \delta_Z \right)^2 + \left( c^S_{g^\prime} \delta_S \right)^2 + \left( c^a_{g^\prime} \delta_a \right)^2 + \left(c^{GF}_{g^\prime} \delta_{GF} \right)^2 }\\
    \Delta g_s &= g_{s0} \sqrt{ \left(c^h_{g_s} \delta_h \right)^2 + \left( c^t_{g_s} \delta_t \right)^2 + \left(c^Z_{g_s} \delta_Z \right)^2 + \left(c^S_{g_s} \delta_S \right)^2 }\\
    \Delta y_t &= y_{t0} \sqrt{ \left( c^h_{y_t} \delta_h \right)^2 + \left(c^t_{y_t} \delta_t \right)^2 + \left(c^Z_{y_t} \delta_Z \right)^2 + \left( c^S_{y_t} \delta_S \right)^2 }\\
    \Delta y_b &= y_{b_0} \sqrt{ \left( c^b_{y_b} \delta_b \right)^2  + \left( c^h_{y_b} \delta_h \right)^2  + \left( c^t_{y_b} \delta_t \right)^2 + \left( c^Z_{y_t} \delta_Z \right)^2 + \left(c^S_{y_t} \delta_S \right)^2 }\\
    \Delta \lambda &= \lambda_0 \sqrt{ \left( c^h_{\lambda} \delta_h \right)^2 + \left( c^t_{\lambda} \delta_t \right)^2 + \left( c^Z_{\lambda} \delta_Z \right)^2 + \left( c^S_{\lambda} \delta_S \right)^2 + \left( c^a_{\lambda} \delta_a \right)^2 + \left( c^{GF}_{\lambda} \delta_{GF} \right)^2 + \left( c^b_{\lambda} \delta_b \right)^2 } \\
\end{align*}
and 
\begin{align*}
    \Delta a_1 &= \dfrac{5 g^\prime_0 \Delta g^\prime}{24\pi^2} & \Delta a_2  &= \dfrac{g_0 \Delta g}{8\pi^2} & \Delta a_3  &= \dfrac{g_{s0}\Delta g_s}{8\pi^2} \\[6pt] 
    \Delta a_t &= \dfrac{y_{t0}\Delta y_t}{8\pi^2} &  \Delta a_b &= \dfrac{y_{b0}\Delta y_b}{8\pi^2} & \Delta a_\lambda &= \dfrac{\Delta\lambda}{16\pi^2}\\
\end{align*}
By performing calculations, we obtain the initial conditions at the electroweak scale $\mu_0=173.22~\GeV$ together with their corresponding parametric uncertainties presented in Table.~\ref{tab:paramerr}.

\begin{table}[ht]
    \begin{center}
\begin{tabular}{|c|c|c|c|c|c|c|}
  \hline \rowcolor[HTML]{EFEFEF}
   & $111$ & $222$ & $333$ & $321$ & $432$ & $433$ \\\hline
  $a_1,\,\pow{-3}$ & $1.35636(13)$ & $1.35748(13)$ & $1.35717(13)$ & $1.35710(13)$ & $1.35723(13)$ & $1.35723(13)$ \\
 $a_2,\,\pow{-3}$ & $2.66051(18)$ & $2.65408(21)$ & $2.65602(20)$ & $2.65650(20)$ & $2.65568(20)$ & $2.65569(20)$ \\
 $a_3,\,\pow{-3}$ & $8.58(6)$ & $8.59(6)$ & $8.57(6)$ & $8.57(6)$ & $8.57(6)$ & $8.57(6)$ \\
 $a_t,\,\pow{-3}$ & $6.222(21)$ & $5.669(21)$ & $5.543(21)$ & $5.671(21)$ & $5.543(21)$ & $5.543(21)$ \\
 $a_b,\,\pow{-6}$ & $1.786(14)$ & $1.780(14)$ & $1.548(17)$ & $1.784(14)$ & $1.548(17)$ & $1.548(17)$ \\
 $a_\lambda,\,\pow{-4}$ & $8.187(14)$ & $8.032(14)$ & $7.999(14)$ & $8.160(14)$ & $8.022(14)$ & $7.999(14)$ \\
  \hline
\end{tabular}
\end{center}
\caption{Initial conditions of renormalization group equations with parametric uncertainty (values at the scale  $173.22 \,\GeV$) depending on loops\label{tab:paramerr}}
\end{table}

As it turned out, parametric uncertainties ended up to be much larger than $\delta_\text{err}$ of the fits discussed in the previous subsections. This shows that the fits can actually be used to transform \onshell\ parameters into \MSbar\ quantities, as was performed above. As we already mentioned above, we have checked that the ``statistical'' approach provides the same results.

\begin{figure}[!t]
\centering
\begin{tabular}{cc}
	\includegraphics[width = 0.475\textwidth]{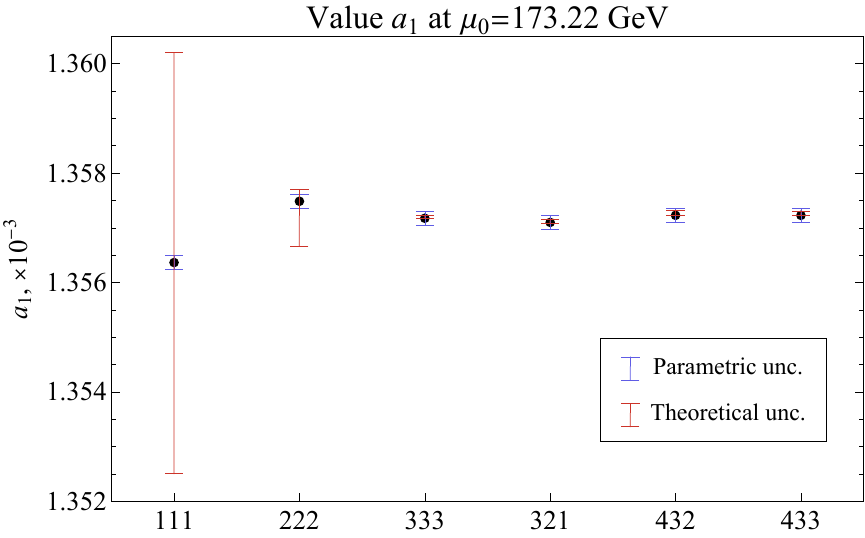} &
	\includegraphics[width = 0.475\textwidth]{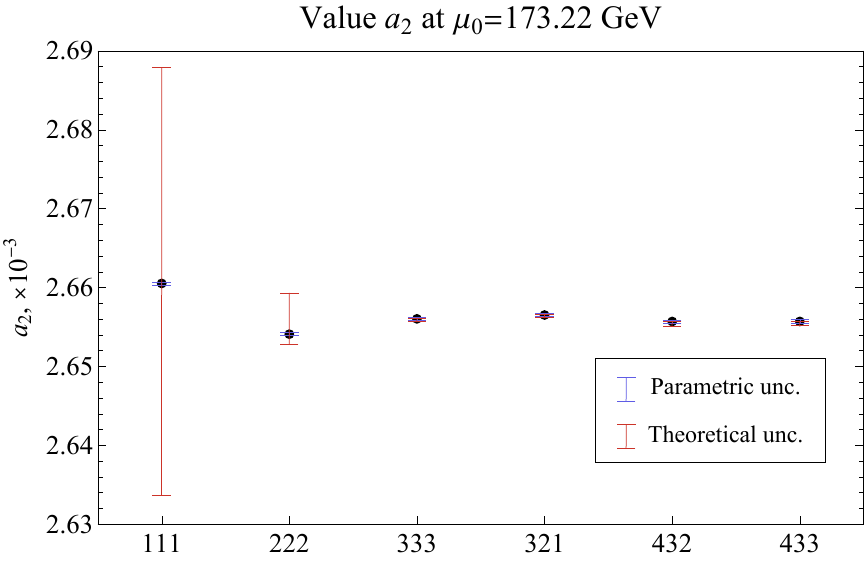} \\
	\includegraphics[width = 0.475\textwidth]{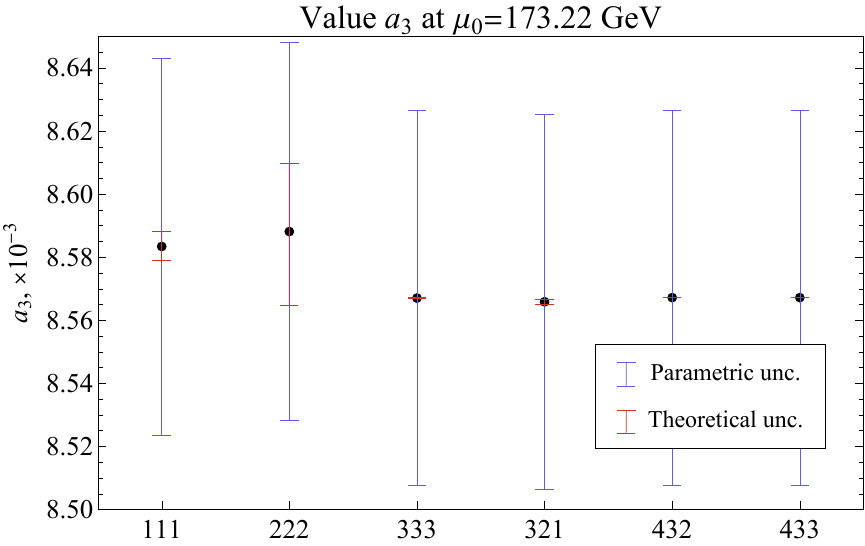} &
	\includegraphics[width = 0.475\textwidth]{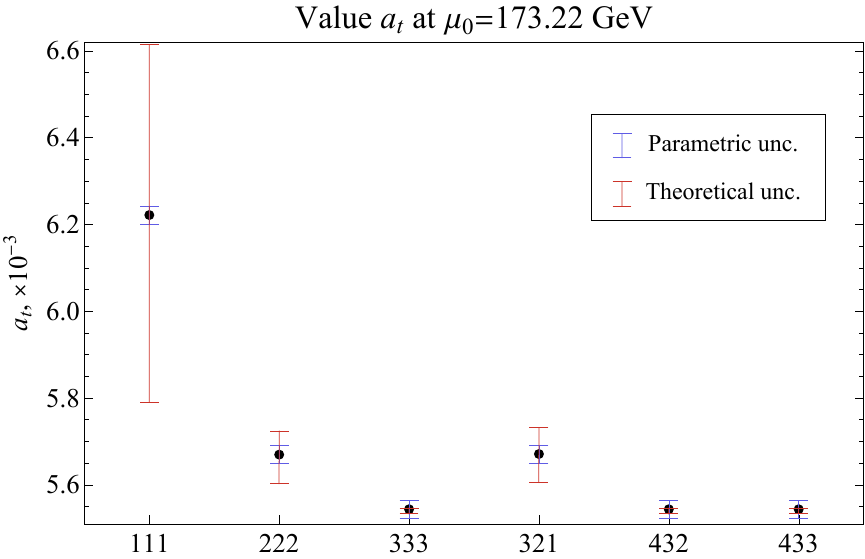} \\
	\includegraphics[width = 0.475\textwidth]{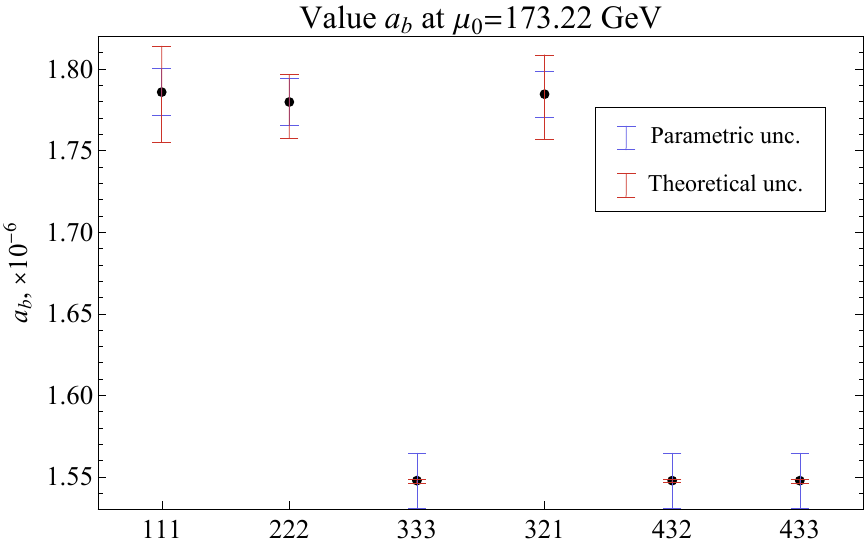} &
	\includegraphics[width = 0.475\textwidth]{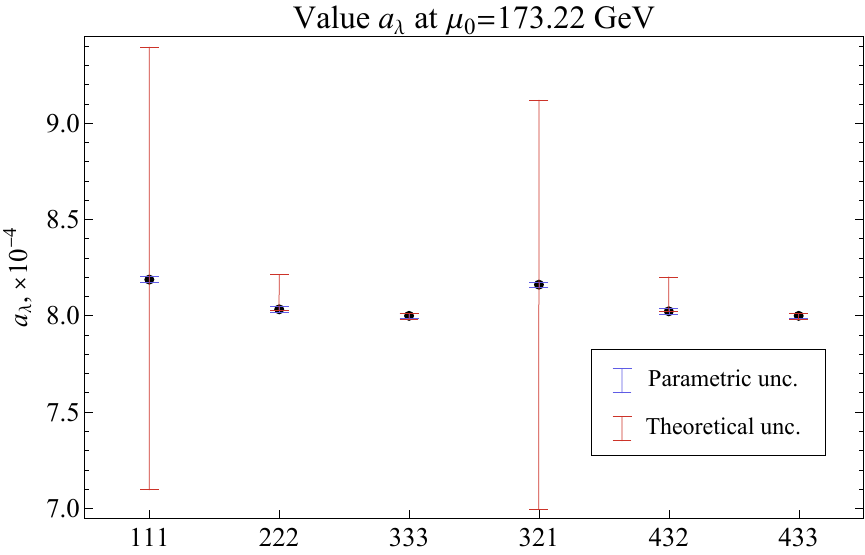} \\    
\end{tabular}
\caption{\MSbar\ quantities at $\mu_0 = 173.22~\GeV$ with parametric and theoretical uncertainties\label{fig:bc_uncertanties}}
\end{figure}

We summarize our study of the initial condition and our uncertainty estimates for different loop configurations in Fig.~\ref{fig:bc_uncertanties}. In what follows we use these results when studying solutions of the SM RGE equations beyond the one-loop level. Our main focus will be on the running of the Higgs boson self-coupling $\lambda(\mu)$ due to its relation to the electroweak vacuum stability discussed in the next section.

\section{Problem of (meta)stability of the SM vacuum\label{sec:stability}}

It is expected that taking into account the quantum corrections $\Delta V$ to the tree-level potential will not change $v$ much compared to the tree value $v_\text{tree}$, which justifies the calculation of $\Delta V(\phi)$ using perturbation theory. However, quantum effects can also lead to the appearance of an additional minimum in $V_\text{eff}(\phi)$ at $\phi = \tilde v \gg v$. Depending on the values of the SM parameters, the additional minimum can be either false $V_{\text{eff}}(\tilde v) \geq V_{\text{eff}}(v)$ or true $V_{\text{eff}}(\tilde v) < V_{\text{eff}}(v)$. In the latter case, there is a possibility of decay of the electroweak vacuum, and an important criterion for the self-consistency of the SM is the stability of the latter. It is obvious that the boundary of absolute stability is the degeneracy condition $V_{\text{eff}}(\tilde v) = V_{\text{eff}}(v)$. In the case of the SM, calculations show that the electroweak vacuum may be unstable. However, the lifetime of the latter is quite large and the parameters of the model lie quite far from the boundary of the metastability region, corresponding to the unit probability of the decay of the electroweak vacuum throughout the history of the Universe.

Computing $V_{\text{eff}}(\phi)$ in higher orders of the PT is a non-trivial task and it is based on calculating the path integral using the background-field formalism proposed in Ref.~\cite{Jackiw:1974cv}. The main technical difficulties are related to the gauge dependence of $V_{\text{eff}}(\phi)$ (see, for example, Refs.~\cite{Jackiw:1974cv,Martin:2018emo,Andreassen:2014eha}) as well as ``spurious'' infrared divergences arising in calculations in the finite order of perturbation theory \cite{Einhorn:2007rv,Elias-Miro:2014pca,Martin:2014bca,Braathen:2016cqe,Espinosa:2016uaw,Espinosa:2017aew}. This complicates obtaining physical, gauge-independent results and requires re-summation of various contributions in all orders of PT.

Moreover, attempts to extend the effective potential $V_{\text{eff}}(\phi)$ calculated in the final PT order to large field values lead to a violation of the PT applicability conditions. This problem is also solved by rearrangement of the PT series using the renormalization group equation satisfied by $V_\text{eff}(\phi)$ (see, e.g., Refs.~\cite{Ford:1992pn,Martin:2017lqn}).

In Refs.~\cite{Buttazzo:2013uya,Bednyakov:2015sca}, the stability of the SM vacuum at the three-loop level was investigated. It was shown that the values of the key SM parameters affecting the analysis and related to the top quark mass, the Higgs boson mass, and the strong constant are very close to the absolute stability boundary\footnote{Recent discussion on the absolute stability in the SM and some of its extensions can be found in Ref.~\cite{Hiller:2024zjp}.}  in the metastability region. In this paper, we assume that the electroweak vacuum is false and investigate how the high-order terms in the SM RGE affect the calculated rate for the decay.

The lifetime of the false vacuum is related to the rate of bubble formation of the true vacuum in a unit volume per unit time (see, for example, the review \cite{Devoto:2022qen}):
\begin{align}
	\gamma = \mathcal{A} e^{-\mathcal{B}},
	\label{eq:decay_rate_def}
\end{align}
where $\mathcal{B}$ is the action for the bounce solution \cite{Coleman:1977py} and the pre-exponential factor describes the quantum corrections \cite{Callan:1977pt}. The calculation is based on using an approximate expression for the potential for the neutral component of the Higgs field $\phi$ in the region $\phi \gg v$
\begin{align}
	V_{\mathrm{eff}}(\phi) \simeq \dfrac{1}{4} \lambda(\phi) \phi^4. 
	\label{eq:V_eff_simple}
\end{align}
In Eq.\eqref{eq:V_eff_simple} we neglected the mass term, and the running self-coupling constant $\lambda(\mu)$ is calculated at $\mu \simeq \phi$. The spherically symmetric solution in four-dimensional Euclidean space $\phi_b(r)$ of the "bounce" type satisfies equation
\begin{align}
	\dfrac{d^2\phi_b}{d r^2} + \dfrac{3}{r} \dfrac{d \phi_b}{d r} = V_{\mathrm{eff}}'(\phi_b) \simeq \phi_b^3 \left[ \lambda(\phi_b) + \dfrac{1}{4} \bar \beta_\lambda(\phi_b)\right], \qquad \lim\limits_{r\to\infty} \phi_b = v \simeq 0, \quad \dfrac{d \phi}{dr}|_{r=0} = 0,
	\label{eq:bounce_equation}
\end{align}
where
\begin{align}
\bar \beta_\lambda(\mu) = \dfrac{d \lambda(\mu)}{d \ln \mu} = 2 \dfrac{d \lambda(\mu)}{d \ln \mu^2}.
\end{align}
Equation \eqref{eq:bounce_equation} can be simplified by approximating the expression in square brackets with a constant:
\begin{align}
   \left[\lambda(\phi_b) + \dfrac{1}{4} \bar \beta_\lambda(\phi_b)\right] \simeq  \lambda^*<0, \quad   \lambda^* \equiv \lambda(\mu^*):\ \bar \beta_\lambda(\mu^*) = 0
\end{align}
Thus, we obtain 
\begin{align}
   \dfrac{d^2\phi_b}{d r^2} + \dfrac{3}{r} \dfrac{d \phi_b}{d r} = - |\lambda^*| \phi_b^3,
   \label{eq:bounce_equation_simple}
\end{align}
with the solution (Fubini-Lipatov instanton
\cite{Fubini:1976jm,Lipatov:1977hj})
\begin{align}
	\phi_b(r) = \sqrt{\dfrac{8}{|\lambda^*|}} \dfrac{R}{r^2 + R^2}.
	\label{eq:fubini_lipatov_instanton}
\end{align}
The latter depends on the arbitrary dimensional parameter $R$ related to the initial condition 
\begin{align}
    \phi_b(0) = \sqrt{\dfrac{8}{|\lambda^*|}} \dfrac{1}{R}, \quad \phi_b(R) = \sqrt{\dfrac{8}{|\lambda^*|}} \dfrac{1}{2R} = \dfrac{\phi_b(0)}{2}, 
\end{align}
and the radius of the instanton. The scale invariance of the classical potential (in the limit of zero mass) leads to the fact that the value of the action for the "bounce" does not depend on the parameter $R$
\begin{align}
	\mathcal{B} & = S[\phi_b] = 2 \pi^2 \int r^3 d r \left( \dfrac{1}{2} [\partial_r \phi_b(r)]^2 + \dfrac{\lambda^*}{4} \phi_b^4(r)	\right) 
		= \dfrac{8\pi^2}{3|\lambda^*|}
	\equiv \frac{1}{6 |a^*_\lambda|} 
		.
\end{align}
However, when quantum corrections are taken into account, the degeneracy is lifted and, roughly speaking, the characteristic instanton size $R^*$, which gives the main contribution to the lifetime, is related to the scale $\mu^* \sim 1/R^*$, at which $|\lambda^*|$ is maximal. Thus, from dimensional considerations (see, for a more accurate treatment, e.g., Ref.~\cite{Andreassen:2017rzq}), the width of the vacuum decay per unit volume is given by
\begin{align}
\frac{\Gamma}{V} \equiv \gamma \approx (\mu^{*})^4 \mathcal{\tilde A} e^{- \frac{1}{6 |a_\lambda^*|}}, \qquad \beta_\lambda(\mu^*) = 0,
	\label{eq:gamma_approx}
\end{align}
where $\mathcal{\tilde A}$ is a dimensionless quantity determined by fluctuations \cite{Isidori:2001bm} of quantum fields near the ``bounce'' solution (see, e.g., recent \cite{Baratella:2024hju} and references therein). The probability of the electroweak vacuum decay can be estimated as the probability of forming a bubble of true vacuum in the visible part of the Universe over the entire history of the latter \cite{Buttazzo:2013uya}, i.e., it is necessary to multiply $\gamma$ by the four-volume of the today's past light-cone $(VT)_{\text{light-cone}} \approx 0.15/H_0^4$, where $H_0 \approx 70~(\text{km/s})\cdot\text{Mpc}^{-1} \approx 1.44 \pow{-42}\,\GeV$, giving rise to
\begin{align}
	\mathcal{P} = \gamma \cdot (V T)_{\text{light-cone}} & \approx  0.15 \cdot \mathcal{\tilde A} \cdot \frac{\mu^{*\,4}}{H_0^4}  e^{-\frac{1}{6 |a_\lambda^*|}}, \qquad
	(V T)_{\text{light-cone}} \approx 3.5 \pow{166} \cdot \GeV^{-4}. 
	\label{eq:vacuum_decay_prop}
\end{align}

It what follows, we first study the running of $\lambda$ for different loop configurations and compute $t^* = 2\ln (\mu^*/\mu_0)$ together with $\lambda^*$. Then we evaluate $\log_{10} \mathcal{P}$ and analyze its dominant uncertainties.

\subsection{Running of $\lambda$: from electroweak to Planck scale}

Let us fix again the initial condition \eqref{eq:fixed_initial_condition} at the electroweak scale $\mu_0 = 173.22$ GeV and demonstrate the influence of higher PT orders\footnote{Ignoring for the moment the fact that the boundary conditions also depend on the loop configuration}  in various beta functions $\beta_a$ on the behavior of $a_\lambda(\mu)$ near the scale\footnote{This scale is sometimes associated with the gauge-dependent ``instability scale of the SM electroweak vacuum'' (see., e.g., Ref.\cite{DiLuzio:2015iua})} $\mu_r$ at which $a_\lambda(\mu_r) = 0$. Since the UV asymptotics of the non-Abelian and Abelian gauge coupling constants in the SM are different, we consider separately the contributions to each of the gauge beta functions. In this section we use the refined $|L_{a_1},L_{a_2},L_{a_3}|L_{a_t},L_{a_b}|L_{a_\lambda}|$ notation to specify loop orders in RGEs for each considered coupling. 

\begin{figure}[!t]
	\centering
	\begin{tabular}{cc}
		\includegraphics{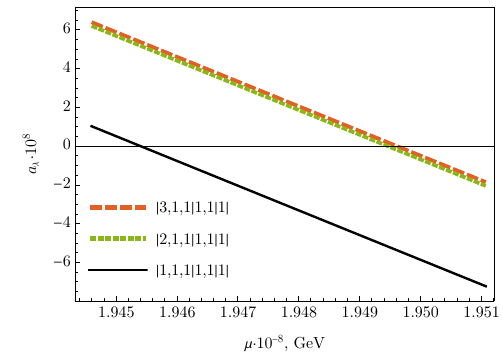} & \includegraphics{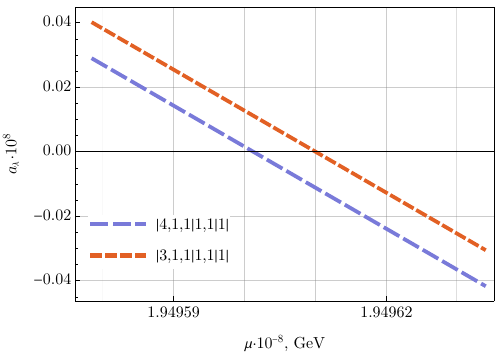} \\
		\includegraphics{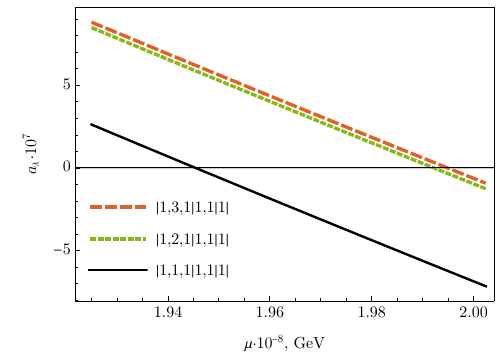} & \includegraphics{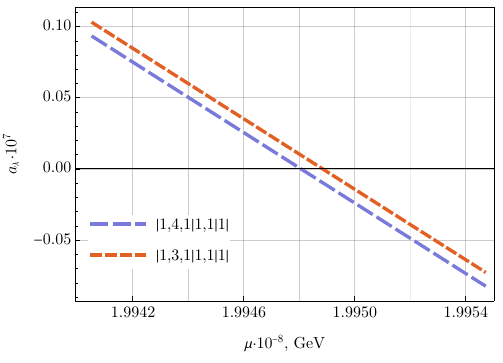} \\
		\includegraphics{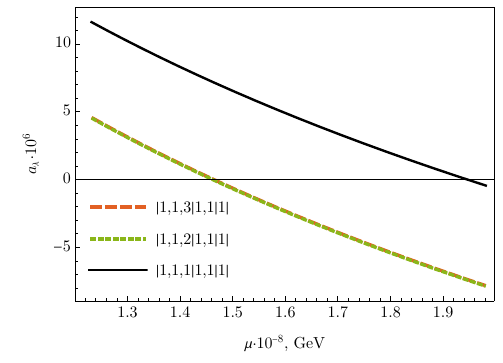} & \includegraphics{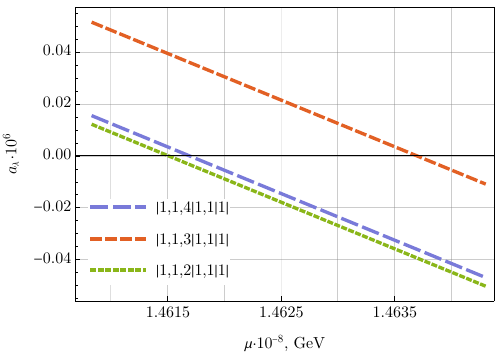} 
\end{tabular}
	
\caption{Behavior of $a_\lambda(\mu)$ near $a_\lambda = 0$ taking into account different orders in the gauge beta functions. The initial values are taken to be the same \eqref{eq:fixed_initial_condition} and only the PT orders are changed in the case of $a_1$ (upper plots), $a_2$ (middle plots), and $a_3$ (lower plots). The left column compares the one-loop running with two and higher loops.
	The right column demonstrates the difference between the latter.   
	For the remaining coupling constants the one-loop result is used.
}\label{fig:x123contrib}
\end{figure}

In Fig.~\ref{fig:x123contrib} we show how different contributions to gauge-coupling beta functions affect the running of $\alpha_\lambda(\mu)$ near the $\mu_r$ scale.  
The effect of adding the higher-order contributions to the beta functions of $a_1$ and $a_2$ is that $\dot a_\lambda$, negative at the electroweak scale, decreases in absolute value with $\mu$, thus
pushing the scale $\mu_r: a_\lambda(\mu_r) = 0$ to the region of larger $\mu$. Even though $a_3$ does not enter into $\dot a_\lambda$ at the one-loop level, its value affects the running of the Yukawa couplings and indirectly $a_\lambda$. Adding higher orders to $\beta_{3}$ ``accelerates'' the approach of $a_3$ to zero in the UV. This in turn makes $a_t$ decrease more slowly with $\mu$. This increases the negative contribution $-3 a_t^2$ to one-loop $\beta_\lambda$ so that the scale  $\mu_r$ decreases compared to the case $|1,1,1|1,1|1|$. As can be seen, the transition to three loops somewhat strengthens the two-loop effect in the cases of $a_1$ and $a_2$, and weakens it in the case of $a_3$. The four-loop contribution to the gauge beta functions in all cases gives an effect opposite to the three-loop one.

\begin{figure}[!t]
	\centering
	\begin{tabular}{cc}
		\includegraphics{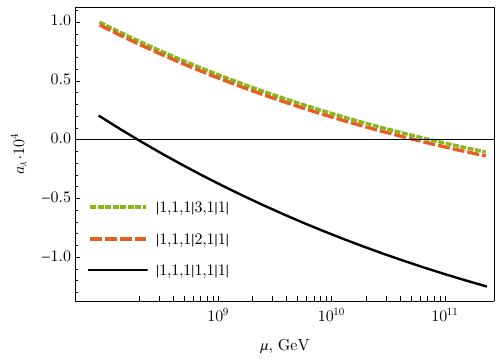}&
		\includegraphics{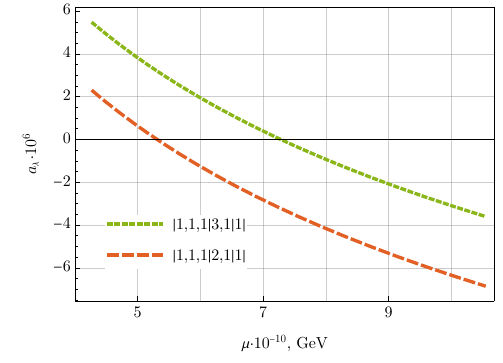} \\
		\includegraphics{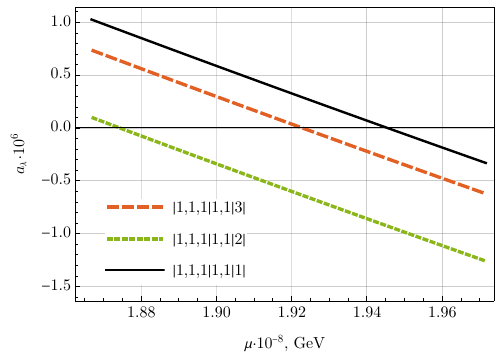}&
		\includegraphics{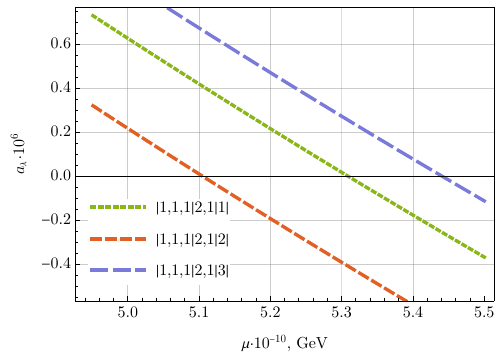} 
\end{tabular}
\caption{Behavior of $a_\lambda(\mu)$ near $a_\lambda = 0$ taking into account different orders in the beta functions for $a_t$ (upper left and upper right) and $a_\lambda$ (lower left). A combined effect of high-order terms in $\beta_t$ and $\beta_\lambda$ (lower right) is demonstrated in the right plot.  
	The initial values are taken fixed \eqref{eq:fixed_initial_condition} for all cases. 
}\label{fig:y1zcontrib}
\end{figure}

Figure~\ref{fig:y1zcontrib} shows the running of $\alpha_\lambda(\mu)$ for $\mu \approx \mu_r$, but now we separately turn on high-order terms in RGE for top-quark Yukawa (upper plots), $a_\lambda$ itself (lower left panel), and in both beta functions (lower right panel). All other couplings run at one loop. 
	Adding more loops to the beta function $a_t$ leads to $a_\lambda$ decreasing more slowly ($a_t$ tends to zero more quickly). 
	Taking into account the second loop in $\dot a_\lambda$ decreases $\mu_r$ and the third one increases $\mu_r$ again. 
	It is interesting to note that when using one-loop expressions for $\dot a_t$, the curve corresponding to the three-loop $\beta_\lambda$ lies between the curves for one and two loops. 
	When using two- or three-loop $\beta_t$, the three-loop $\beta_\lambda$ leads to the largest $\mu_r$.

\begin{figure}[!ht]
	\centering
	\begin{tabular}{ccc}
		\includegraphics[width=0.33\textwidth]{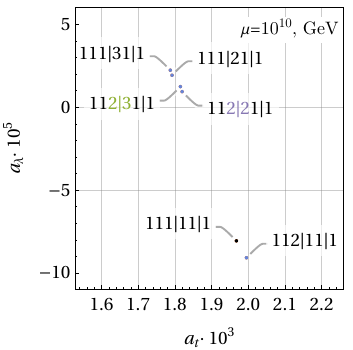} &
		\includegraphics[width=0.33\textwidth]{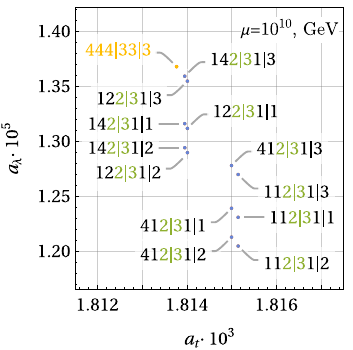} &
		\includegraphics[width=0.33\textwidth]{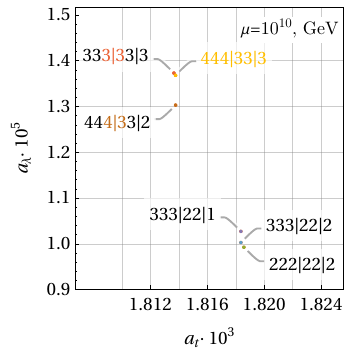} 
\end{tabular}
	
    \caption{Dependence of the running $a_\lambda$ and $a_t$ at the $10^{10}$ GeV scale on the used beta function orders. The initial values are taken to be the same at the $\mu_0$ scale.
The contributions to $\beta_{3}$ and $\beta_t$ have the greatest influence. 
Taking into account higher corrections in the remaining beta functions does not change the result much (middle and right plots). 
}\label{fig:y1z_other_contribs}
\end{figure}

Finally,   we solve SM RGEs with the initial conditions \eqref{eq:fixed_initial_condition} 
and provide the values of running $a_\lambda$ and $a_t$ at fixed $\bar \mu = 10^{10}$ \GeV. 
We start from the $|1,1,1|1,1|1|$ configuration and turn on different loop corrections one by one.
In Fig.~\ref{fig:y1z_other_contribs} we demonstrate typical shifts that we obtain in this kind of study.  The strongest effect comes from high-order corrections to $\beta_{3}$ and $\beta_t$.
The one-loop $\beta_t$ leads to negative values of $a_\lambda(\bar\mu)$. The latter becomes even more negative if the two-loop $\beta_{3}$ is used, since in this case $a_t$ decreases more slowly in the UV. 
Adding higher orders to $\beta_t$ improves the situation, reduces $a_t(\bar\mu)$ and leads to positive $a_\lambda(\bar\mu)$ (left plot in Fig.~\ref{fig:y1z_other_contribs}).

At the same time, the considered couplings exhibit a less pronounced dependence on high-order corrections to the remaining beta functions. If we stick to the three-loop $\beta_t$ and the two-loop $\beta_{3}$ and consider different PT orders in other beta functions, we get the shifts given in the middle panel of Fig.~\ref{fig:y1z_other_contribs}. Higher corrections to the electroweak beta functions tend to decrease $a_t(\bar \mu)$ and increase $a_\lambda(\bar\mu)$ . The two-loop contributions to $\beta_\lambda$ decrease $a_\lambda(\bar\mu)$ compared to the one-loop case, while adding three-loop terms to $\beta_\lambda$ makes $a_\lambda(\bar \mu)$ larger (see, the middle and right parts of Fig.~\ref{fig:y1z_other_contribs}).

Let us now take into account the uncertainties in the boundary conditions given in Table~\ref{tab:paramerr}. 
Our main focus will be on the computation of two characteristic scales $\mu_r$ and $\mu^*$ corresponding to zero ($a_\lambda(\mu_r) = 0$) and minimal ($a_\lambda(\mu^*) = a^*_\lambda$) values of the Higgs boson coupling.
Given the results of the previous section, we estimate the parametric and theoretical uncertainties of $t_r \equiv 2\ln\mu_r/\mu_0$, $t^* \equiv  2 \ln \mu^*/\mu_0$, and $a_\lambda^*$, together with that of the vacuum decay probability \eqref{eq:vacuum_decay_prop} .
\begin{figure}[!ht]
  \centering
  \includegraphics[width = 10cm]{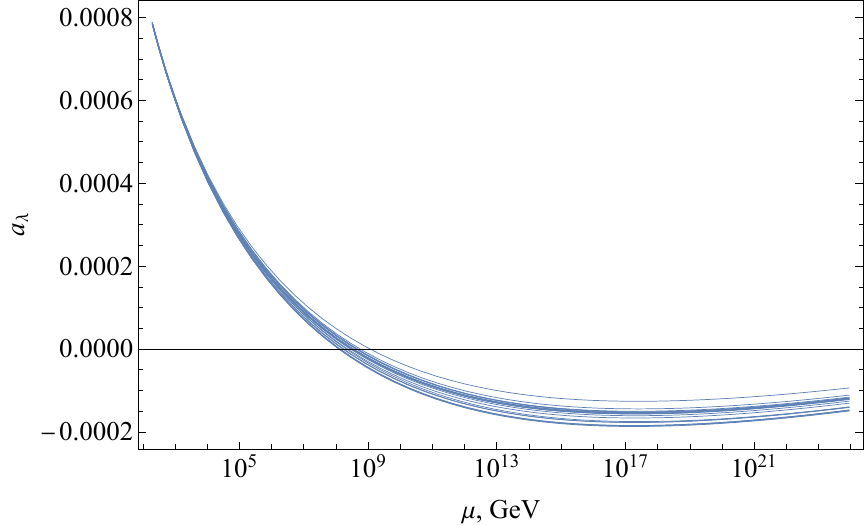}
  \caption{An example of 20 curves for $222$ with randomly-generated initial conditions\label{fig:exampleofrandom}}
  
\end{figure}

\begin{figure}[h]
\centering
  \includegraphics[width = 0.31\linewidth]{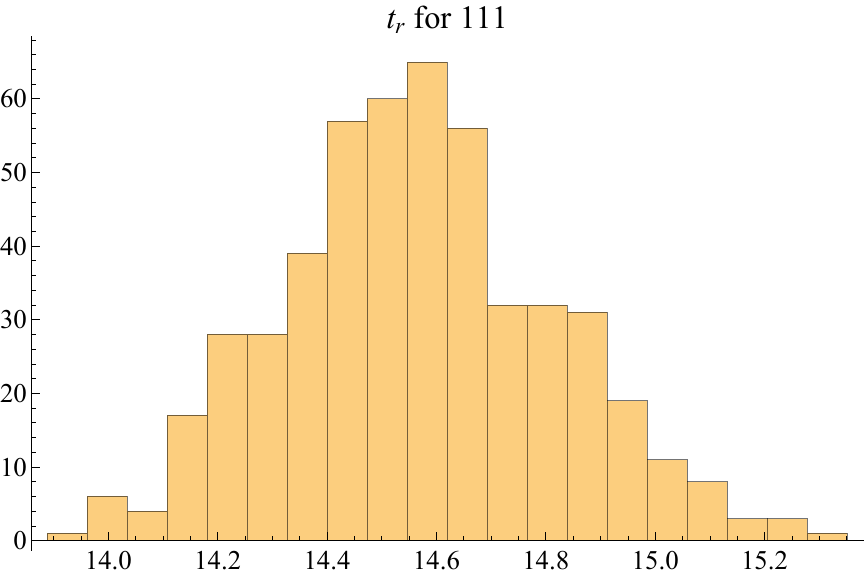}
  \includegraphics[width = 0.31\linewidth]{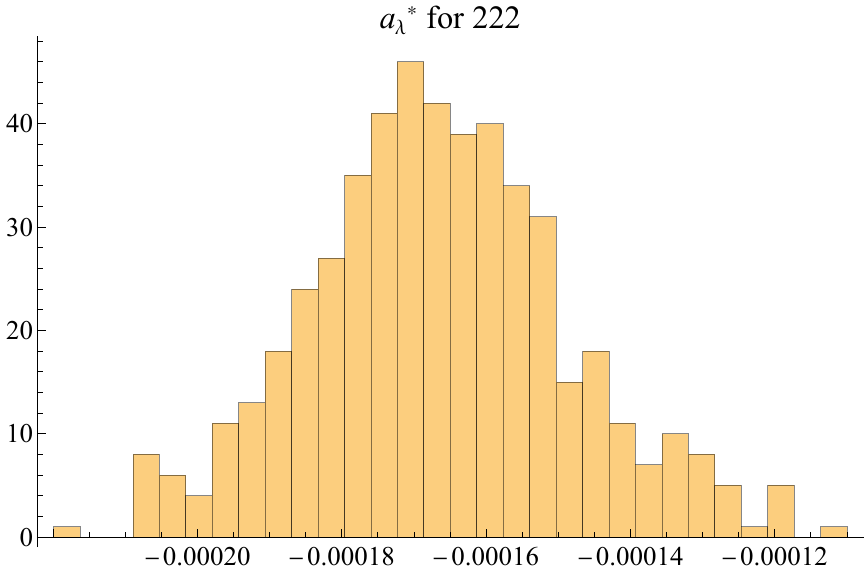}
  \includegraphics[width = 0.31\linewidth]{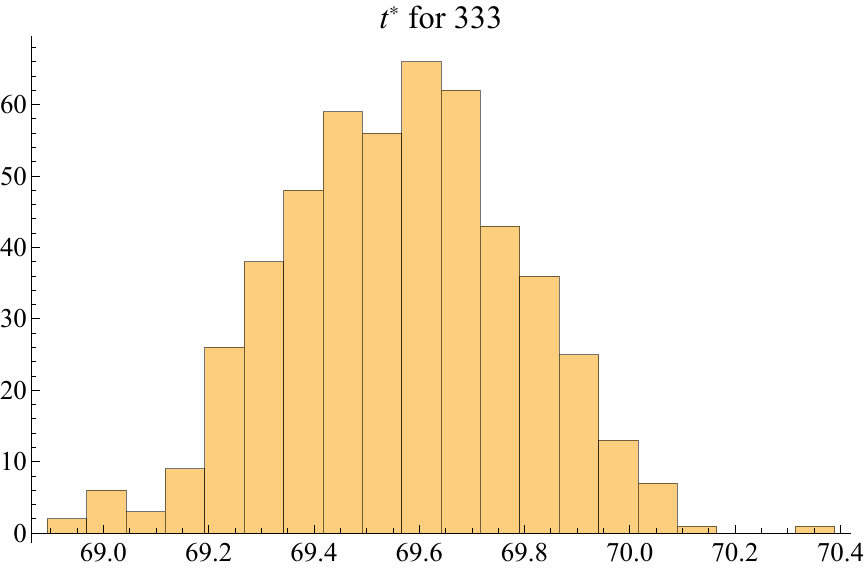}
  \caption{Distributions of $t_r$ for $111$, $a_\lambda^*$ for $222$, and  $t^*$ for $333$ generated 
  for a sample of 500 random boundary conditions.\label{fig:tr_ts_alam_distributions}}
\end{figure}

The \emph{parametric error} will be estimated by generating random curves with the distribution of the initial conditions according to Table~\ref{tab:paramerr}. 
We considered a sample of $500$ curves for each configuration (see,e.g., Fig.~\ref{fig:exampleofrandom}). The distributions of the resulting $t_r,\, t^*$, and $a_\lambda^*$ turn out to be very close to the Gaussian one (Fig.~\ref{fig:tr_ts_alam_distributions}).

By processing the generated random data, we obtain the mean values and the parametric uncertainties for each considered loop configuration and provide them in Table.~\ref{tab:paramresult}.
The values of $\mu_r$ and $\mu^*$ are given in \GeV. They have asymmetric errors, since they depend exponentially on $t_r$ and $t^*$, respectively, which are normally distributed.

\begin{table}[!h]
\begin{center}
\begin{tabular}{|c|c|c|c|}
  \hline \rowcolor[HTML]{EFEFEF}
   & $111$ & $222$ & $333$ \\[6pt] \hline
 $t_r$ & $14.57(25)$ & $28.3(12)$ & $35.3(22)$ \\[6pt]
 $\mu_r, \GeV$ & $\left(2.53_{-0.30}^{+0.34}\right)\pow5$ & $\left(2.4_{-1.1}^{+2.1}\right)\pow8$ & $\left(0.8_{-0.5}^{+1.6}\right)\pow{10}$ \\[6pt] \hline
 $a_\lambda^*, \pow{-4}$ & $-6.51(20)$ & $-1.65(18)$ & $-0.91(17)$ \\[6pt] \hline
 $t^*$ & $63.23(27)$ & $68.80(26)$ & $69.61(22)$ \\[6pt]
 $\mu^*, \GeV$ & $\left(9.3_{-1.2}^{+1.4}\right)\pow{15}$ & $\left(1.51_{-0.18}^{+0.21}\right)\pow{17}$ & $\left(2.26_{-0.24}^{+0.27}\right)\pow{17}$ \\[6pt]
  \hline\hline \rowcolor[HTML]{EFEFEF}
  & $321$ & $432$ & $433$ \\[6pt] \hline
 $t_r$ & $29.9(26)$ & $35.8(26)$ & $35.3(28)$ \\[6pt]
 $\mu_r, \GeV$ & $\left(0.5_{-0.4}^{+1.5}\right)\pow9$ & $\left(1.0_{-0.8}^{+2.8}\right)\pow{10}$ & $\left(0.8_{-0.6}^{+2.5}\right)\pow{10}$ \\[6pt] \hline
 $a_\lambda^*, \pow{-4}$ & $-1.51(17)$ & $-0.88(16)$ & $-0.91(17)$ \\[6pt] \hline
 $t^*$ & $69.62(23)$ & $69.74(23)$ & $69.61(23)$ \\[6pt]
 $\mu^*, \GeV$ & $\left(2.27_{-0.24}^{+0.27}\right)\pow{17}$ & $\left(2.41_{-0.26}^{+0.29}\right)\pow{17}$ & $\left(2.26_{-0.25}^{+0.28}\right)\pow{17}$ \\[6pt]
 \hline
\end{tabular}
\end{center}
\caption{Values of $t_r,\,\mu_r,\,z^*,\,t^*,\,\mu^*$ with parametric uncertainty depending on the loops\label{tab:paramresult}}
\end{table}

\begin{figure}[!h]
\centering
  \includegraphics[width = 0.48\textwidth]{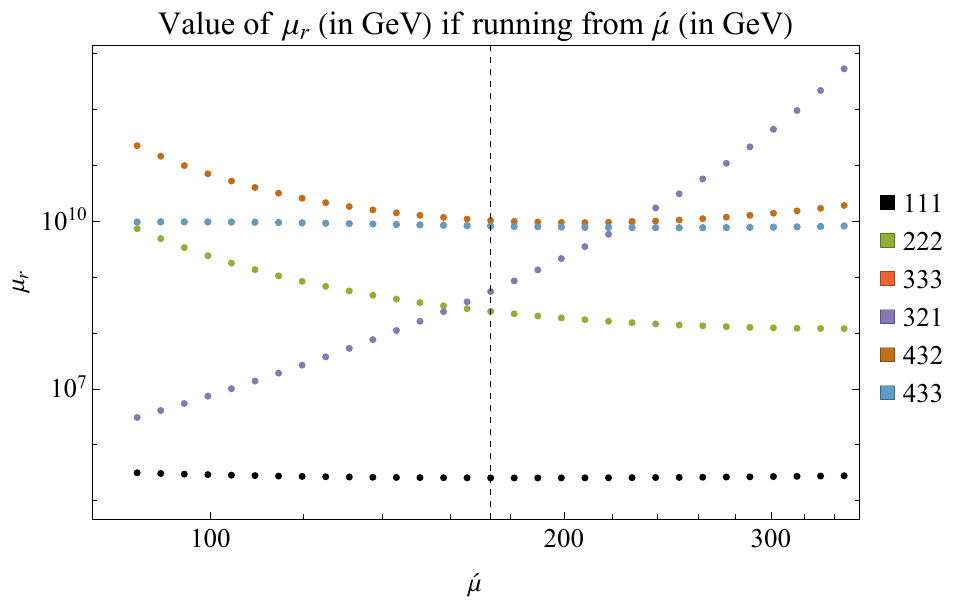}
  \includegraphics[width = 0.48\textwidth]{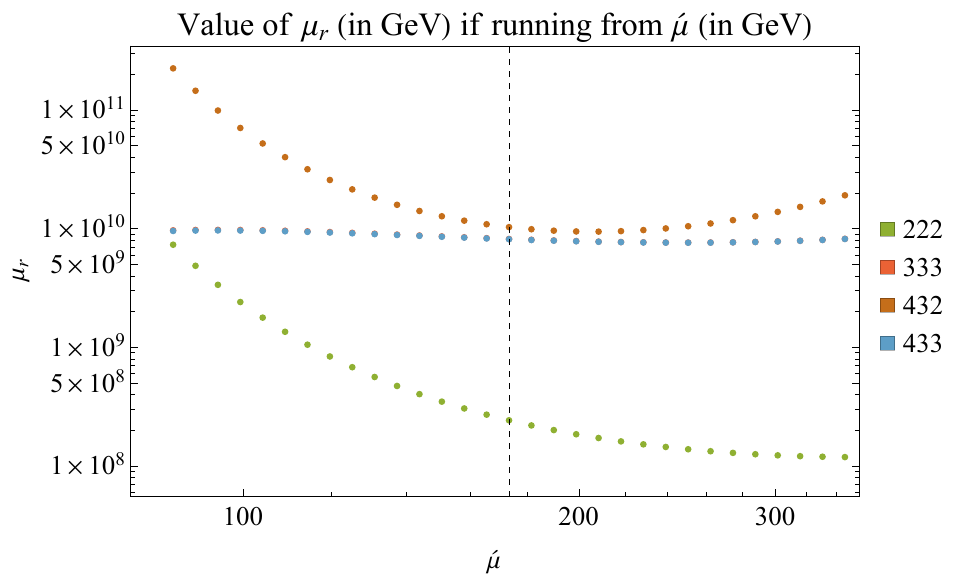}
  \caption{Dependence of $\mu_r$ on the matching scale $\acute\mu$ for different loop configurations\label{fig:th_mu_r}}
\end{figure}

\begin{figure}[!h]
\centering
  \includegraphics[width = 0.47\textwidth]{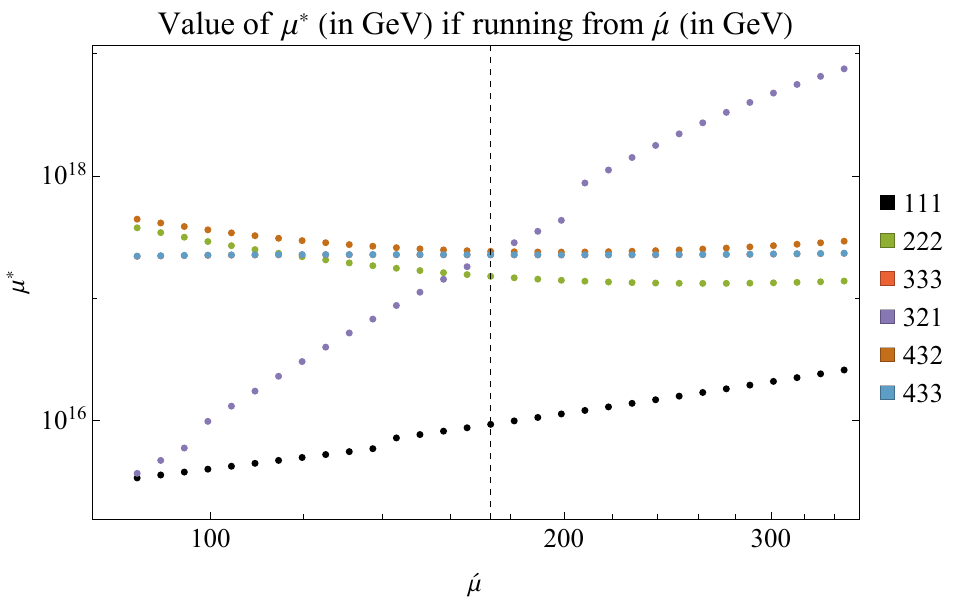}
  \includegraphics[width = 0.49\textwidth]{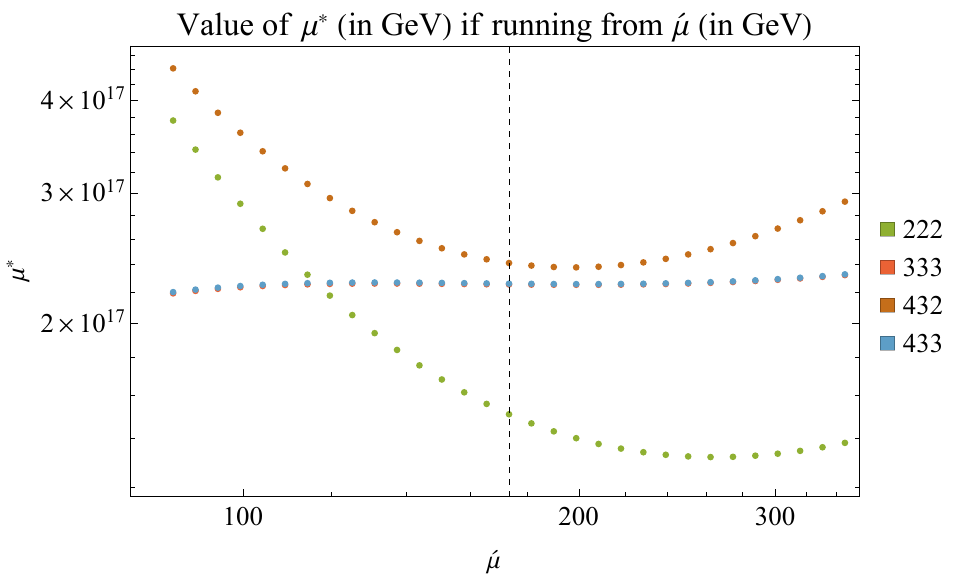}
  \caption{Dependence of $\mu^*$ on the matching scale $\acute\mu$ for different loop configurations \label{fig:th_mu_star}}
\end{figure}

\begin{figure}[!h]
\centering
  \includegraphics[width = 0.47\textwidth]{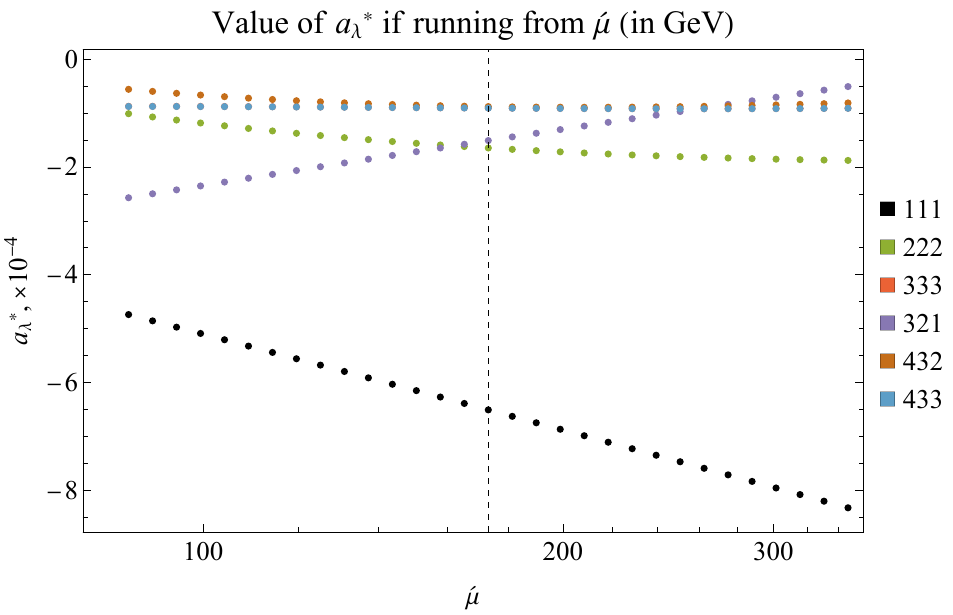}
  \includegraphics[width = 0.49\textwidth]{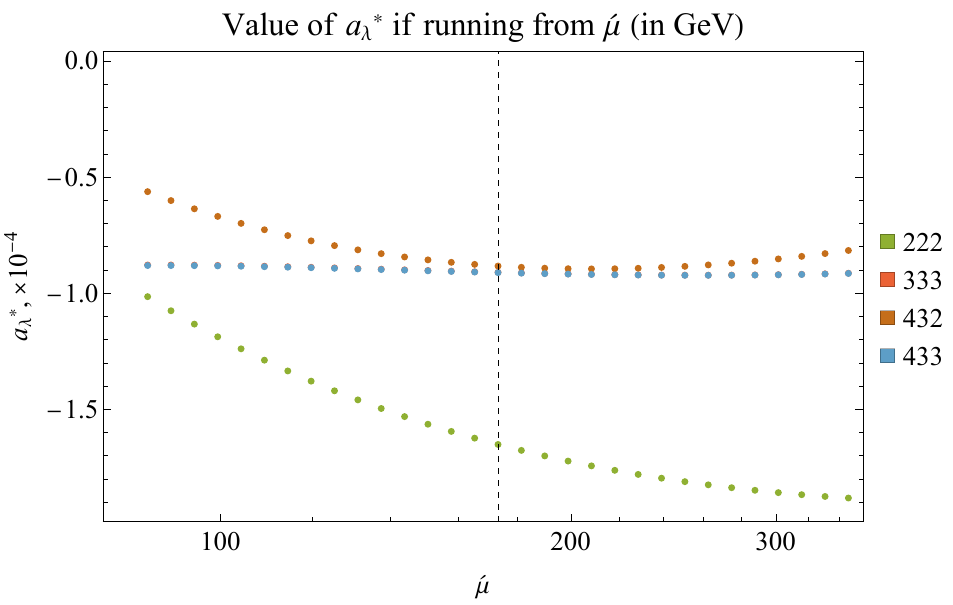}
  \caption{Dependence of $a_\lambda^*$ on the matching scale $\acute\mu$  for different loop configurations\label{fig:th_z_star}}
\end{figure}

Let us proceed to theoretical uncertainties. 
We follow the same procedure that we used in Sec.~\ref{sec:th_unc_couplings}. We consider $\mu_r(\acute \mu)$, $\mu^*(\acute \mu)$ and $a^*_\lambda(\acute \mu)$ as functions of the matching scale $\acute \mu$  and compute the corresponding values on the same grid of points (see Figs.~\ref{fig:th_mu_r},\ref{fig:th_mu_star},\ref{fig:th_z_star}). 
To estimate theoretical uncertainties, we take minimal and maximal values of $\mu_r$, $\mu^*$, and $a^*_\lambda$ on this interval $\acute\mu \in (\mu_0/2, 2\mu_0)$ for each loop configuration (see Table.~\ref{tab:th_mu_r_mu_star_z_star_sum}).

\begin{table}[!h]
    \centering
    \begin{tabular}{|c|c|c|c|c|}
  \hline \rowcolor[HTML]{EFEFEF}
  \multicolumn{2}{|c|}{} & $111$ & $222$ & $333$ \\[6pt] \hline
 & $\text{max}_\text{th}$ & $3.1224 \pow{5}$ & $7.2999 \pow{9}$ & $9.7304 \pow{9}$ \\[6pt]
 \multirow{-2}{*}{$\mu_r, \GeV$} & $\text{min}_\text{th}$ & $2.5259 \pow{5}$ & $1.1859 \pow{8}$ & $7.6057 \pow{9}$ \\[6pt] \hline
  & $\text{max}_\text{th}$ & $-4.74192$ & $-1.0157$ & $-0.8794$ \\[6pt]
 \multirow{-2}{*}{$a_\lambda^*, \pow{-4}$} & $\text{min}_\text{th}$ & $-8.32633$ & $-1.88267$ & $-0.92327$ \\[6pt] \hline
  & $\text{max}_\text{th}$ & $2.5861 \pow{16}$ & $3.7545 \pow{17}$ & $2.3223 \pow{17}$ \\[6pt]
 \multirow{-2}{*}{$\mu^*, \GeV$} & $\text{min}_\text{th}$ & $3.3925 \pow{15}$ & $1.3191 \pow{17}$ & $2.1944 \pow{17}$ \\[6pt] 
\hline \hline \rowcolor[HTML]{EFEFEF}
   \multicolumn{2}{|c|}{} & $321$ & $432$ & $433$ \\[6pt] \hline
    & $\text{max}_\text{th}$ & $5.3054 \pow{12}$ & $2.2344 \pow{11}$ & $9.6287 \pow{9}$ \\[6pt]
 \multirow{-2}{*}{$\mu_r, \GeV$} & $\text{min}_\text{th}$ & $3.0499 \pow{6}$ & $9.4124 \pow{9}$ & $7.5906 \pow{9}$ \\[6pt] \hline
  & $\text{max}_\text{th}$ & $-0.51275$ & $-0.56361$ & $-0.88162$ \\[6pt]
 \multirow{-2}{*}{$a_\lambda^*, \pow{-4}$} & $\text{min}_\text{th}$ & $-2.57486$ & $-0.89618$ & $-0.9238$ \\[6pt] \hline
  & $\text{max}_\text{th}$ & $7.474 \pow{18}$ & $4.4154 \pow{17}$ & $2.3278 \pow{17}$ \\[6pt]
 \multirow{-2}{*}{$\mu^*, \GeV$} & $\text{min}_\text{th}$ & $3.6987 \pow{15}$ & $2.3787 \pow{17}$ & $2.2028 \pow{17}$ \\[6pt]
 \hline
    \end{tabular}
    \label{theorresult}
    \caption{Theoretical uncertainties of $\mu_r,,\mu^*$, and $a^*_\lambda$ for different loops 
    \label{tab:th_mu_r_mu_star_z_star_sum}}
\end{table}

\begin{figure}[!h]
\centering
  \includegraphics[width = 0.475\textwidth]{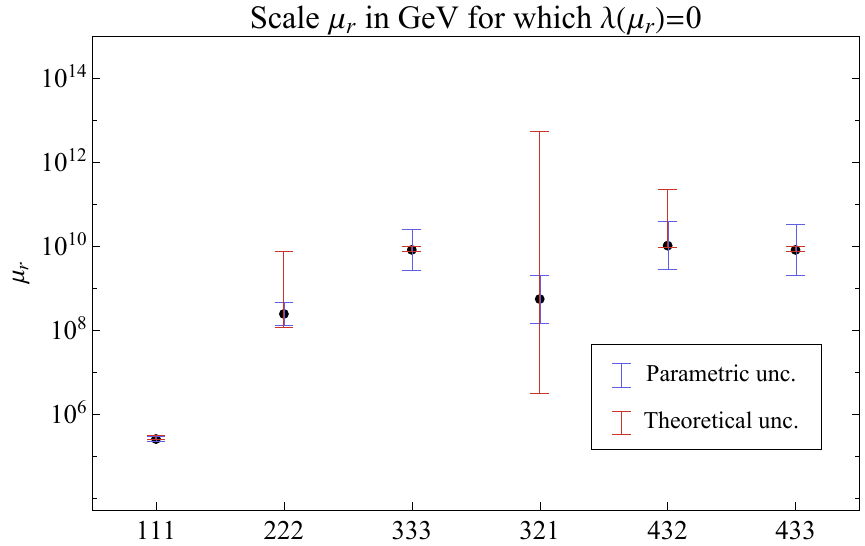}
  \includegraphics[width = 0.485\textwidth]{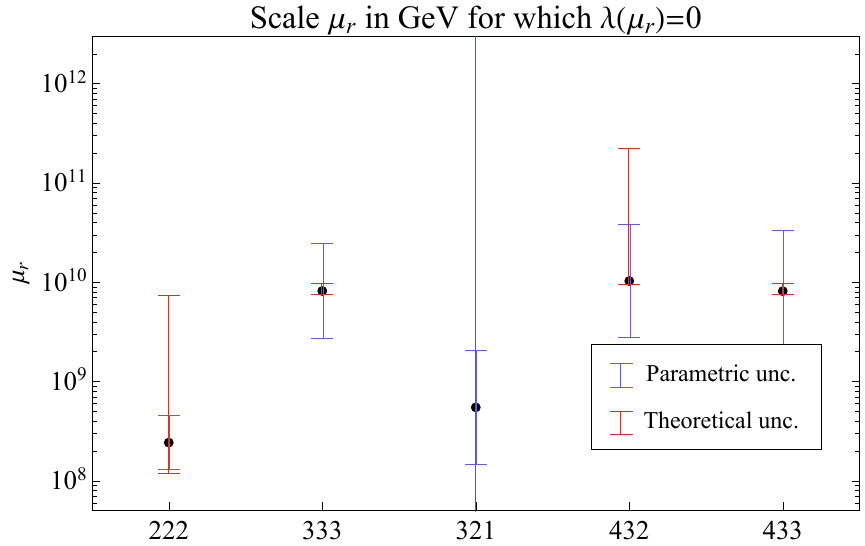}
  \caption{Dependence of $\mu_r$ for which $a_\lambda(\mu_r) = 0$ on the considered loop configurations\label{fig:mur_loops}}
\end{figure}

\begin{figure}[!h]
\centering
  \includegraphics[width = 0.48\textwidth]{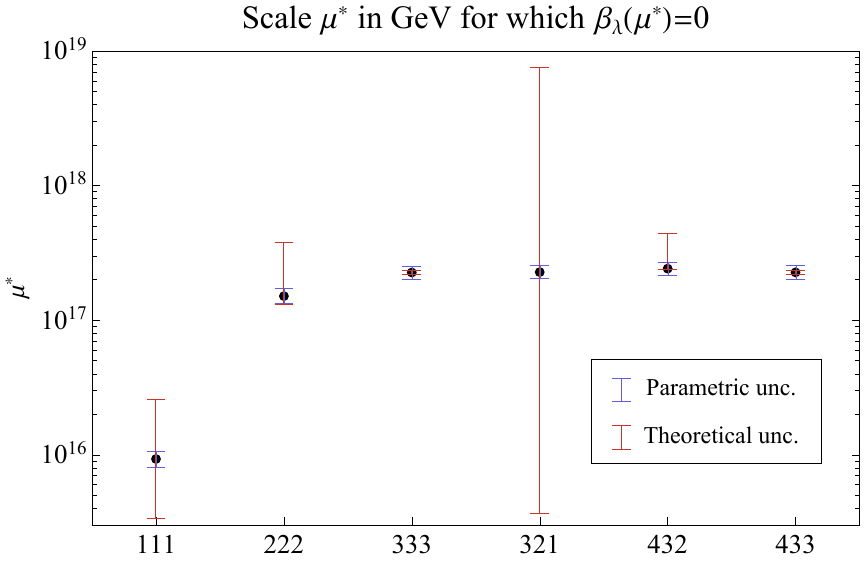}
  \includegraphics[width = 0.48\textwidth]{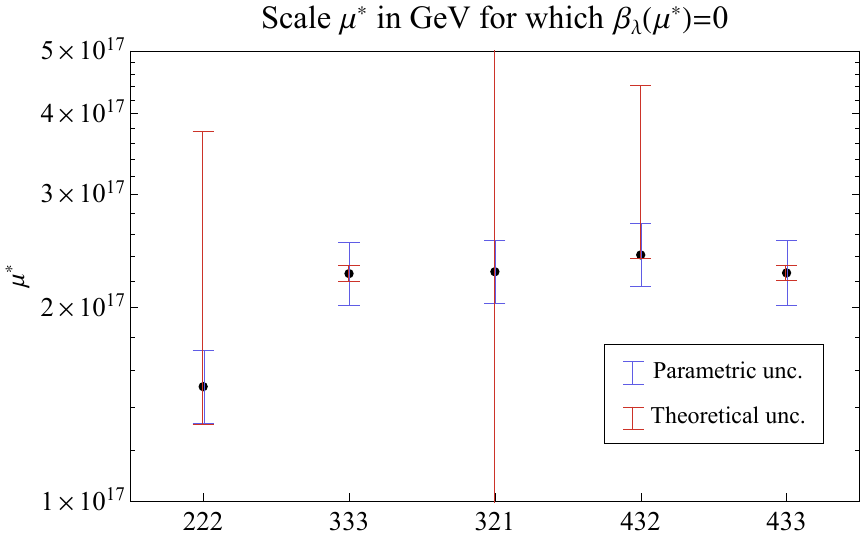}
  \caption{Dependence of $\mu^*$, for which $a_\lambda(\mu^*)$ is minimal, on the considered loop configurations\label{fig:mu_star_loops}}
\end{figure}

\begin{figure}[!h]
\centering
  \includegraphics[width = 0.475\textwidth]{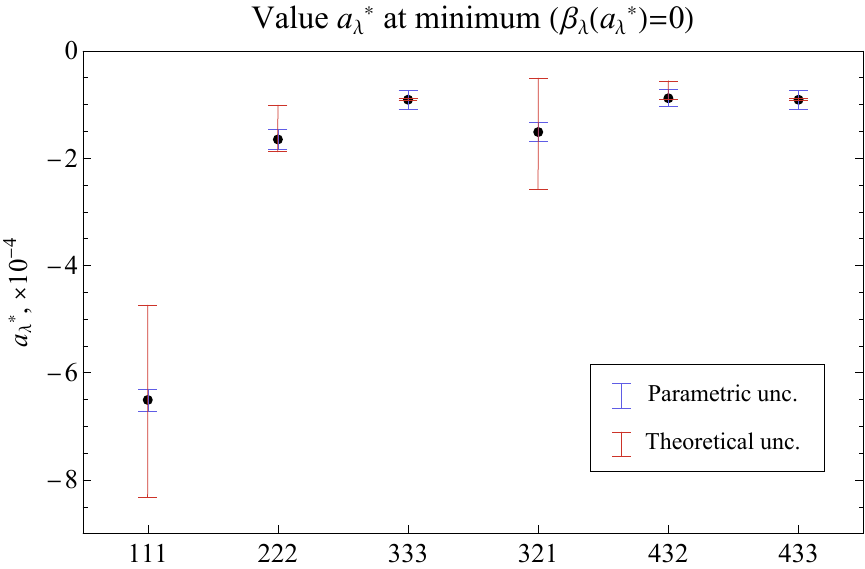}
  \includegraphics[width = 0.485\textwidth]{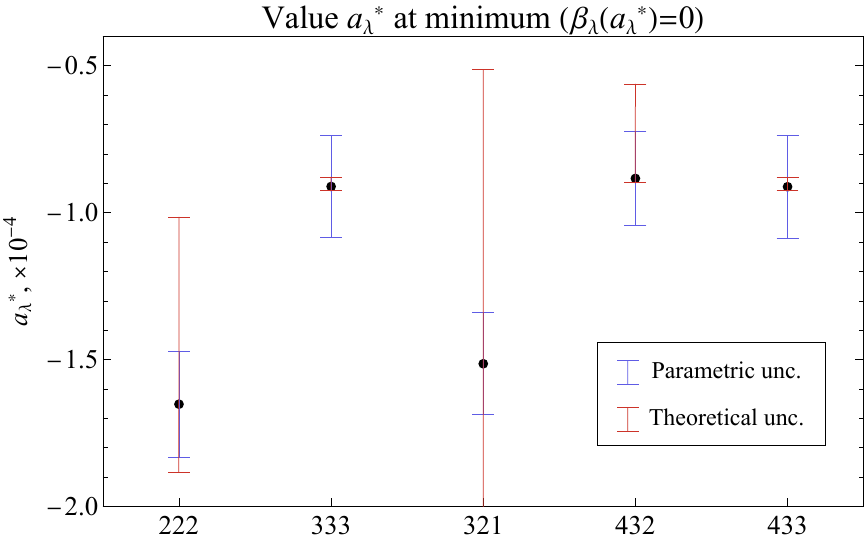}
  \caption{Dependence of minimal value $a^*_\lambda \equiv a_\lambda(\mu^*)$ on the considered loop configurations\label{fig:a_lambda_star_loops}}
\end{figure}

Summing up, we obtain the following dependencies of $\mu_r,\, \mu^*,\, a_\lambda^*$ on loop configurations. 
The higher PT corrections tend to increase both $\mu_r$ (Fig.~\ref{fig:mur_loops}) and $\mu^*$ (Fig.~\ref{fig:mu_star_loops}) comparing to the leading-order result. 
In Fig.~\ref{fig:a_lambda_star_loops} one sees a drastic change in $a^*_\lambda$ when going from the 111 configuration to the 222 one and further from two loops to three and partially four-loop configurations.  
The parametric uncertainties do not change much from order to order. On the contrary, our estimation of theoretical uncertainties clearly shows that the ``diagonal'' loop configurations have smaller ``errors'' than the ``non-diagonal''  ones (compare, e.g.,  333 and 321) with obvious shrinking of the uncertainty for higher loops.   
Starting from three loops theoretical uncertainties of the ``diagonal'' configurations become smaller than the parametric ones. 

\subsection{Vacuum decay probability}

Now we are ready to compute the probability of the decay of the electroweak vacuum during the past history of the Universe 
$$
\mathcal{P} \approx  0.15 \cdot \mathcal{\tilde A} \cdot \frac{\mu^{*\,4}}{H_0^4}  \exp\left\{-\frac{1}{6 |a_\lambda^*|}\right\}
$$
and provide naive estimates of theoretical and parametric uncertainties of the latter, originating from the uncertainties of $\mu^*$ and $a^*_\lambda$.
It will be convenient for us to consider the value $\log_{10}\mathcal P$ and express it in terms of $a^*_\lambda$ and $t^* = 2\ln \mu^*/\mu_0$.
For simplicity, we set the value of $\mathcal{\tilde A}$ to be constant\footnote{see Ref.~\cite{Baratella:2024hju} for most recent reevaluation.} and equal to $10^{-16}$ \cite{Andreassen:2017rzq}. 
With $H_0=1.44\pow{-42} \,\GeV$ for $\log_{10}\mathcal P$ we obtain:

\begin{equation}\label{eq:log10P}
    \log_{10}\mathcal P \approx 159.497 + 0.434294 \left( 2t^* - \dfrac{1}{6 |a_\lambda^*|} \right)
\end{equation}

\begin{table}[!h]
    \centering
    \begin{tabular}{|c|c|c|c|c|c|c|}
  \hline \rowcolor[HTML]{EFEFEF}
      & $111$ & $222$ & $333$ & $321$ & $432$ & $433$ \\[6pt] \hline
  $\text{param}$ & $103.2\pm 3.5$ & $-219\pm 48$ & $-574\pm 152$ & $-258\pm 55$ & $-599\pm 148$ & $-574\pm 152$ \\
 $\text{max}_\text{th}$ & $60$ & $-492$ & $-603$ & $-1186$ & $-1063$ & $-601$ \\
 $\text{min}_\text{th}$ & $129$ & $-165$ & $-564$ & $-68$ & $-588$ & $-564$ \\
 \hline
    \end{tabular}
    \caption{Logarithm of the vacuum decay probability $\log_{10}\mathcal P$ depending on loop configurations. Parametric (``param'' row) uncertainties and our estimates for theoretical intervals are indicated. Large positive value of $\log_{10} \mathcal{P}$ for the 111 case corresponds to the fact that the decay probability reaches one at much earlier time.}
    \label{tab:epicP}
\end{table} 

\begin{figure}[!h]
\centering
  \includegraphics[width = 0.48\textwidth]{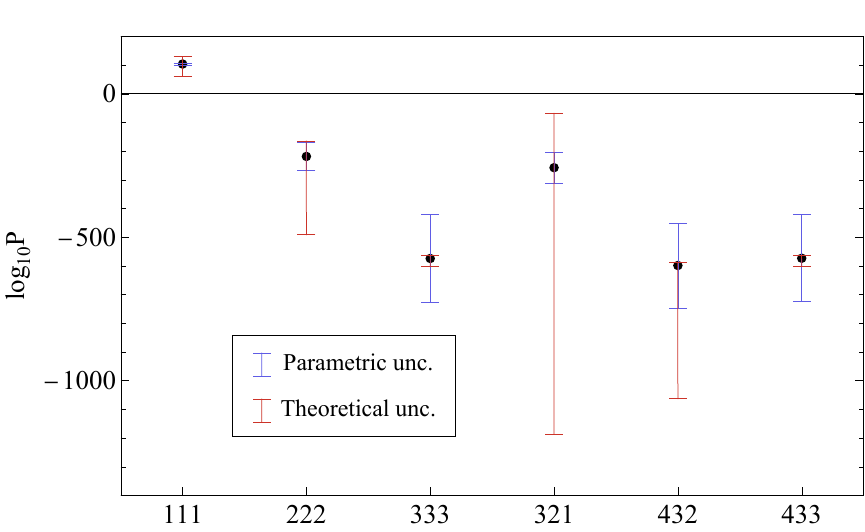}
  \caption{Dependence of $\log_{10}\mathcal P$ on loop configurations\label{fig:logP}}
\end{figure}

In Table.~\ref{tab:epicP}, the value $\text{max}_\text{th}$ corresponds to \textit{maximum} vacuum lifetime, while $\text{min}_\text{th}$; to the minimum, i.e., each value of $\text{min}_\text{th}$ here is greater than $\text{max}_\text{th}$. These values were calculated along the same lines as in previous sections, i.e. we maximize/minimize the value of $\log_{10} \mathcal{P}$ as a function of the matching scale  $\acute\mu$ in the interval $\acute \mu \in (\mu_0/2, 2\mu_0)$ ($\mu_0 = 173.22 \,\GeV$). Obviously, the probability can not be larger than one, so the  value of $\log_{10} \mathcal{P} \simeq 100$  just means that the vacuum decay probability would have reached 100\% much, much earlier in the history of the Universe. We illustrate the dependence of $\log_{10} \mathcal{P}$ on loop configurations in Fig.~\ref{fig:logP}. 

We can see again a reduction of the theoretical uncertainties for higher loop ``diagonal'' loop configurations. The same seems to be true for the ``non-diagonal'' cases. One can notice 
that the 321 configuration with the one-loop running of the self-coupling gives rise to a larger decay probability than that obtained with the three-loop running and is compatible with the 222 prediction.  However, the estimated theoretical uncertainty of 321 is much larger. The same is true if we compare 432 with 333 or 433. Thus, we argue that a more consistent treatment requires ``diagonal'' beta functions together with ``diagonal'' matching\footnote{see a discussion of the subtleties in the ``non-diagonal'' matching in Appendix~\ref{app:smdr}} to be used in the analysis. 
\section{Conclusions \label{sec:conclusions}}

We have reviewed the renormalization-group equations for the most important SM dimensionless couplings that describe gauge interactions of the SM particles, Yukawa interactions of the heavy top and bottom quarks, and the Higgs boson self-interactions. We provided an illustrative analysis of the one-loop RGE and highlighted regions in space of couplings with different asymptotic behavior. 
Beyond the leading order we discussed various loop configurations and carried out a refined analysis of the ``non-diagonal'' cases when gauge, Yukawa, and self-coupling beta functions are used at different loops.
Following \cite{Alam:2022cdv}, we used \SMDR\ code and derived approximate expressions for the running couplings at the electroweak scale chosen to be $\mu_0 = 173.22$ \GeV. The latter were utilized in our study of the parametric uncertainties for the running parameters. We showed that these uncertainties are almost independent of the considered loop configurations (see Table.~\ref{tab:paramerr}).

Our analysis of the SM renormalization-group equations reveals that the choice of loop order for the beta functions and matching conditions significantly impacts the theoretical uncertainty estimated from matching scale variation. We confirm that this uncertainty decreases when higher loop orders are included. However, employing theoretically motivated “non‑diagonal” loop configurations—such as $L,(L-1),(L-2)$ for some $L>2$, which are suggested by Weyl consistency conditions—introduces a larger theoretical uncertainty in key predictions (e.g., the vacuum decay probability) compared with the traditional “diagonal” approach of using $L$-loop RGEs for all couplings. We therefore stress that a consistent treatment of loop orders in both the renormalization-group equations and their initial conditions is required.

Different observables can obviously be most sensitive to different parameters\footnote{For example, in our vacuum decay study, we are primarily interested in $a_\lambda$}. Theoretical uncertainties of the former are impacted by the lowest available loop order for the latter.
Nevertheless, the computation of the necessary ingredients for the full $L$-loop analysis can be technically challenging and time-consuming. As a consequence, we appreciate a common approach (like in Refs.~\cite{Kniehl:2016enc,Martin:2019lqd}) that utilizes all state-of-the-art results available in literature.

\section{Acknowledgments}
We thank A.~Kataev, T.~Steudtner and Y.~Schr{\"o}der for their comments on the manuscript.
\appendix

\section{Modification of \SMDR \label{app:smdr}}

To convert \onshell\ parameters into \MSbar\ variables by means of \SMDR\, we used the internal variable {\tt loop}, which is passed as an argument to various \SMDR\ functions to specify the required order of perturbation theory. As already mentioned, by default the user has no control over these values and they are set to maximum.

Therefore, we used a modified version\footnote{Available at \url{https://github.com/arturfedoruk/RG}} of the {\tt SMDR\_Fit\_Inputs()} function, which iteratively fits the \MSbar\ parameters to the given \onshell\ values, to which the ability to adjust specific loops during recalculation was added\footnote{In addition, we also excluded from consideration the Yukawa constants for quarks and leptons lighter than the $b$ quark.}. The specific choice of the {\tt loop} parameter for the \SMDR\ functions called inside {\tt SMDR\_Fit\_Inputs()} to calculate pseudoobservables is presented in Table~\ref{tab:loopchoice} for the various loop configurations. It is worth noting that not all the necessary higher-order corrections are currently known. Therefore, we used approximations that correspond as closely as possible to the rule: ``$L$-loop RG equations + $(L-1)$ - loop recalculation formulas''. Moreover, as can be easily seen from \eqref{eq:SM_rel_pars_obs}, to determine the coupling constants for given particle masses, knowledge of the vacuum average ``extracted'' from $G_F$ is necessary.
For the cases $L_g = L_y = L_\lambda=1,2,3$ considered within the framework of standard perturbation theory, there is no question about how many loops $\bar \delta_r$ should be taken in. However, in the cases where different orders of the RG equations are taken for the gauge, Yukawa, and Higgs constants, the formal RG independence of particle masses can be spoiled if an insufficient number of terms are taken in the expression for $\bar \delta_r$. Therefore, in the orders $321$, $432$, and $433$ considered here we use the two-loop\footnote{The highest order available in \SMDR.} expression for $\bar \delta_r$.

In the original version of the program, formulas of the form \eqref{eq:SM_rel_pars_obs} are used at fixed but in some cases different $\mu_{th}$ scales. For example, by default, for the top quark mass, the $\bar \delta_t$ corrections are calculated at $\mu_{th_1}=M_t$, and for the $W$, $Z$, and $H$ masses the renormalization scale is chosen to be $\mu_{th_2}=160$ GeV (see \cite{Martin:2019lqd}). As a consequence, the iterative procedure for extracting running parameters necessarily uses the RG equations in the SM, which relate the running parameters on $\mu_{th_1}$ and $\mu_{th_2}$. In this paper, following \cite{Bednyakov:2015sca}, we fixed $\mu_{th} = \mu_0$ for all pseudoobservables of interest to us, thereby avoiding the use of the RG equations in the SM when calculating the running parameters for a given $\mu_0$. This allows us to quantitatively estimate theoretical uncertainties in the extraction of $a(\mu_0)$ by comparing the values obtained directly from the inversion of formulas of the form \eqref{eq:SM_rel_pars_obs} at the $\mu_0$ scale with the results of solving the RG equations whose initial values are found from \eqref{eq:SM_rel_pars_obs} at the $\acute\mu \in (\mu_0/2, \mu_0 \cdot 2)$ scale (see Section \ref{sec:th_unc_couplings}).

\begin{table}[!ht]
\begin{center}
\begin{tabular}{|c|c|c|c|c|c|c|c|}
  \hline
    function \SMDR\ & pseudoobservable & {\tt 111} & {\tt 222} & {\tt 333} & {\tt 321} & {\tt 432} & {\tt 433} \\\hline
  {\tt SMDR\_Eval\_Mt\_pole} & $M_t$ & {\tt 0} & {\tt 1} & {\tt 2} & {\tt 1} & {\tt 2} &  {\tt 2} \\
  {\tt SMDR\_Eval\_Mh\_pole} & $M_h$ & {\tt 0} & {\tt 1} & {\tt 2} & {\tt 0} & {\tt 1} & {\tt 2} \\
  {\tt SMDR\_Eval\_MZ\_pole} & $M_Z$ & {\tt 0} & {\tt 1} & {\tt 2} & {\tt 2} & {\tt 2.5} & {\tt 2.5} \\
  {\tt SMDR\_Eval\_MW\_pole} & $M_W$ & {\tt 0} & {\tt 1} & {\tt 2} & {\tt 2} & {\tt 2.5} & {\tt 2.5} \\
  {\tt SMDR\_Eval\_GFermi} & $G_F$ & {\tt 0} & {\tt 1} & {\tt 2} & {\tt 2} & {\tt 2} & {\tt 2} \\
  {\tt SMDR\_Eval\_QCDQED\_at\_MZ} & {\small$\alpha^{(5)}(M_Z), \alpha_s^{(5)}(M_Z)$} & {\tt 1} & {\tt 1} & {\tt 2} & {\tt 2} & {\tt 3} & {\tt 3} \\
  {\tt SMDR\_Eval\_mbmb} & $m_b(m_b)$ & {\tt 1} & {\tt 1} & {\tt 2} & {\tt 1} & {\tt 2} & {\tt 2} \\
  \hline
\end{tabular}
\end{center}
\caption{Choice of the {\tt loop} parameter for functions called inside the modified {\tt Fit\_Inputs()} for different loop configurations. \SMDR\ does not implement the {\tt 0} values for loops in {\tt SMDR\_Eval\_QCDQED\_at\_MZ} and {\tt SMDR\_Eval\_mbmb}. Therefore, the minimum {\tt loop} value for them is unity. In addition, the use of four-loop RG equations for the gauge constants (cases $432$ and $443$) implies the calculation of $M_Z$ (and $M_W$) with an accuracy of three loops. In \SMDR\, in addition to the full two-loop effects, only the leading three-loop QCD corrections for these masses are implemented. This corresponds to the fact that when calling {\tt SMDR\_Eval\_MZ\_pole} and {\tt SMDR\_Eval\_MW\_pole} the maximum value of the variable ${\tt loop} = 2.5$.
\label{tab:loopchoice}}
\end{table}

It is also important to mention the {\tt err\_target} parameter of {\tt Fit\_Inputs()}, the \emph{relative} error value allowed by the fitting. The program performs the fit by processing a specific sort of descent, which halts when {\tt err\_target} is bigger than the sum of the relative deviations of \onshell\ quantities $\mathcal O_i$ from the target value, which you specify when calling a function:
$$
\text{\tt err\_target} > \sum_{i=1}^7\left|\dfrac{\mathcal O_i(a^\text{\tt Fit\_Inputs()}) }{ \mathcal O_i^\text{target}} - 1\right|
$$

In all our calculations we use\footnote{Default value in the original \SMDR} $\text{\tt err\_target} = 10^{-6}$. Greater precision requires more computation. Nevertheless, the precision of the obtained \MSbar\ quantities $a_i^\text{\tt Fit\_Inputs()}$ turns to be higher. 
We estimated the precision  of $a_i^\text{\tt Fit\_Inputs()}$ as a difference between the last two steps $(j_f-1)$ and $j_f$ of the fitting procedure implemented in \SMDR\:

$$
\delta_{a_i}^\text{\tt Fit\_Inputs()} = \left| \dfrac{a_i^\text{\tt Fit\_Inputs()}(\text{step} \;j_f)}{a_i^\text{\tt Fit\_Inputs()}(\text{step} \;j_f - 1)} - 1 \right|
$$
and got approximately $(2\pow{-9},\, 2\pow{-9},\, 6\pow{-11},\, 2\pow{-9},\, 3\pow{-7},\, 6\pow{-9})$ for $(a_1,a_2,a_3,a_t,a_b,a_\lambda)$ at $111$ and $(2\pow{-8},\, 4\pow{-9},\, 3\pow{-9},\, 2\pow{-8},\, 2\pow{-8},\, 2\pow{-7})$ at $433$. Those observations are relevant in \ref{sec:parametric_uncertainties}.

\section{State-of-the-art RGE \label{app:bigbigformulas}}

The state-of-the-art RG equations of loop order $433$ that were used in the paper are of the following form (here $\zeta_3 = \zeta(3)$ stands for the Riemann zeta-function):

\begingroup
\allowdisplaybreaks
\small
\begin{align}
&    \frac{da_1}{d\ln\mu^2} = \frac{41}{10} a_1^2  + a_1^2 \bigg( \frac{199}{50} a_1 +\frac{27}{10} a_2 +\frac{44}{5} a_3 -\frac{17}{10} a_t -\frac{1}{2} a_b\bigg)\nonumber\\
&    + a_1^2 \bigg( -\frac{388613}{24000} a_1^2 +\frac{123}{160} a_1 a_2 +\frac{789}{64} a_2^2 -\frac{137}{75} a_1 a_3 -\frac{3}{5} a_2 a_3 +\frac{297}{5} a_3^2 -\frac{2827}{800} a_1 a_t -\frac{471}{32} a_2 a_t -\frac{29}{5} a_3 a_t +\frac{189}{16} a_t^2 -\frac{1267}{800} a_1 a_b\nonumber\\
&-\frac{1311}{160} a_2 a_b -\frac{17}{5} a_3 a_b +\frac{369}{40} a_t a_b +\frac{57}{16} a_b^2 +\frac{27}{50} a_1 a_\lambda +\frac{9}{10} a_2 a_\lambda -\frac{9}{5} a_\lambda^2 \bigg)\nonumber\\
&    + a_1^2 \bigg( -\frac{143035709}{4320000} a_1^3 -\frac{1638851}{22500} a_1^3 \zeta_3 -\frac{3819731}{96000} a_1^2 a_2 +\frac{16529}{500} a_1^2 a_2 \zeta_3 +\frac{572059}{57600} a_1 a_2^2 -\frac{6751}{300} a_1 a_2^2 \zeta_3 -\frac{117923}{11520} a_2^3 -\frac{3109}{20} a_2^3 \zeta_3 \nonumber\\
&-\frac{3629273}{27000} a_1^2 a_3 +\frac{180076}{1125} a_1^2 a_3 \zeta_3 -\frac{69}{100} a_1 a_2 a_3 -\frac{41971}{360} a_2^2 a_3 +\frac{1868}{15} a_2^2 a_3 \zeta_3 +\frac{83389}{675} a_1 a_3^2 -\frac{68656}{225} a_1 a_3^2 \zeta_3 -\frac{437}{3} a_2 a_3^2\nonumber\\
&+\frac{736}{5} a_2 a_3^2 \zeta_3 +\frac{1529}{15} a_3^3 -\frac{4640}{9} a_3^3 \zeta_3 +\frac{8978897}{288000} a_1^2 a_t +\frac{1299}{250} a_1^2 a_t \zeta_3 -\frac{42841}{3200} a_1 a_2 a_t -\frac{561}{50} a_1 a_2 a_t \zeta_3 -\frac{439841}{3840} a_2^2 a_t\nonumber\\
&+\frac{154}{5} a_2^2 a_t \zeta_3 -\frac{503}{75} a_1 a_3 a_t +\frac{102}{25} a_1 a_3 a_t \zeta_3 +\frac{367}{5} a_2 a_3 a_t -\frac{474}{5} a_2 a_3 a_t \zeta_3 -\frac{5731}{90} a_3^2 a_t +\frac{796}{5} a_3^2 a_t \zeta_3 +\frac{29059}{640} a_1 a_t^2\nonumber\\
&-\frac{357}{100} a_1 a_t^2 \zeta_3 +\frac{71463}{640} a_2 a_t^2 +\frac{639}{20} a_2 a_t^2 \zeta_3 +\frac{1429}{20} a_3 a_t^2 -60 a_3 a_t^2 \zeta_3 -\frac{13653}{160} a_t^3 -\frac{51}{10} a_t^3 \zeta_3 +\frac{3876629}{288000} a_1^2 a_b -\frac{249}{250} a_1^2 a_b \zeta_3\nonumber\\
&-\frac{7461}{640} a_1 a_2 a_b -\frac{3}{10} a_1 a_2 a_b \zeta_3 -\frac{80311}{1280} a_2^2 a_b +\frac{62}{5} a_2^2 a_b \zeta_3 -\frac{79}{15} a_1 a_3 a_b -\frac{18}{5} a_1 a_3 a_b \zeta_3 -\frac{69}{5} a_2 a_3 a_b -\frac{138}{5} a_2 a_3 a_b \zeta_3\nonumber\\
&-\frac{83}{18} a_3^2 a_b -4 a_3^2 a_b \zeta_3 +\frac{306401}{14400} a_1 a_t a_b -\frac{5}{3} a_1 a_t a_b \zeta_3 +\frac{6131}{64} a_2 a_t a_b +\frac{381}{10} a_2 a_t a_b \zeta_3 +\frac{6901}{90} a_3 a_t a_b -\frac{1208}{15} a_3 a_t a_b \zeta_3\nonumber\\
&-\frac{29281}{480} a_t^2 a_b +\frac{14}{5} a_t^2 a_b \zeta_3 +\frac{1919}{128} a_1 a_b^2 +\frac{9}{20} a_1 a_b^2 \zeta_3 +\frac{44931}{640} a_2 a_b^2 +\frac{99}{20} a_2 a_b^2 \zeta_3 +\frac{497}{20} a_3 a_b^2 -\frac{12}{5} a_3 a_b^2 \zeta_3 -\frac{32269}{480} a_t a_b^2\nonumber\\
&+\frac{14}{5} a_t a_b^2 \zeta_3 -\frac{957}{32} a_b^3 -\frac{3}{2} a_b^3 \zeta_3 +\frac{3627}{2000} a_1^2 a_\lambda +\frac{1917}{200} a_1 a_2 a_\lambda +\frac{889}{80} a_2^2 a_\lambda -\frac{963}{50} a_1 a_t a_\lambda -\frac{81}{10} a_2 a_t a_\lambda -\frac{237}{10} a_t^2 a_\lambda\nonumber\\
&-\frac{531}{50} a_1 a_b a_\lambda -\frac{153}{10} a_2 a_b a_\lambda -\frac{57}{10} a_b^2 a_\lambda -\frac{1269}{100} a_1 a_\lambda^2 -\frac{981}{20} a_2 a_\lambda^2 +\frac{297}{5} a_t a_\lambda^2 +\frac{189}{5} a_b a_\lambda^2 +\frac{156}{5} a_\lambda^3\bigg), 
\label{eq:a1bigbigformula}
\end{align}
\begin{align}
&    \frac{da_2}{d\ln\mu^2} = -\frac{19}{6} a_2^2 + a_2^2 \bigg( \frac{35}{6} a_2+\frac{9}{10} a_1+12 a_3-\frac{3}{2} a_t-\frac{3}{2} a_b \bigg) \nonumber\\
&    + a_2^2 \bigg( \frac{324953}{1728} a_2^2+\frac{873}{160} a_1 a_2-\frac{5597}{1600} a_1^2+39 a_2 a_3-\frac{1}{5} a_1 a_3+81 a_3^2-\frac{729}{32} a_2 a_t-\frac{593}{160} a_1 a_t-7 a_3 a_t+\frac{147}{16} a_t^2-\frac{729}{32} a_2 a_b\nonumber\\
&    -\frac{533}{160} a_1 a_b-7 a_3 a_b+\frac{117}{8} a_t a_b+\frac{147}{16} a_b^2+\frac{3}{2} a_2 a_\lambda+\frac{3}{10} a_1 a_\lambda-3 a_\lambda^2 \bigg) \nonumber\\
&    + a_2^2 \bigg( \frac{124660945}{62208} a_2^3-\frac{78803}{36} a_2^3 \zeta_3-\frac{375767}{11520} a_1 a_2^2+\frac{4631}{60} a_1 a_2^2 \zeta_3-\frac{787709}{19200} a_1^2 a_2+\frac{659}{100} a_1^2 a_2 \zeta_3-\frac{6418229}{288000} a_1^3+\frac{21173}{1500} a_1^3 \zeta_3 \nonumber\\
& -\frac{72881}{72} a_2^2 a_3+\frac{4108}{3} a_2^2 a_3 \zeta_3+\frac{161}{20} a_1 a_2 a_3-\frac{52297}{1800} a_1^2 a_3+\frac{508}{15} a_1^2 a_3 \zeta_3+\frac{2587}{3} a_2 a_3^2-640 a_2 a_3^2 \zeta_3-\frac{437}{9} a_1 a_3^2+\frac{736}{15} a_1 a_3^2 \zeta_3\nonumber\\
&    +\frac{257}{3} a_3^3-\frac{1760}{3} a_3^3 \zeta_3-\frac{500665}{2304} a_2^2 a_t+\frac{239}{6} a_2^2 a_t \zeta_3-\frac{102497}{1920} a_1 a_2 a_t-7 a_1 a_2 a_t \zeta_3+\frac{465089}{19200} a_1^2 a_t-\frac{249}{50} a_1^2 a_t \zeta_3-\frac{361}{3} a_2 a_3 a_t\nonumber\\
&    -14 a_2 a_3 a_t \zeta_3+\frac{199}{15} a_1 a_3 a_t-\frac{94}{5} a_1 a_3 a_t \zeta_3-\frac{307}{6} a_3^2 a_t+84 a_3^2 a_t \zeta_3+\frac{30213}{128} a_2 a_t^2-\frac{63}{4} a_2 a_t^2 \zeta_3+\frac{3161}{128} a_1 a_t^2+\frac{153}{20} a_1 a_t^2 \zeta_3\nonumber\\
&    +\frac{239}{4} a_3 a_t^2-36 a_3 a_t^2 \zeta_3-\frac{2143}{32} a_t^3-\frac{9}{2} a_t^3 \zeta_3-\frac{500665}{2304} a_2^2 a_b+\frac{239}{6} a_2^2 a_b \zeta_3-\frac{90029}{1920} a_1 a_2 a_b-1 a_1 a_2 a_b \zeta_3+\frac{17903}{1280} a_1^2 a_b\nonumber\\
&    -\frac{143}{50} a_1^2 a_b \zeta_3-\frac{361}{3} a_2 a_3 a_b-14 a_2 a_3 a_b \zeta_3+a_1 a_3 a_b-\frac{78}{5} a_1 a_3 a_b \zeta_3-\frac{307}{6} a_3^2 a_b+84 a_3^2 a_b \zeta_3+\frac{16735}{64} a_2 a_t a_b+\frac{57}{2} a_2 a_t a_b \zeta_3\nonumber\\
&    +\frac{33959}{960} a_1 a_t a_b+8 a_1 a_t a_b \zeta_3+\frac{739}{6} a_3 a_t a_b-152 a_3 a_t a_b \zeta_3-\frac{3265}{32} a_t^2 a_b+6 a_t^2 a_b \zeta_3+\frac{30213}{128} a_2 a_b^2-\frac{63}{4} a_2 a_b^2 \zeta_3+\frac{15937}{640} a_1 a_b^2\nonumber\\
&    +\frac{99}{20} a_1 a_b^2 \zeta_3+\frac{239}{4} a_3 a_b^2-36 a_3 a_b^2 \zeta_3-\frac{3265}{32} a_t a_b^2+6 a_t a_b^2 \zeta_3-\frac{2143}{32} a_b^3-\frac{9}{2} a_b^3 \zeta_3+\frac{2905}{48} a_2^2 a_\lambda+\frac{69}{8} a_1 a_2 a_\lambda+\frac{457}{400} a_1^2 a_\lambda +52 a_\lambda^3  \nonumber\\
&    -\frac{75}{2} a_2 a_t a_\lambda-\frac{27}{10} a_1 a_t a_\lambda-\frac{39}{2} a_t^2 a_\lambda-\frac{75}{2} a_2 a_b a_\lambda-\frac{51}{10} a_1 a_b a_\lambda-\frac{39}{2} a_b^2 a_\lambda-\frac{363}{4} a_2 a_\lambda^2-\frac{327}{20} a_1 a_\lambda^2+75 a_t a_\lambda^2+75 a_b a_\lambda^2
\bigg), 
\label{eq:a2bigbigformula}
    \end{align}
\begin{align}
&    \frac{da_3}{d\ln\mu^2} = - 7 a_3^2 + a_3^2 \bigg( -26 a_3+\frac{11}{10} a_1+\frac{9}{2} a_2-2 a_t-2 a_b \bigg) \nonumber\\
&     + a_3^2 \bigg( \frac{65}{2} a_3^2+\frac{77}{15} a_1 a_3-\frac{523}{120} a_1^2+21 a_2 a_3-\frac{3}{40} a_1 a_2+\frac{109}{8} a_2^2-40 a_3 a_t-\frac{101}{40} a_1 a_t-\frac{93}{8} a_2 a_t+15 a_t^2-40 a_3 a_b\nonumber\\
&	-\frac{89}{40} a_1 a_b-\frac{93}{8} a_2 a_b+18 a_t a_b+15 a_b^2 \bigg) \nonumber\\
&     + a_3^2 \bigg( \frac{63559}{18} a_3^3-\frac{44948}{9} a_3^3 \zeta_3-\frac{57739}{540} a_1 a_3^2+\frac{8119}{45} a_1 a_3^2 \zeta_3-\frac{17771}{270} a_1^2 a_3+\frac{451}{18} a_1^2 a_3 \zeta_3-\frac{6085099}{216000} a_1^3+\frac{17473}{900} a_1^3 \zeta_3\nonumber\\
&	-\frac{5969}{12} a_2 a_3^2+869 a_2 a_3^2 \zeta_3+\frac{69}{20} a_1 a_2 a_3-\frac{46951}{4800} a_1^2 a_2+\frac{973}{100} a_1^2 a_2 \zeta_3+\frac{953}{9} a_2^2 a_3-\frac{475}{6} a_2^2 a_3 \zeta_3-\frac{37597}{2880} a_1 a_2^2+\frac{691}{60} a_1 a_2^2 \zeta_3\nonumber\\
&	-\frac{176815}{1728} a_2^3-\frac{935}{4} a_2^3 \zeta_3-\frac{6709}{9} a_3^2 a_t+272 a_3^2 a_t \zeta_3-\frac{1283}{60} a_1 a_3 a_t-\frac{8}{5} a_1 a_3 a_t \zeta_3+\frac{362287}{14400} a_1^2 a_t-\frac{19}{100} a_1^2 a_t \zeta_3-\frac{473}{4} a_2 a_3 a_t\nonumber\\
&	-72 a_2 a_3 a_t \zeta_3+\frac{77}{160} a_1 a_2 a_t-\frac{27}{2} a_1 a_2 a_t \zeta_3-\frac{12887}{192} a_2^2 a_t+\frac{117}{4} a_2^2 a_t \zeta_3+427 a_3 a_t^2-96 a_3 a_t^2 \zeta_3+\frac{3641}{160} a_1 a_t^2+\frac{21}{10} a_1 a_t^2 \zeta_3\nonumber\\
&	+\frac{3201}{32} a_2 a_t^2+\frac{45}{2} a_2 a_t^2 \zeta_3-\frac{423}{4} a_t^3-6 a_t^3 \zeta_3-\frac{6709}{9} a_3^2 a_b+272 a_3^2 a_b \zeta_3-\frac{1487}{60} a_1 a_3 a_b-\frac{104}{5} a_1 a_3 a_b \zeta_3+\frac{210847}{14400} a_1^2 a_b\nonumber\\
&	-\frac{7}{100} a_1^2 a_b \zeta_3-\frac{473}{4} a_2 a_3 a_b-72 a_2 a_3 a_b \zeta_3-\frac{155}{32} a_1 a_2 a_b-\frac{9}{2} a_1 a_2 a_b \zeta_3-\frac{12887}{192} a_2^2 a_b+\frac{117}{4} a_2^2 a_b \zeta_3+\frac{4282}{9} a_3 a_t a_b\nonumber\\
&	-\frac{400}{3} a_3 a_t a_b \zeta_3+\frac{19033}{720} a_1 a_t a_b-\frac{5}{3} a_1 a_t a_b \zeta_3+\frac{1895}{16} a_2 a_t a_b+3 a_2 a_t a_b \zeta_3-\frac{1171}{12} a_t^2 a_b-8 a_t^2 a_b \zeta_3+427 a_3 a_b^2\nonumber\\
&	-96 a_3 a_b^2 \zeta_3+\frac{2869}{160} a_1 a_b^2+\frac{81}{10} a_1 a_b^2 \zeta_3+\frac{3201}{32} a_2 a_b^2+\frac{45}{2} a_2 a_b^2 \zeta_3-\frac{1171}{12} a_t a_b^2-8 a_t a_b^2 \zeta_3-\frac{423}{4} a_b^3-6 a_b^3 \zeta_3-30 a_t^2 a_\lambda\nonumber\\
&	-30 a_b^2 a_\lambda+36 a_t a_\lambda^2+36 a_b a_\lambda^2 \bigg),
\label{eq:a3bigbigformula}
    \end{align}
\begin{align}
&    \frac{da_t}{d\ln\mu^2} = a_t \bigg( \frac{9}{2} a_t-\frac{17}{20} a_1-\frac{9}{4} a_2-8 a_3+\frac{3}{2} a_b \bigg) \nonumber\\
&    + a_t \bigg( -12 a_t^2+\frac{393}{80} a_1 a_t+\frac{1187}{600} a_1^2+\frac{225}{16} a_2 a_t-\frac{9}{20} a_1 a_2-\frac{23}{4} a_2^2+36 a_3 a_t+\frac{19}{15} a_1 a_3+9 a_2 a_3-108 a_3^2-\frac{11}{4} a_t a_b\nonumber\\
&	+\frac{7}{80} a_1 a_b+\frac{99}{16} a_2 a_b+4 a_3 a_b-\frac{1}{4} a_b^2-12 a_t a_\lambda+6 a_\lambda^2 \bigg) \nonumber\\
&    + a_t \bigg( \frac{339}{8} a_t^3+\frac{27}{2} a_t^3 \zeta_3-\frac{2437}{80} a_1 a_t^2-\frac{458179}{19200} a_1^2 a_t-\frac{93}{200} a_1^2 a_t \zeta_3+\frac{763523}{24000} a_1^3-\frac{13073}{1000} a_1^3 \zeta_3-\frac{1593}{16} a_2 a_t^2+\frac{8097}{640} a_1 a_2 a_t\nonumber\\
&	+\frac{369}{20} a_1 a_2 a_t \zeta_3+\frac{1227}{320} a_1^2 a_2-\frac{1377}{200} a_1^2 a_2 \zeta_3+\frac{32391}{256} a_2^2 a_t-\frac{729}{8} a_2^2 a_t \zeta_3+\frac{819}{320} a_1 a_2^2-\frac{243}{40} a_1 a_2^2 \zeta_3+\frac{455}{576} a_2^3+\frac{1125}{8} a_2^3 \zeta_3\nonumber\\
&	-157 a_3 a_t^2-\frac{126}{5} a_1 a_3 a_t+36 a_1 a_3 a_t \zeta_3+\frac{2047}{150} a_1^2 a_3-\frac{748}{25} a_1^2 a_3 \zeta_3-168 a_2 a_3 a_t+180 a_2 a_3 a_t \zeta_3-\frac{321}{20} a_1 a_2 a_3\nonumber\\
&	+\frac{435}{4} a_2^2 a_3-108 a_2^2 a_3 \zeta_3+\frac{3827}{6} a_3^2 a_t-228 a_3^2 a_t \zeta_3+\frac{1633}{60} a_1 a_3^2-\frac{176}{5} a_1 a_3^2 \zeta_3+\frac{987}{4} a_2 a_3^2-144 a_2 a_3^2 \zeta_3-\frac{4166}{3} a_3^3\nonumber\\
&	+640 a_3^3 \zeta_3+\frac{739}{16} a_t^2 a_b-\frac{1383}{160} a_1 a_t a_b+\frac{1}{2} a_1 a_t a_b \zeta_3-\frac{40673}{19200} a_1^2 a_b-\frac{199}{200} a_1^2 a_b \zeta_3-\frac{2307}{32} a_2 a_t a_b-\frac{9}{2} a_2 a_t a_b \zeta_3\nonumber\\
&	+\frac{747}{128} a_1 a_2 a_b+\frac{27}{10} a_1 a_2 a_b \zeta_3+\frac{10341}{256} a_2^2 a_b-\frac{225}{8} a_2^2 a_b \zeta_3+27 a_3 a_t a_b-32 a_3 a_t a_b \zeta_3-\frac{457}{30} a_1 a_3 a_b-\frac{28}{5} a_1 a_3 a_b \zeta_3\nonumber\\
&	-\frac{27}{2} a_2 a_3 a_b-108 a_2 a_3 a_b \zeta_3-\frac{305}{2} a_3^2 a_b-44 a_3^2 a_b \zeta_3+\frac{825}{8} a_t a_b^2-48 a_t a_b^2 \zeta_3-\frac{959}{160} a_1 a_b^2+\frac{19}{10} a_1 a_b^2 \zeta_3-\frac{2283}{32} a_2 a_b^2\nonumber\\
&	+\frac{63}{2} a_2 a_b^2 \zeta_3+82 a_3 a_b^2-64 a_3 a_b^2 \zeta_3+\frac{477}{16} a_b^3+\frac{9}{2} a_b^3 \zeta_3+198 a_t^2 a_\lambda-\frac{127}{10} a_1 a_t a_\lambda-\frac{1089}{400} a_1^2 a_\lambda-\frac{135}{2} a_2 a_t a_\lambda\nonumber\\
&	+\frac{117}{40} a_1 a_2 a_\lambda-\frac{171}{16} a_2^2 a_\lambda+16 a_3 a_t a_\lambda+93 a_t a_b a_\lambda+15 a_b^2 a_\lambda+\frac{15}{4} a_t a_\lambda^2+9 a_1 a_\lambda^2+45 a_2 a_\lambda^2-\frac{291}{4} a_b a_\lambda^2-36 a_\lambda^3 \bigg), 
\label{eq:atbigbigformula}
    \end{align}
\begin{align}
&    \frac{da_b}{d\ln\mu^2} = a_b \bigg( \frac{9}{2} a_b-\frac{1}{4} a_1-\frac{9}{4} a_2-8 a_3+\frac{3}{2} a_t \bigg) \nonumber\\
&    + a_b \bigg( -12 a_b^2+\frac{237}{80} a_1 a_b-\frac{127}{600} a_1^2+\frac{225}{16} a_2 a_b-\frac{27}{20} a_1 a_2-\frac{23}{4} a_2^2+36 a_3 a_b+\frac{31}{15} a_1 a_3+9 a_2 a_3-108 a_3^2-\frac{11}{4} a_t a_b\nonumber\\
&	+\frac{91}{80} a_1 a_t+\frac{99}{16} a_2 a_t+4 a_3 a_t-\frac{1}{4} a_t^2-12 a_b a_\lambda+6 a_\lambda^2 \bigg) \nonumber\\
&    + a_b \bigg( \frac{339}{8} a_b^3+\frac{27}{2} a_b^3 \zeta_3-\frac{1981}{80} a_1 a_b^2-\frac{209659}{19200} a_1^2 a_b-\frac{171}{200} a_1^2 a_b \zeta_3+\frac{93241}{8000} a_1^3-\frac{769}{200} a_1^3 \zeta_3-\frac{1593}{16} a_2 a_b^2+\frac{8493}{640} a_1 a_2 a_b\nonumber\\
&	-\frac{9}{5} a_1 a_2 a_b \zeta_3+\frac{2139}{320} a_1^2 a_2-\frac{81}{40} a_1^2 a_2 \zeta_3+\frac{32391}{256} a_2^2 a_b-\frac{729}{8} a_2^2 a_b \zeta_3-\frac{633}{320} a_1 a_2^2-\frac{243}{40} a_1 a_2^2 \zeta_3+\frac{455}{576} a_2^3+\frac{1125}{8} a_2^3 \zeta_3\nonumber\\
&	-157 a_3 a_b^2-18 a_1 a_3 a_b+\frac{132}{5} a_1 a_3 a_b \zeta_3-\frac{337}{75} a_1^2 a_3-\frac{44}{5} a_1^2 a_3 \zeta_3-168 a_2 a_3 a_b+180 a_2 a_3 a_b \zeta_3-\frac{153}{20} a_1 a_2 a_3\nonumber\\
&	+\frac{435}{4} a_2^2 a_3-108 a_2^2 a_3 \zeta_3+\frac{3827}{6} a_3^2 a_b-228 a_3^2 a_b \zeta_3+\frac{833}{12} a_1 a_3^2-\frac{176}{5} a_1 a_3^2 \zeta_3+\frac{987}{4} a_2 a_3^2-144 a_2 a_3^2 \zeta_3-\frac{4166}{3} a_3^3\nonumber\\
&	+640 a_3^3 \zeta_3+\frac{739}{16} a_t a_b^2-\frac{4203}{160} a_1 a_t a_b+\frac{77}{10} a_1 a_t a_b \zeta_3-\frac{104729}{19200} a_1^2 a_t-\frac{13}{200} a_1^2 a_t \zeta_3-\frac{2307}{32} a_2 a_t a_b-\frac{9}{2} a_2 a_t a_b \zeta_3\nonumber\\
&	+\frac{3267}{640} a_1 a_2 a_t+\frac{189}{20} a_1 a_2 a_t \zeta_3+\frac{10341}{256} a_2^2 a_t-\frac{225}{8} a_2^2 a_t \zeta_3+27 a_3 a_t a_b-32 a_3 a_t a_b \zeta_3-\frac{161}{6} a_1 a_3 a_t+4 a_1 a_3 a_t \zeta_3\nonumber\\
&	-\frac{27}{2} a_2 a_3 a_t-108 a_2 a_3 a_t \zeta_3-\frac{305}{2} a_3^2 a_t-44 a_3^2 a_t \zeta_3+\frac{825}{8} a_t^2 a_b-48 a_t^2 a_b \zeta_3-\frac{363}{160} a_1 a_t^2-\frac{17}{10} a_1 a_t^2 \zeta_3-\frac{2283}{32} a_2 a_t^2\nonumber\\
&	+\frac{63}{2} a_2 a_t^2 \zeta_3+82 a_3 a_t^2-64 a_3 a_t^2 \zeta_3+\frac{477}{16} a_t^3+\frac{9}{2} a_t^3 \zeta_3+198 a_b^2 a_\lambda-\frac{139}{10} a_1 a_b a_\lambda-\frac{9}{16} a_1^2 a_\lambda-\frac{135}{2} a_2 a_b a_\lambda\nonumber\\
&	-\frac{27}{40} a_1 a_2 a_\lambda-\frac{171}{16} a_2^2 a_\lambda+16 a_3 a_b a_\lambda+93 a_t a_b a_\lambda+15 a_t^2 a_\lambda+\frac{15}{4} a_b a_\lambda^2+9 a_1 a_\lambda^2+45 a_2 a_\lambda^2-\frac{291}{4} a_t a_\lambda^2-36 a_\lambda^3 \bigg), 
\label{eq:abbigbigformula}
    \end{align}
\begin{align}
&    \frac{da_\lambda}{d\ln\mu^2} = 12 a_\lambda^2-\frac{9}{10} a_1 a_\lambda+\frac{27}{400} a_1^2-\frac{9}{2} a_2 a_\lambda+\frac{9}{40} a_1 a_2+\frac{9}{16} a_2^2+6 a_t a_\lambda-3 a_t^2+6 a_b a_\lambda-3 a_b^2 \nonumber\\
&    -156 a_\lambda^3+\frac{54}{5} a_1 a_\lambda^2+\frac{1887}{400} a_1^2 a_\lambda-\frac{3411}{4000} a_1^3+54 a_2 a_\lambda^2+\frac{117}{40} a_1 a_2 a_\lambda-\frac{1677}{800} a_1^2 a_2-\frac{73}{16} a_2^2 a_\lambda-\frac{289}{160} a_1 a_2^2+\frac{305}{32} a_2^3\nonumber\\
&	-72 a_t a_\lambda^2+\frac{17}{4} a_1 a_t a_\lambda-\frac{171}{200} a_1^2 a_t+\frac{45}{4} a_2 a_t a_\lambda+\frac{63}{20} a_1 a_2 a_t-\frac{9}{8} a_2^2 a_t+40 a_3 a_t a_\lambda-\frac{3}{2} a_t^2 a_\lambda-\frac{4}{5} a_1 a_t^2-16 a_3 a_t^2\nonumber\\
&	+15 a_t^3-72 a_b a_\lambda^2+\frac{5}{4} a_1 a_b a_\lambda+\frac{9}{40} a_1^2 a_b+\frac{45}{4} a_2 a_b a_\lambda+\frac{27}{20} a_1 a_2 a_b-\frac{9}{8} a_2^2 a_b+40 a_3 a_b a_\lambda-21 a_t a_b a_\lambda-3 a_t^2 a_b\nonumber\\
&	-\frac{3}{2} a_b^2 a_\lambda+\frac{2}{5} a_1 a_b^2-16 a_3 a_b^2-3 a_t a_b^2+15 a_b^3 \nonumber\\
&    +3588 a_\lambda^4+2016 a_\lambda^4 \zeta_3-\frac{474}{5} a_1 a_\lambda^3+\frac{72}{5} a_1 a_\lambda^3 \zeta_3-\frac{3762}{25} a_1^2 a_\lambda^2-\frac{729}{25} a_1^2 a_\lambda^2 \zeta_3+\frac{88639}{2000} a_1^3 a_\lambda-\frac{13437}{1000} a_1^3 a_\lambda \zeta_3-\frac{839889}{128000} a_1^4\nonumber\\
&	+\frac{336339}{80000} a_1^4 \zeta_3-474 a_2 a_\lambda^3+72 a_2 a_\lambda^3 \zeta_3-\frac{999}{5} a_1 a_2 a_\lambda^2-\frac{486}{5} a_1 a_2 a_\lambda^2 \zeta_3+\frac{13941}{200} a_1^2 a_2 a_\lambda-\frac{1323}{200} a_1^2 a_2 a_\lambda \zeta_3-\frac{237787}{32000} a_1^3 a_2\nonumber\\
&	+\frac{19593}{4000} a_1^3 a_2 \zeta_3-\frac{1389}{8} a_2^2 a_\lambda^2-513 a_2^2 a_\lambda^2 \zeta_3+\frac{18411}{160} a_1 a_2^2 a_\lambda-\frac{1179}{40} a_1 a_2^2 a_\lambda \zeta_3-\frac{81509}{9600} a_1^2 a_2^2+\frac{19953}{1600} a_1^2 a_2^2 \zeta_3+\frac{58031}{288} a_2^3 a_\lambda\nonumber\\
&	+\frac{4419}{8} a_2^3 a_\lambda \zeta_3-\frac{33133}{1152} a_1 a_2^3-\frac{243}{32} a_1 a_2^3 \zeta_3+\frac{228259}{3072} a_2^4-\frac{20061}{128} a_2^4 \zeta_3+\frac{297}{10} a_1^2 a_3 a_\lambda-\frac{792}{25} a_1^2 a_3 a_\lambda \zeta_3-\frac{5049}{1000} a_1^3 a_3\nonumber\\
&	+\frac{594}{125} a_1^3 a_3 \zeta_3-\frac{1683}{200} a_1^2 a_2 a_3+\frac{198}{25} a_1^2 a_2 a_3 \zeta_3+\frac{405}{2} a_2^2 a_3 a_\lambda-216 a_2^2 a_3 a_\lambda \zeta_3-\frac{459}{40} a_1 a_2^2 a_3+\frac{54}{5} a_1 a_2^2 a_3 \zeta_3-\frac{459}{8} a_2^3 a_3\nonumber\\
&	+54 a_2^3 a_3 \zeta_3+873 a_t a_\lambda^3-\frac{117}{4} a_1 a_t a_\lambda^2-\frac{144}{5} a_1 a_t a_\lambda^2 \zeta_3-\frac{203887}{4800} a_1^2 a_t a_\lambda-\frac{1347}{50} a_1^2 a_t a_\lambda \zeta_3+\frac{376509}{32000} a_1^3 a_t-\frac{27}{50} a_1^3 a_t \zeta_3\nonumber\\
&	+\frac{639}{4} a_2 a_t a_\lambda^2-432 a_2 a_t a_\lambda^2 \zeta_3-\frac{19527}{160} a_1 a_2 a_t a_\lambda+\frac{531}{5} a_1 a_2 a_t a_\lambda \zeta_3+\frac{76323}{6400} a_1^2 a_2 a_t-\frac{27}{25} a_1^2 a_2 a_t \zeta_3-\frac{6957}{64} a_2^2 a_t a_\lambda\nonumber\\
&	-\frac{351}{2} a_2^2 a_t a_\lambda \zeta_3+\frac{10461}{1280} a_1 a_2^2 a_t+\frac{81}{20} a_1 a_2^2 a_t \zeta_3-\frac{6849}{256} a_2^3 a_t+\frac{297}{4} a_2^3 a_t \zeta_3-1224 a_3 a_t a_\lambda^2+1152\, a_3 a_t a_\lambda^2 \zeta_3\nonumber\\
&	-\frac{2419}{30} a_1 a_3 a_t a_\lambda+\frac{408}{5} a_1 a_3 a_t a_\lambda \zeta_3+\frac{1761}{200} a_1^2 a_3 a_t-\frac{162}{25} a_1^2 a_3 a_t \zeta_3-\frac{489}{2} a_2 a_3 a_t a_\lambda+216 a_2 a_3 a_t a_\lambda \zeta_3+\frac{747}{20} a_1 a_2 a_3 a_t\nonumber\\
&	-\frac{108}{5} a_1 a_2 a_3 a_t \zeta_3+\frac{651}{8} a_2^2 a_3 a_t-54 a_2^2 a_3 a_t \zeta_3+\frac{1244}{3} a_3^2 a_t a_\lambda-48 a_3^2 a_t a_\lambda \zeta_3+\frac{1719}{2} a_t^2 a_\lambda^2+756 a_t^2 a_\lambda^2 \zeta_3-\frac{497}{8} a_1 a_t^2 a_\lambda\nonumber\\
&	+\frac{171}{5} a_1 a_t^2 a_\lambda \zeta_3+\frac{67793}{9600} a_1^2 a_t^2+\frac{2957}{400} a_1^2 a_t^2 \zeta_3-\frac{4977}{8} a_2 a_t^2 a_\lambda+513 a_2 a_t^2 a_\lambda \zeta_3-\frac{1079}{320} a_1 a_2 a_t^2-\frac{2229}{40} a_1 a_2 a_t^2 \zeta_3\nonumber\\
&	+\frac{9909}{128} a_2^2 a_t^2-\frac{819}{16} a_2^2 a_t^2 \zeta_3+895 a_3 a_t^2 a_\lambda-1296 a_3 a_t^2 a_\lambda \zeta_3+\frac{931}{30} a_1 a_3 a_t^2-\frac{56}{5} a_1 a_3 a_t^2 \zeta_3-\frac{31}{2} a_2 a_3 a_t^2+24 a_2 a_3 a_t^2 \zeta_3\nonumber\\
&	-\frac{266}{3} a_3^2 a_t^2+32 a_3^2 a_t^2 \zeta_3+\frac{117}{8} a_t^3 a_\lambda-198 a_t^3 a_\lambda \zeta_3+\frac{3467}{160} a_1 a_t^3+\frac{51}{5} a_1 a_t^3 \zeta_3+\frac{3411}{32} a_2 a_t^3-27 a_2 a_t^3 \zeta_3-38 a_3 a_t^3\nonumber\\
&	+240 a_3 a_t^3 \zeta_3-\frac{1599}{8} a_t^4-36 a_t^4 \zeta_3+873 a_b a_\lambda^3+\frac{1251}{20} a_1 a_b a_\lambda^2-\frac{576}{5} a_1 a_b a_\lambda^2 \zeta_3-\frac{149623}{4800} a_1^2 a_b a_\lambda-\frac{141}{50} a_1^2 a_b a_\lambda \zeta_3\nonumber\\
&	+\frac{145569}{32000} a_1^3 a_b+\frac{27}{100} a_1^3 a_b \zeta_3+\frac{639}{4} a_2 a_b a_\lambda^2-432 a_2 a_b a_\lambda^2 \zeta_3-\frac{9027}{160} a_1 a_2 a_b a_\lambda+\frac{36}{5} a_1 a_2 a_b a_\lambda \zeta_3+\frac{68427}{6400} a_1^2 a_2 a_b\nonumber\\
&	+\frac{81}{50} a_1^2 a_2 a_b \zeta_3-\frac{6957}{64} a_2^2 a_b a_\lambda-\frac{351}{2} a_2^2 a_b a_\lambda \zeta_3+\frac{18297}{1280} a_1 a_2^2 a_b+\frac{27}{5} a_1 a_2^2 a_b \zeta_3-\frac{6849}{256} a_2^3 a_b+\frac{297}{4} a_2^3 a_b \zeta_3\nonumber\\
&	-1224 a_3 a_b a_\lambda^2+1152 a_3 a_b a_\lambda^2 \zeta_3-\frac{991}{30} a_1 a_3 a_b a_\lambda+24 a_1 a_3 a_b a_\lambda \zeta_3+\frac{2049}{200} a_1^2 a_3 a_b-\frac{162}{25} a_1^2 a_3 a_b \zeta_3-\frac{489}{2} a_2 a_3 a_b a_\lambda\nonumber\\
&	+216 a_2 a_3 a_b a_\lambda \zeta_3+\frac{699}{20} a_1 a_2 a_3 a_b-\frac{108}{5} a_1 a_2 a_3 a_b \zeta_3+\frac{651}{8} a_2^2 a_3 a_b-54 a_2^2 a_3 a_b \zeta_3+\frac{1244}{3} a_3^2 a_b a_\lambda-48 a_3^2 a_b a_\lambda \zeta_3\nonumber\\
&	+117 a_t a_b a_\lambda^2-864 a_t a_b a_\lambda^2 \zeta_3-\frac{929}{20} a_1 a_t a_b a_\lambda-\frac{6}{5} a_1 a_t a_b a_\lambda \zeta_3-\frac{6381}{1600} a_1^2 a_t a_b-\frac{9}{25} a_1^2 a_t a_b \zeta_3-\frac{531}{4} a_2 a_t a_b a_\lambda\nonumber\\
&	+54 a_2 a_t a_b a_\lambda \zeta_3+\frac{1001}{160} a_1 a_2 a_t a_b+\frac{93}{10} a_1 a_2 a_t a_b \zeta_3-\frac{2655}{64} a_2^2 a_t a_b+\frac{117}{2} a_2^2 a_t a_b \zeta_3+82 a_3 a_t a_b a_\lambda-96 a_3 a_t a_b a_\lambda \zeta_3\nonumber\\
&	-8 a_2 a_3 a_t a_b+96 a_2 a_3 a_t a_b \zeta_3+192 a_3^2 a_t a_b+\frac{6399}{8} a_t^2 a_b a_\lambda+144 a_t^2 a_b a_\lambda \zeta_3+\frac{1337}{160} a_1 a_t^2 a_b-\frac{84}{5} a_1 a_t^2 a_b \zeta_3\nonumber\\
&	+\frac{477}{32} a_2 a_t^2 a_b-2 a_3 a_t^2 a_b-48 a_3 a_t^2 a_b \zeta_3-\frac{717}{8} a_t^3 a_b-36 a_t^3 a_b \zeta_3+\frac{1719}{2} a_b^2 a_\lambda^2+756 a_b^2 a_\lambda^2 \zeta_3-\frac{5737}{40} a_1 a_b^2 a_\lambda\nonumber\\
&	+\frac{747}{5} a_1 a_b^2 a_\lambda \zeta_3-\frac{104383}{9600} a_1^2 a_b^2-\frac{407}{80} a_1^2 a_b^2 \zeta_3-\frac{4977}{8} a_2 a_b^2 a_\lambda+513 a_2 a_b^2 a_\lambda \zeta_3-\frac{3239}{320} a_1 a_2 a_b^2-\frac{933}{40} a_1 a_2 a_b^2 \zeta_3\nonumber\\
&	+\frac{9909}{128} a_2^2 a_b^2-\frac{819}{16} a_2^2 a_b^2 \zeta_3+895 a_3 a_b^2 a_\lambda-1296 a_3 a_b^2 a_\lambda \zeta_3-\frac{641}{30} a_1 a_3 a_b^2+\frac{136}{5} a_1 a_3 a_b^2 \zeta_3-\frac{31}{2} a_2 a_3 a_b^2\nonumber\\
&	+24 a_2 a_3 a_b^2 \zeta_3-\frac{266}{3} a_3^2 a_b^2+32 a_3^2 a_b^2 \zeta_3+\frac{6399}{8} a_t a_b^2 a_\lambda+144 a_t a_b^2 a_\lambda \zeta_3-\frac{2299}{160} a_1 a_t a_b^2+\frac{78}{5} a_1 a_t a_b^2 \zeta_3+\frac{477}{32} a_2 a_t a_b^2\nonumber\\
&	-2 a_3 a_t a_b^2-48 a_3 a_t a_b^2 \zeta_3+72 a_t^2 a_b^2 \zeta_3+\frac{117}{8} a_b^3 a_\lambda-198 a_b^3 a_\lambda \zeta_3+\frac{5111}{160} a_1 a_b^3-15 a_1 a_b^3 \zeta_3+\frac{3411}{32} a_2 a_b^3\nonumber\\
&	-27 a_2 a_b^3 \zeta_3-38 a_3 a_b^3+240 a_3 a_b^3 \zeta_3-\frac{717}{8} a_t a_b^3-36 a_t a_b^3 \zeta_3-\frac{1599}{8} a_b^4-36 a_b^4 \zeta_3.
\label{eq:alambdabigbigformula}
\end{align}
\endgroup

\bibliography{review_rge_sm_v3}
\end{document}